\documentclass[final,number,sort&compress,5p,times,twocolumn]{elsarticle}

\journal{Nucl.\ Instr.\ and Meth.\ A}

\usepackage{color} 
\usepackage{subcaption}
\usepackage{wasysym}
\usepackage{graphicx}
\usepackage{hyperref}
\usepackage{xcolor}

\usepackage{url}

\begin{document}
\begin{frontmatter}

\title{Texas Active Target (TexAT) Detector for Experiments With Rare Isotope Beams}
    
 \author[Cyclotron]{E.~Koshchiy}
 \ead{koshchiy@tamu.edu}
  
 \author[Cyclotron,TAMU] {G.V.~Rogachev}
 \ead{rogachev@tamu.edu}
 
 \author[IRFU]{E.~Pollacco}

 \author[Cyclotron]{S.~Ahn}
 
 \author[Cyclotron]{E.~Uberseder}
 
 \author[Cyclotron,TAMU]{ J.~Hooker\fnref{fn1}}
 \fntext[fn1] {Present address: Department of Physics and Astronomy, University of Tennessee, Knoxville, Tennessee 37996, USA}

 \author[Cyclotron,TAMU]{J.~Bishop}
 
 \author[Cyclotron,TAMU]{E.~Aboud}
 
 \author[Cyclotron]{M.~Barbui}
 
 \author[Cyclotron]{V.Z.~Goldberg}
 
 \author[Cyclotron,TAMU]{C.~Hunt}
  
 \author[Cyclotron,TAMU]{H.~Jayatissa}
 
 \author[Cyclotron,TAMU]{C.~Magana}
 
 \author[Cyclotron,TAMU]{R.~O'Dwyer}
 
 \author[Cyclotron]{B.T.~Roeder}
 
 \author[Cyclotron]{A.~Saastamoinen}
  
 \author[Cyclotron,TAMU]{S.~Upadhyayula}

  \address[Cyclotron]{Cyclotron Institute, Texas A\&M University, College Station, TX 77843, USA}
  
  \address[TAMU]{Department of Physics \& Astronomy, Texas A\&M University, College Station, TX 77843, USA}
  
  \address[IRFU]{IRFU, CEA, Saclay, Gif-Sur-Ivette, France}

  \begin{abstract}
  
The TexAT (Texas Active Target) detector is a new active-target time projection chamber (TPC) that was built at the Cyclotron 
Institute Texas A$\&$M University. The detector is designed to be of general use for nuclear structure and nuclear astrophysics 
experiments with rare isotope beams. TexAT combines a highly segmented Time Projection Chamber (TPC) with two layers of 
solid state detectors. It provides high efficiency and flexibility for experiments with low intensity exotic beams, allowing for 
the 3D track reconstruction of the incoming and outgoing particles involved in nuclear reactions and decays.



\end{abstract}
\end{frontmatter}

\section{Introduction}

Over the last 20-30 years, experiments with rare isotope beams (RIBs) developed from being
exotic undertakings in a select few laboratories into the main stream of nuclear science. RIBs provide a
pathway to venture far beyond the constrains typically encountered in experiments with stable beams.
RIBs open up an opportunity to study very exotic nuclei using relatively simple and well understood
reactions, such as elastic and inelastic scattering, one/two nucleon transfer, Coulomb excitation, etc. They
allow key reaction rates to be measured that are relevant for explosive processes in astrophysics with
radioactive nuclei. All these benefits come at a price. The typical intensity of RIBs is many orders of
magnitude lower than that of stable beams. Therefore, efficiency of the experimental setup becomes
the determining factor for RIBs experiments. One of the most efficient experimental approaches that can be
used with RIBs is utilizing an active target detector. These detectors can be designed to have almost 4$\pi$ solid angle
coverage, and they naturally allow the use of thick targets without loss of energy resolution. Thick targets also
allows the measurement of excitation functions without the need to change the beam energy \cite{Goldberg}. 
The versatility
of these devices for RIB experiments has been recognized and many nuclear physics laboratories around
the world are in the process of constructing and using the general purpose devices (TACTIC at U of York/TRIUMF \cite{TACTIC},
MAYA at GANIL/TRIUMF \cite{MAYA}, AT-TPC at NSCL \cite{AT-TPC}, ACTAR at GANIL/GSI \cite{ACTAR}, ANASEN at
FSU/LSU/NSCL \cite{ANASEN}, MUSIC \cite{MUSIC}, MSTPC  \cite{MSTPC}  and others). Some TPCs aimed at specific processes
e.g., 2p - decay  \cite{Cwiok}, or the only optical readout active target TPC for studies with gamma-ray beams \cite{Gai}. 
A comprehensive overview on the subject can be found in \cite{Beceiro},  \cite{TPC_Review},   \cite{Ayyad2018}.

	While the concept is similar for all active target detectors (the target material is spatially extended and ``active'' to allow
for tracking of the reaction products), the specific implementation may be very different depending on the energy range, 
type and quality of RIBs characteristic of the particular facility.
We designed and constructed a general-purpose active target detector (Texas Active Target, TexAT) aimed at low energy  nuclear 
physics experiments with rare isotope beams produced by either the MARS recoil separator \cite{MARS}, 
 or the new reaccelerated beam facility currently under development at the Cyclotron Institute Texas A$\&$M University. 
TexAT can be used for a wide variety of experiments to detect the charged products of nuclear reactions with rare isotope beams. 
In an ``active'' target mode, the nuclei from the detector gas mixture act as a target. Resonant elastic and inelastic scattering
of protons and $\alpha$ -particles, ($\alpha$,p) and (p,$\alpha$) reactions, nucleon-transfer reactions, such as (d,p), 
(d,$^3$He), (p,d), (p,t), ($\alpha$,t), as well as a fusion-evaporation reactions are examples of the experiments that can be performed 
with TexAT. TexAT can also be used for $\beta$-delayed charged particle emission experiments with beams of radioactive ions.

\section{Overview of TexAT design and main components}
	
\subsection{General design}
		
TexAT consists of a gas-filled Time Projection Chamber (TPC) and two layers of solid states detectors which surround the TPC in Cartesian geometry.	
  	 	
TPCs with a high level of segmentation in the readout plane provide 3D- vertex and tracking capabilities for both incoming beam ions 
and charged reaction products and allows for information on the energy losses for each track in the gas.

Solid state detectors contribute an additional particle ID and the total energy information for the ions that escape the gas volume. 
TexAT has been designed to be able to operate in a rather broad energy range: from sub-MeV to tens MeV/nucleon and for some 
experiments the energy of protons can be as high as $\sim$100 MeV. 

The typical thickness of high-quality  transient Si detectors available on the market is about 1 mm.
 They have been used to form an inner ($\Delta$E)- layer of the solid- state detector wall.
 
To stop the light ions with energies greater 
than the punch-through energies for the Si detectors, a thick detector is required. When choosing the total absorption detectors we focused
 on using commercially available and affordable thick detectors of large area. Currently, the thickest  semiconductor detectors for charged 
 particles are manufactured using the so called lithium-drift technology. For example, 30 mm in diameter, 12 mm  thick coaxial type Si(Li)- 
  detector for charged  particle detection have been manufactured  and investigated in \cite{SiLi}.

 Still, large (50mm x 50mm) Si(Li)- detectors of the required thickness ($\sim$10 mm) are very expensive and require special care.
  Moreover, in some of the experiments envisioned with TexAT, protons with energies as high as 50 MeV may need to be detected. Even 10 mm 
  Si would not be sufficient for these applications. An additional consideration is that CsI(Tl) scintillators are efficient $\gamma$- detectors, that can 
  be used in experiments that require charged particle-$\gamma$ coincidence with TexAT. Therefore, for both practical and scientific reasons
   we have decided to use  the 40 mm thick scintillation CsI(Tl) to form the outer wall backing  Si detectors.

\begin{figure}[htb!]
\centering
  \includegraphics[width=1.0\columnwidth]{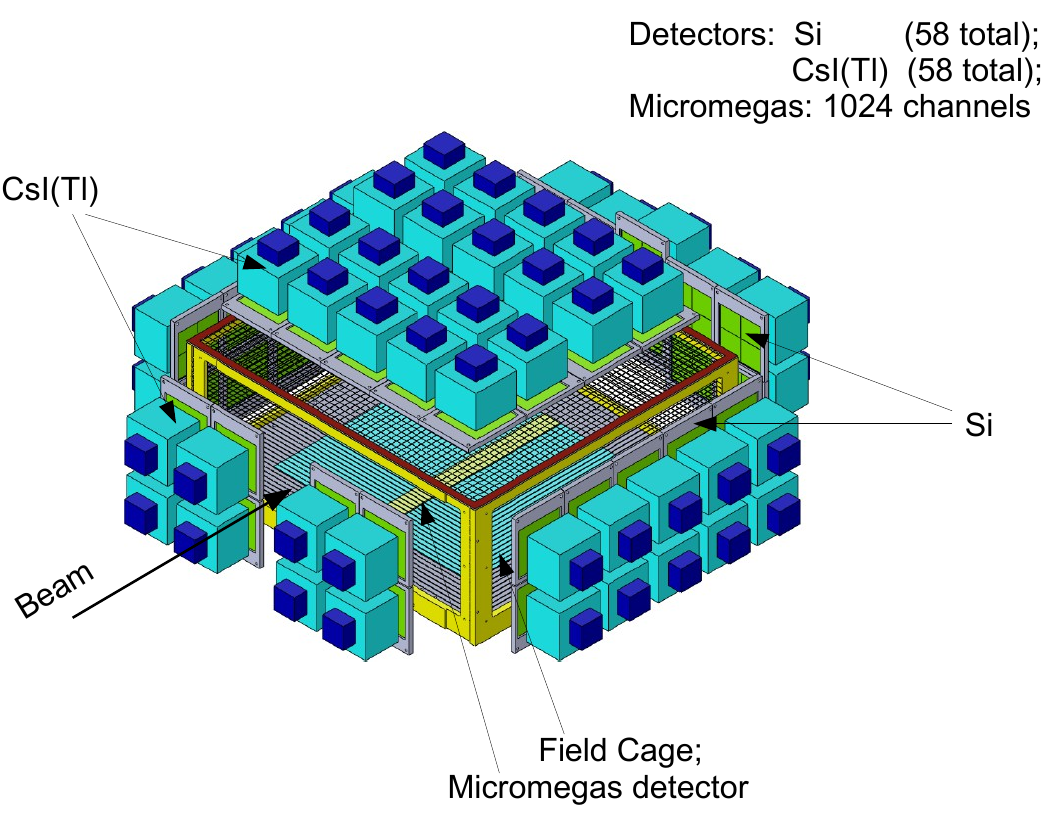}
    \caption{Sketch of the TexAT Assembly. Detailed description of the individual components is provided in the text.} 
   \label{fig:TexAT_Assembly}
\end{figure}

The 3D-rendered representation of TexAT is shown in Fig.\ref{fig:TexAT_Assembly}. In the full configuration, the TPC is surrounded 
by an array of 58 Si-CsI(Tl) telescope detectors: 8 at the ``upstream''/``beam'' - side, 10 at the ``downstream'' - and both 
``beam- left'' and ``beam- right''- sides (30 total)  and 20 at the ``bottom'' -  of TPC side (shown on top in Fig. \ref{fig:TexAT_Assembly}).
		
The length of the active area of TPC is 224 mm, an active volume is about 5,000 cm$^3$, and the total solid angle covered by Si-CsI(Tl) 
telescopes is about 3$\pi$, providing high efficiency for experiments with low intensity exotic beams. The detailed description of 
individual components is provided below.

\subsubsection{Time Projection Chamber}
	
The two main components of the TexAT TPC are the Micromegas detector and a field cage.

The key element is the Micromegas (Micro-MEsh GAseous Structures) detector \cite{Micromegas_1, Micromegas_2}. It consists of a highly segmented
 anode readout board with a woven  stainless steel mesh stretched above it. Such geometry creates a parallel- plate gas- electron
  amplification volume, where the primary electrons, generated by ionization in the active ``drift'' volume of the detector are amplified. 
  It is basically an ``ideal'' parallel- plate gas-electron proportional counter,  where the micro-mesh side is rendered approximately  95$\%$
   transparent to electrons.  This is because the field between  the micromesh and the readout plane is of the order of 30kV/cm and considerably
    higher than the drift field above the mesh.
A field cage provides a uniform electric field in the active volume of a TPC. To ensure an operation in “drift” mode  
a reduced field strength value E/P has to be smaller than 20 V/cm/Torr (see, for example \cite{Blum}). 
For all of the experiments performed with TexAT the reduced field strength  is between  0.05 V/cm/Torr and 5 V/cm/Torr.

A detailed description of the field cage is given in Sec. \ref{FieldCage}.

The custom Micromegas detector for TexAT was designed in collaboration with IRFU/DPhN Saclay (France) and fabricated at CERN using
``bulk''- technology: the readout board, a photo-resist layer (128 $\mu$m thick) and the cloth mesh (stainless- steel wires of 18 $\mu$m diameter
 interwoven at a pitch of 63 $\mu$m) were encapsulated together and then etched, producing the embedded 300 $\mu$m photoresist 
 pillars in the mesh every 5000 $\mu$m in one single process \cite{GIOMATARIS2006}. 

The amplification (avalanche) gap of 128 $\mu$m between anode pads and embedded mesh allows for gas gains of up to 
the 10$^{4}$-10$^{5}$. For the stable operation of detector in proportional mode the total charge of avalanche should be  below the so-called
Raether limit (\cite{Raether})  before the gas discharge occures. The Raether limit depends on detector geometry and gas composition.
It was shown (see \cite{Peskov}) that the Raether limit for Micromegas detectors and majority of gases  is at the level of 10$^{7}$ electrons. For 
TexAT detector the high gain is required to  reliably detect protons, deutrons and tritons at “side”- areas (see Section \ref{PCB}) of TPC.
The minimum  possible energy losses of protons for typical experiments with TexAT  are   $\sim$keV/mm, producing few dozens of electron-ion pairs per mm on average. Thus, with typical gains of about 10$^{4}$-10$^{5}$, a proton can create a total charge of avalanche at the level of 10$^{6}$-10$^{7}$ electrons, which is below (or close to) the Raether limit, but sufficient to observe proton tracks.

\subsubsection{Micromegas Detector Readout Board Configuration \label{PCB}}
	
\begin{figure}[hbt!]
    \centering
     \includegraphics[width=1.0\columnwidth]{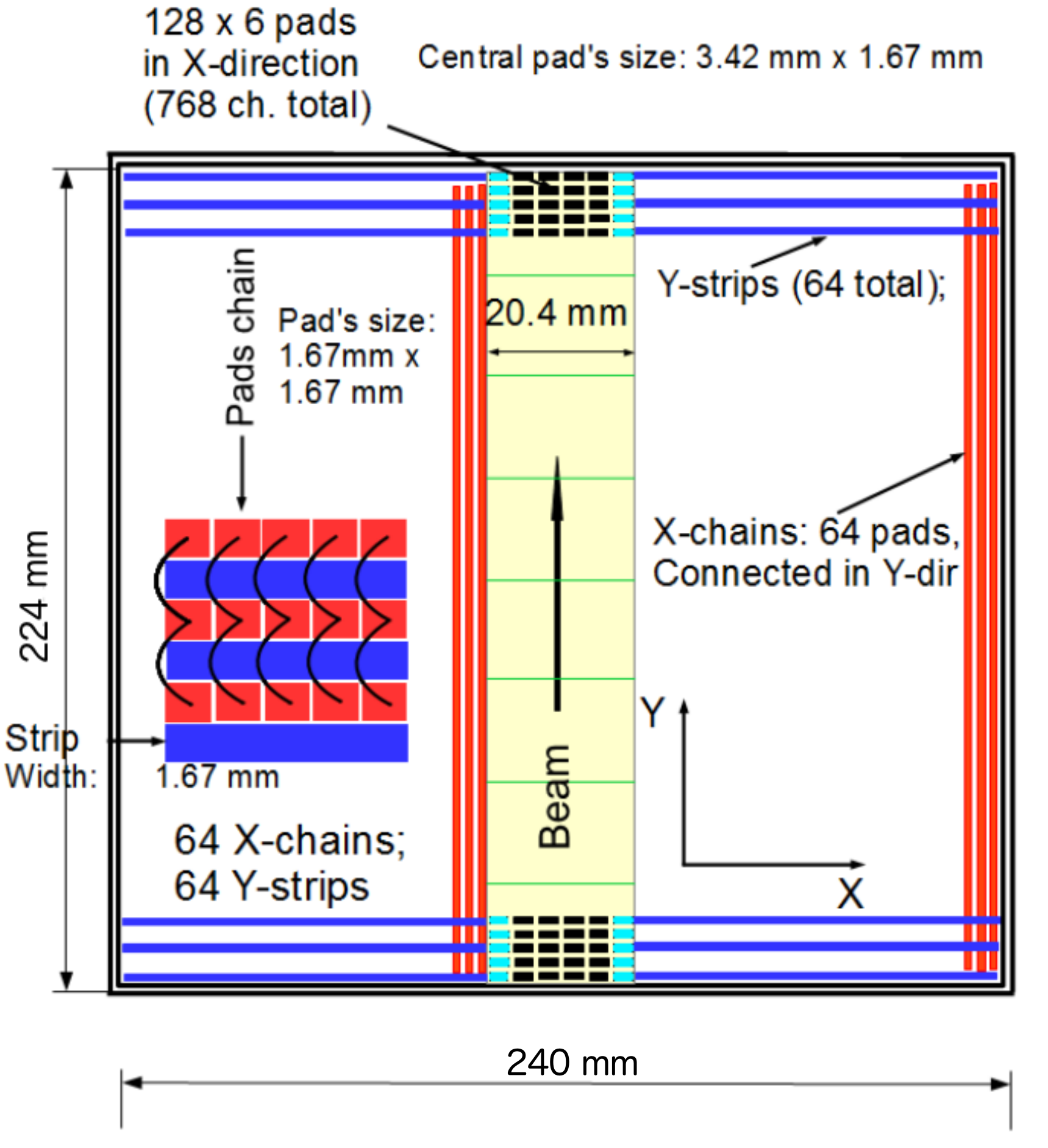}
    \includegraphics[width=1.0\columnwidth]{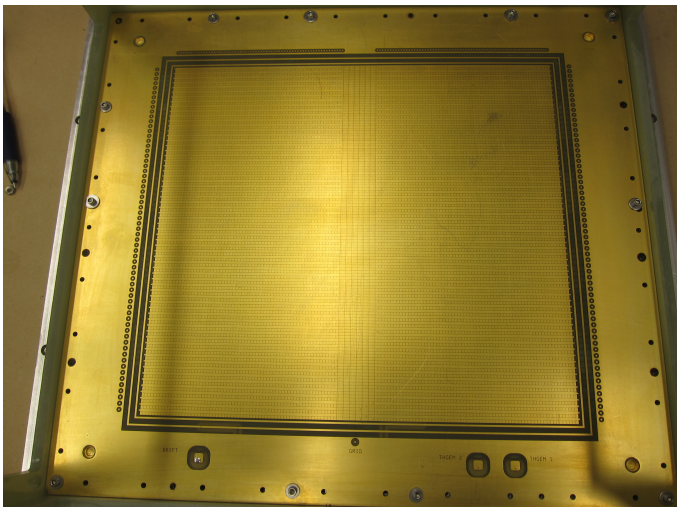}
    \caption{Top panel: The design of an active area of Micromegas PCB. The central region consists of 6 columns with the cyan pads 
    on either side corresponding to side-pads. The strips in the side planes are the solid readout pads that are perpendicular to the 
    beam axis while the chains are the square readout pads in the side regions that go along the beam axis. Bottom panel: Picture of 
    the readout board.}
        \label{fig:MMBoardLayout}
\end{figure}

A rectangular (346$\times$316 mm$^2$) Micromegas PCB has an active area of 224$\times$240 mm$^2$. To reduce the total 
number of channels a unique segmentation/multiplexing scheme was conceived, consisting of rectangular pads in the beam 
region and overlapping strips and chains to the left and right of the central region. The total number of Micromegas readout 
channels is 1,024. The board is divided into three areas as shown in Fig. ~\ref{fig:MMBoardLayout}: the zone to the left of a 
beam axis (L), the beam axis zone (C) and the zone to the right of the beam axis (R).

The central zone (shown in yellow in Fig.~\ref{fig:MMBoardLayout}, top panel) along the beam axis is 20.4 mm wide and has high 
segmentation: 128$\times$6=768 individual pads (the pad size is 1.67$\times$3.42 mm$^2$). The pads are arranged so that there 
are 128 pads in the direction of the beam (along 224 mm side) and 6 pads in the direction perpendicular to the beam.

Both left (L) and right (R) zones are identical and have dimensions of 224$\times$101.5 mm$^2$. Each consists of 64 strips, 
perpendicular to the beam axis. The strip is 1.67$\times$101.5 mm$^2$. There is 1.75 mm of space between strips to allow for
 an individual pad to be placed between the strips. The distance between the centers of the strips is 3.5 mm. There are 64 individual 
 1.67$\times$1.67 mm$^2$ pads between each two strips. The pads located at the same distance from the beam axis are connected 
 into chains, so that there are 64 ``chains'' of 64 interconnected pads arranged in the direction parallel to the beams axis in each of 
 the side zones. Although the strips and chains individually cannot provide an accurate location for tracking (because each strip and 
 chain run along either the beam axis or perpendicular to the beam axis) they can be matched together to make a 2D point (see 
 Section \ref{CSmatching}). A picture of the TexAT Micromegas readout board without the mesh (the blank) is shown in the bottom panel 
 of Fig.~\ref{fig:MMBoardLayout}.
	
The signal readout of Micromegas board is carried out through the 80-pin (only 64 channels are active) high density (0.8 mm pitch) Edge
 Rate\textsuperscript{\textregistered} Rugged High-Speed {SAMTEC} connectors. An optimized signal readout channel map has been 
 designed. According to this map, the PCB has been segmented into different geometric zones of 64 channels:
\begin{itemize}
  \item the central (``beam") area (4 most central pads in each of 128 rows, 512 pads total) is divided to eight zones (shown by 
  green lines in the top panel of Fig.~\ref{fig:MMBoardLayout}.
  \item the left and right side pads of central area form two columns of 2 $\times$64 pads each (shown by cyan in the top panel of 
  Fig.~\ref{fig:MMBoardLayout}).
  \item zone (L) and (R) are parted in the following way:  the group of 32 upstream strips and 32 far from the beam area chains combine 
  into single readout zone; the 32 downstream (top at  Fig.~\ref{fig:MMBoardLayout} strips and 32 close to the center chains make another one.
  \end{itemize}

Following the form-factor of the readout electronics, described in Section \ref{GET}, the zone's readout is organized in a specified
 way: 2$\times$256 channels from the most central area; 256 channels from the left zone (2 x 64 strips and chains) and left side 
 columns of the central zone (2 x 64 pads), and 256 channels from the right zone (2 x 64 chips and chains) and right side columns 
 of the central zone (2 x 64 pads).

Such segmentation facilitates the signal routing and allows to individually bias pads in the separate zones, giving a possibility to create 
avalanche areas with different gas gains within the single Micromegas detector. It will be shown below (see Sec. \ref{lowANDhigh}) 
how useful this feature is.

\subsection{Field cage \label{FieldCage}}

Transparent for the beam ions and reaction products, the field cage maintains the uniform electric field inside a TPC volume with 
sufficient strength to provide a steady drift of electrons. The basic geometry of the cage surrounding the TPC is constrained by 
the Micromegas' readout board design and the solid state detector's wall design: 316 mm (L) x 346 mm (W) x 135mm (H). A 50 
$\mu$m (diameter) gold plated tungsten wire has been chosen as the main element for the cage. The conversion drift volume is 
located between a transparent cathode, which is normally set to the high ``negative'' potential ($\>$ -1000  $\div$   - 3000 V) 
and a grounded mesh, transparent for electrons. 

The electrons released in a process of gas ionization by charged products drift  toward the mesh and amplified at the Micromegas in avalanches  (the mechanism of gas ionization is described in detail,  for example in \cite{Blum}.)                         
 
The homogeneous electric field is supported by the set of guide wires, 
stretched along the walls perimeter, stepped down from the negative potential of the cathode to the anode continuously by voltage 
division using a series of 25 M$\Omega$ resistors to create a uniform electric field. The resistors were hand-picked within a tolerance of 0.5 $\%$.		 	 
	 
\begin{figure}[hbt!]
    \centering
    \includegraphics[width=1.0\columnwidth]{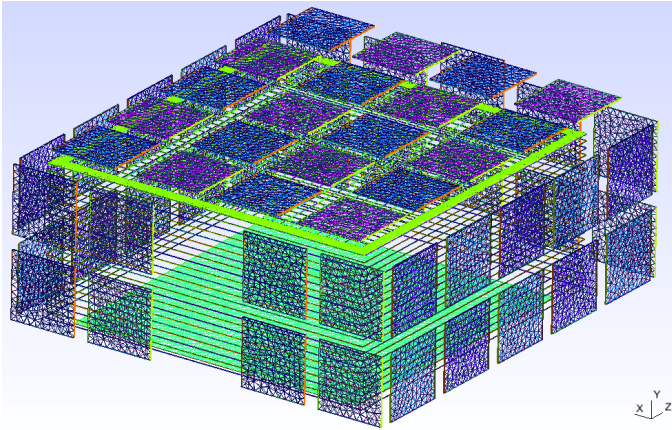}	
    \caption{Simulated model of field cage for TexAT.}
     \label{fig:FieldCageSim} 
\end{figure}

A 3D- model of the TexAT field cage (shown in Figure~\ref{fig:FieldCageSim}) has been created using a finite element mesh generator 
GMSH\cite{GMSH}. To estimate the level of electric field uniformity and optimize wires spacing to create the best condition for electron 
drift inside the TPC, detailed simulations with computer codes GARFIELD \cite{GARFIELD} interfaced with Elmer \cite{Elmer} finite element 
analysis code have been performed.

\begin{figure}[hbt!]
    \centering
 \includegraphics[width=1.0\columnwidth]{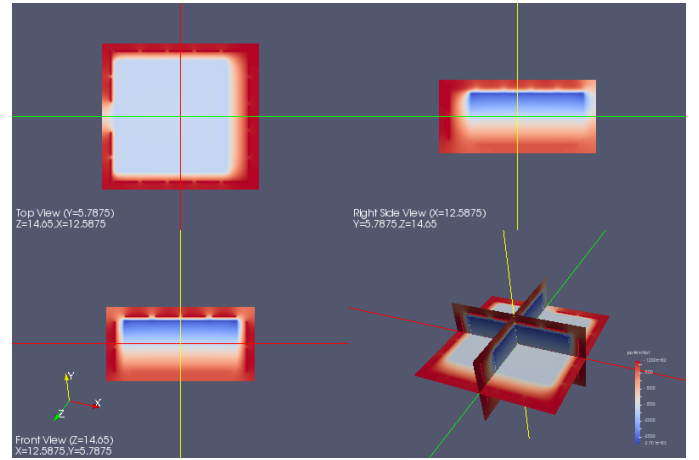}	
    \caption{Electric field simulation.}
     \label{fig:Field3D} 
\end{figure}

The Elmer Solver's Finite Element Analysis provides a map of the electric field at every point in the TPC. A simulated 3D-plot of the 
electrostatic field map inside TexAT TPC (the front of surrounding Si detectors is grounded) is shown in Figure ~\ref{fig:Field3D}. One 
can see that computed field appears to be relatively steady inside the active TPC volume while some deviations are manifested in the 
areas near boundaries of the cage and close to Micromegas mesh. The aberrations of field uniformity are apparently induced by the 
leak of the electric field through the transparent cage and mesh. 

\begin{figure}[hbt!]
    \centering
 \includegraphics[width=1.0\columnwidth]{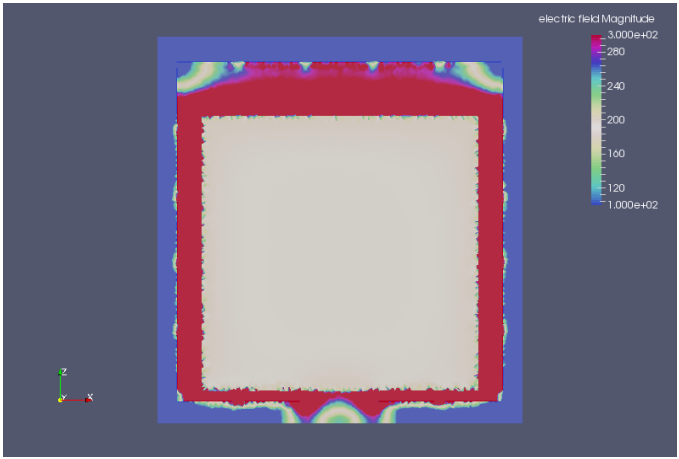}	
    \caption{Electric field simulation: Top  (XZ plane) view.}
     \label{fig:FieldXZ} 
\end{figure}

\begin{figure}[hbt!]
    \centering
 \includegraphics[width=1.0\columnwidth]{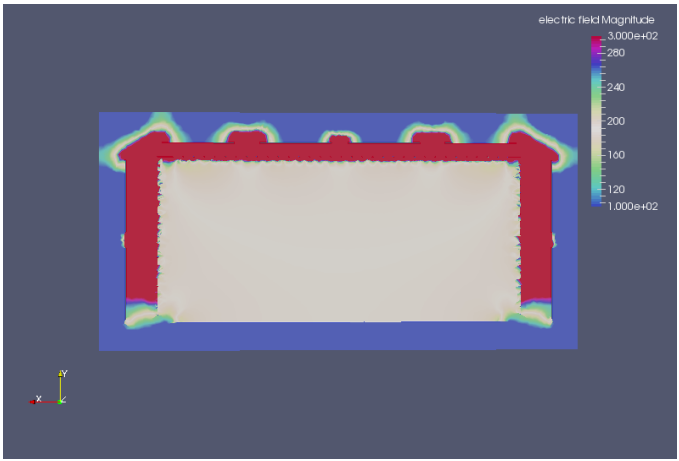}	
    \caption{Electric field simulation: Side (XY plane) view.}
     \label{fig:FieldXY} 
\end{figure}

The simulation of ``realistic'' configuration with the front side of Si detectors is under a full depletion potential of -200 V is shown 
in Figs. ~\ref{fig:FieldXZ} and \ref{fig:FieldXY}. A particular geometry of surrounding detectors enhances non-uniformity effect, especially 
at the ``downstream"-side wall, that has a non-symmetric geometry (see Section \ref{Si} below).

The ``edge'' effects distort the electric force lines near the  walls and create a dependence on the position of the track. These deviations 
can be reduced to some extend by increasing the guide wires density, but a compromise between a transparency factor, acceptable level 
of uniformity, and a practical fabrication has to be established. It has been determined, that the wire spacing of 5 mm provides a sufficient 
level of electric field uniformity inside the active volume of TPC. The 5 mm spacing allows for wire to be mounted on extension springs to 
avoid wire sagging, the technique developed for the ANASEN detector \cite{ANASEN}.

\begin{figure}[hbt!]
    \centering
   \includegraphics[width=1.0\columnwidth]{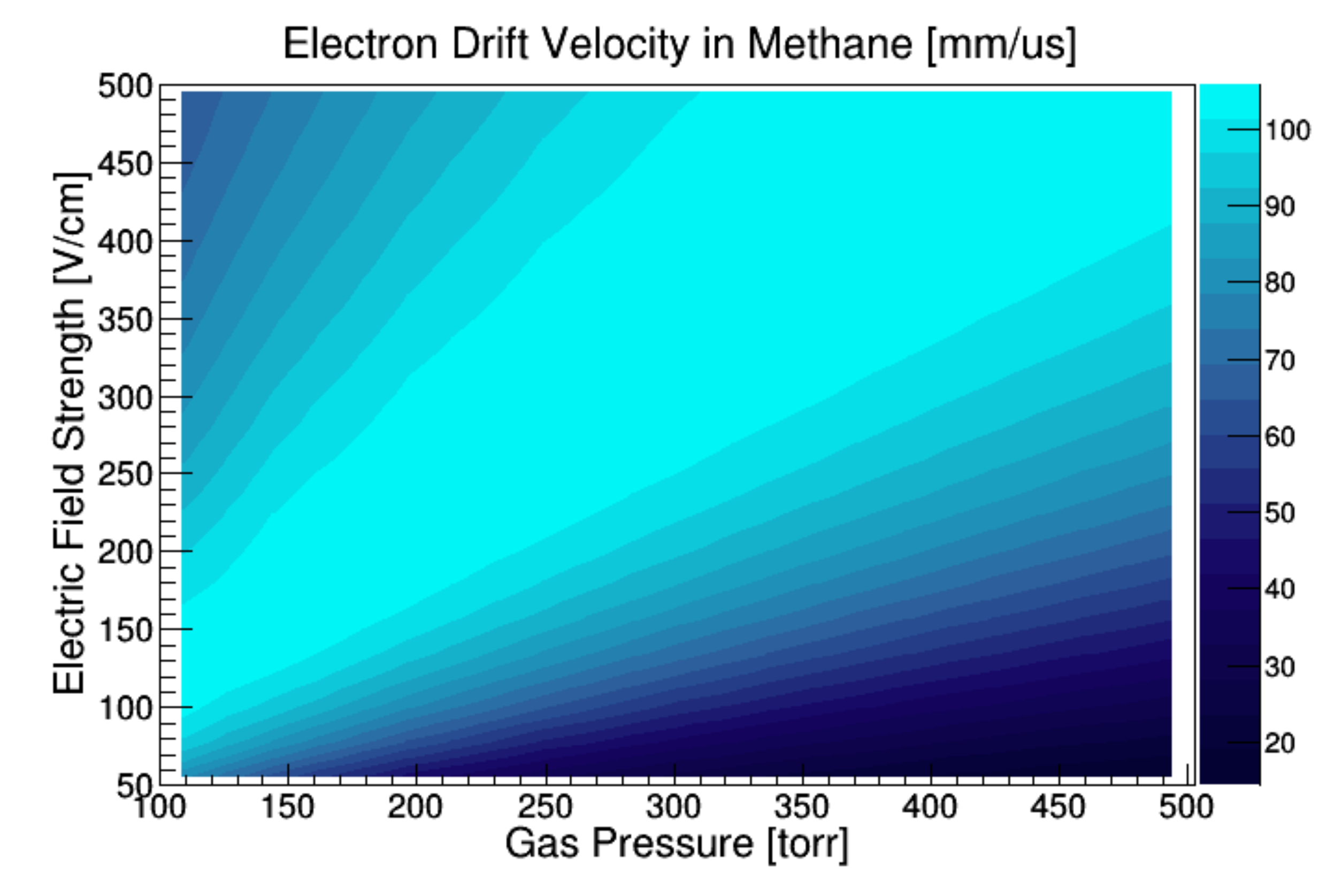}	
    \caption{Electron drift time for different gas pressure and field strength.}
     \label{fig:FieldVsP} 
\end{figure}

The leak of positive potential on the pads through the transparent grounded mesh leads to the perceptible distortion of electric field 
(at the level of 10 $\div$ 15$\%$) near the mesh, which separates the ``drift'' volume from the ``avalanche'' region. This effect is 
unavoidable because of particular geometry, but further computation of electron transport in different gases (methane, butane, Ar/CO$_2$, 
He/CO$_2$ mixtures) using MAGBOLTZ \cite{MAGBOLTZ} have defined the optimal conditions for constant electron drift velocity even in the 
presence of electric field aberrations. One example of such simulations is shown in Fig. ~\ref{fig:FieldVsP} for Methane gas at different pressure. 
The ``navy blue''- colored (shown online) area at the plot represents the optimal conditions for electron drift.
 
\begin{figure}[hbt!]
    \centering
    \includegraphics[width=7cm]{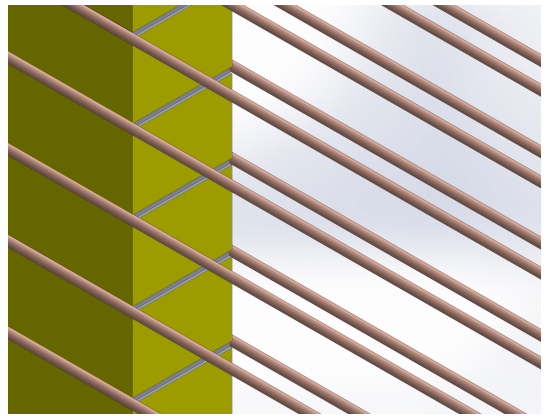}
    \caption{Rendering representation of double layer grid of TexAT field cage.  The grey lines on the field cage frame are  shown to indicate 
    the double-wire "plain". }
  \label{fig:DoubleGrid}
\end{figure}

Another factor, that may deform an electric field is an ``axial" geometry of the guide wires, instead of a classic ``plane" guides. A horizontal 
electric field between the field cage and the walls leaks in the drift region, deforming the electron drift trajectory. To improve linearity near the 
edges the second set of wires was introduced at a horizontal distance of 5 mm from the first set, as shown in Fig.~\ref{fig:DoubleGrid}. This 
second layer establishes the flat iso-potential ``plain'', reducing effects of electric field distortions near the edges.

\begin{figure}[hbt!]
  \centering
\includegraphics[width=1.0\columnwidth]{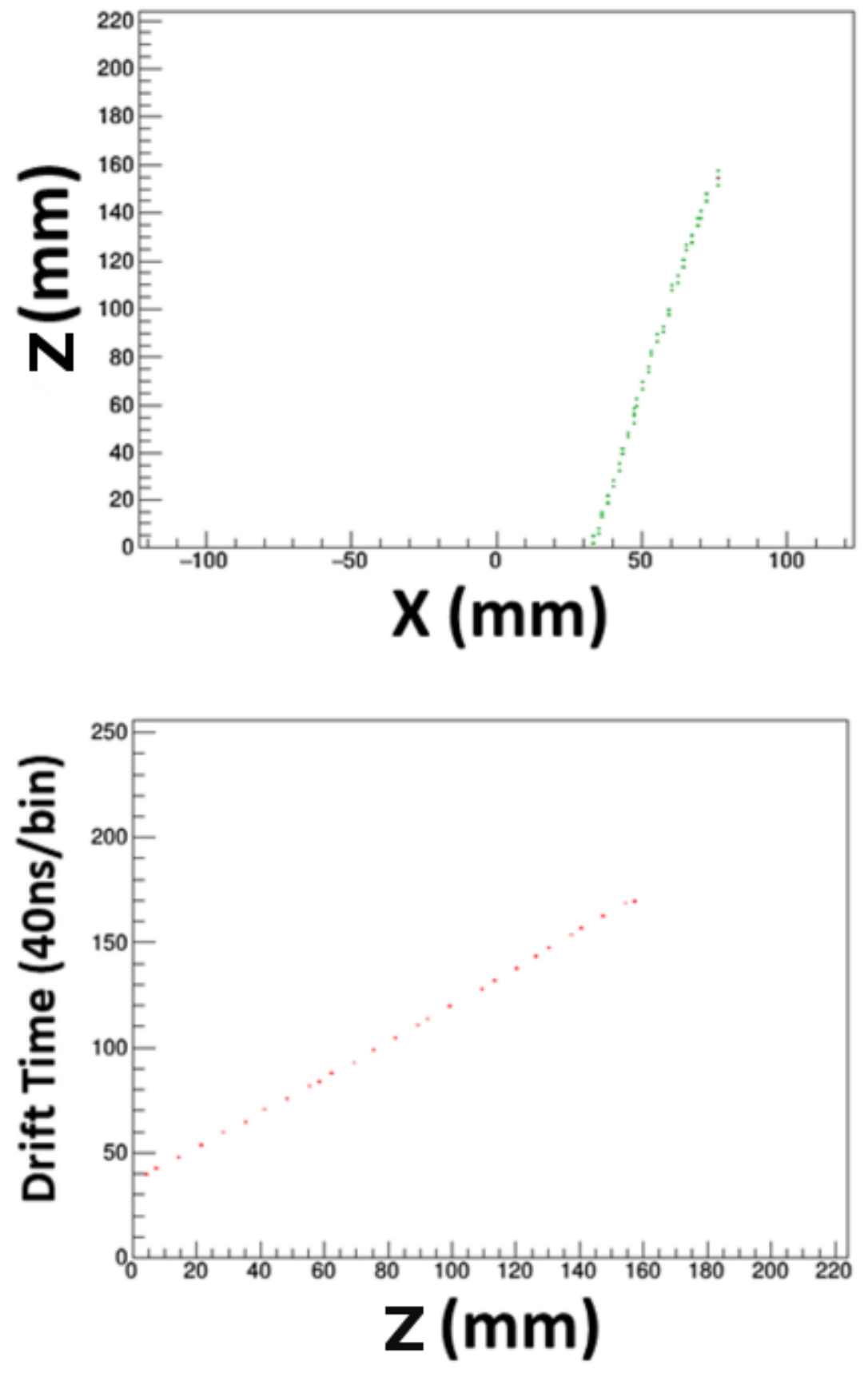}
  \caption{2D track of an $\alpha$- particle from $^{241}$Am source in He+6$\%$CO$_2$ mixture. Top panel: X-axis is the position 
  perpendicular to the beam axis and Y-axis is the position parallel to the beam axis. Bottom panel: X-axis is the position of beam axis 
  and Y-axis is electron drift time. The collimated $\alpha$-source was placed at the upstream of the field cage along the beam axis 
  outside of the Micromegas active area}
    \label{fig:tracks} 
\end{figure}

The two wire cage with 5 mm vertical spacing and 5 mm horizontal distance between the two wires satisfied the condition of minimum 
non-uniformity of the electric field, 98\% transparency and practical fabrication considerations. 
An example of a 2D track of a 5.5 MeV $\alpha$-particle from $^{241}$Am source in He+6$\%$CO$_2$ mixture at the pressure 
of 50 Torr is shown in Fig. \ref{fig:tracks}. 
In the bottom panel of Fig. \ref{fig:tracks} the X-axis is the distance from the entrance of the active area in mm 
and the Y-axis is the TPC electron drift time. The top panel is the projection of the track to the plane parallel to the Micromegas detector. No noticeable 
deviations from linearity are observed in either projection.

\subsection{Solid state detector array \label{Si}}

Arrays of solid state detectors (Si and CsI) are used in TexAT to measure the total energy of particles that have escaped the active target 
volume and to provide an additional PID. Due to 3D tracking in the TPC, high segmentation for the Si wall is not necessary. Double side 
silicon detectors with an active area of 50$\times$50 mm$^2$ were chosen to build a Si detector walls, surrounding the TPC. The principal 
requirements were high energy resolution, reliability and price to quality measure. 

Three different types of Si detectors were used:
\begin{itemize}
  \item MSQ25-1000 from Micron Semiconductor (\cite{Micron}). The geometry is: four quadrant sectors (25$\times$25 mm$^2$) at 
  junction side and single quadrant (50$\times$50 mm$^2$) at ohmic side. The typical energy resolution for $\alpha$ particles from 
  the $^{241}$Am source is 50 - 70 keV, and the total leakage current at full depletion is less than 1 $\mu$A (degraded to 3 $\mu$A 
  after several in-beam experiments);  
  \item W1-500 detector from Micron Semiconductor (\cite{Micron}). This detector is often used downstream at the beam axis. It has 
  16 vertical junction strips at the front and 16 horizontal ohmic strips at the back. The width of the strips is 3 mm. Such geometry 
  provides a position sensitivity of 3 x 3 mm$^2$ allowing a service diagnostic during the secondary beam development and additional 
  position sensitivity at small scattering angles for some experiments at low energy when the beam stops in the gas target.
\item A test set of 12 total silicon four-segment charged particle detectors KDP-1K have been developed at the JSC ``Institute in 
Physical-Technical Problems'', Dubna, Russia (\cite{SiDetDubna}) for TexAT. The detectors were made of n-type Si (660 $\mu$m thick) 
using the ion implantation technology with thermal surface passivation. The input aperture of the detector is 54.1$\times$54.1 mm$^2$. 
The active area of each segment is 25.0$\times$25.0 mm$^2$ with a 50 micron gap between the segments. A system of common guard 
rings was used in the design of the detecting structures. The thickness of the detectors of the delivered batch was 625 micron. Detectors 
were tested both with $\alpha$-source and in-beam. Typical leakage currents of the segment at a full depletion bias voltage of 130 V were 
found to be less than 20 - 30 nA. The spectrometric characteristics of each quadrant of the detectors were measured with an $\alpha$- source 
at the over-depletion bias voltage of 150 V. Typical values of the energy resolution were at the level of 30 - 40 keV FWHM under irradiation 
from the pn junction side. These detectors were in active use for 5 in-beam experiments and only one of 12 has lost the resolution below 
an acceptable level (75 keV). The properties of  KDP-1K detectors are in compliance with technical requirements and it has been decided 
to make them a basic component for TexAT.
\end{itemize}
  
The supporting frames are slightly different but in principle MSQ25-1000 and KDP-1K are interchangeable with some small adjustment. 

The outer layer of TexAT was composed from the thick CsI(Tl) scintillator detectors located at the back of all of the Si detectors. They can 
be used for detection of both charged particles and gamma rays. Cuboid- shaped 40 mm thick CsI(Tl) crystals have been selected to ensure 
that the full energy is measured for particles that can penetrate through the silicon detector. The geometry of crystals was chosen to match 
the Si detectors face: 50$\times$50 mm$^2$. Each crystal is wrapped in 2$\mu$m thick aluminized mylar and has a built-on preamp which 
is read out by a single Hamamatsu 20$\times$20 cm$^2$ S3204 PIN diode. The role of the CsI(Tl) array in the TexAT chamber is primarily to 
identify and measure residual energy of particles that punch-through the Si detectors. Using an $^{241}$Am $\alpha$-source, we have 
determined the energy resolution for the CsI(Tl) detectors to be about 5\%.

\begin{figure}[hbt!]
    \centering
    \includegraphics[width=1.0\columnwidth]{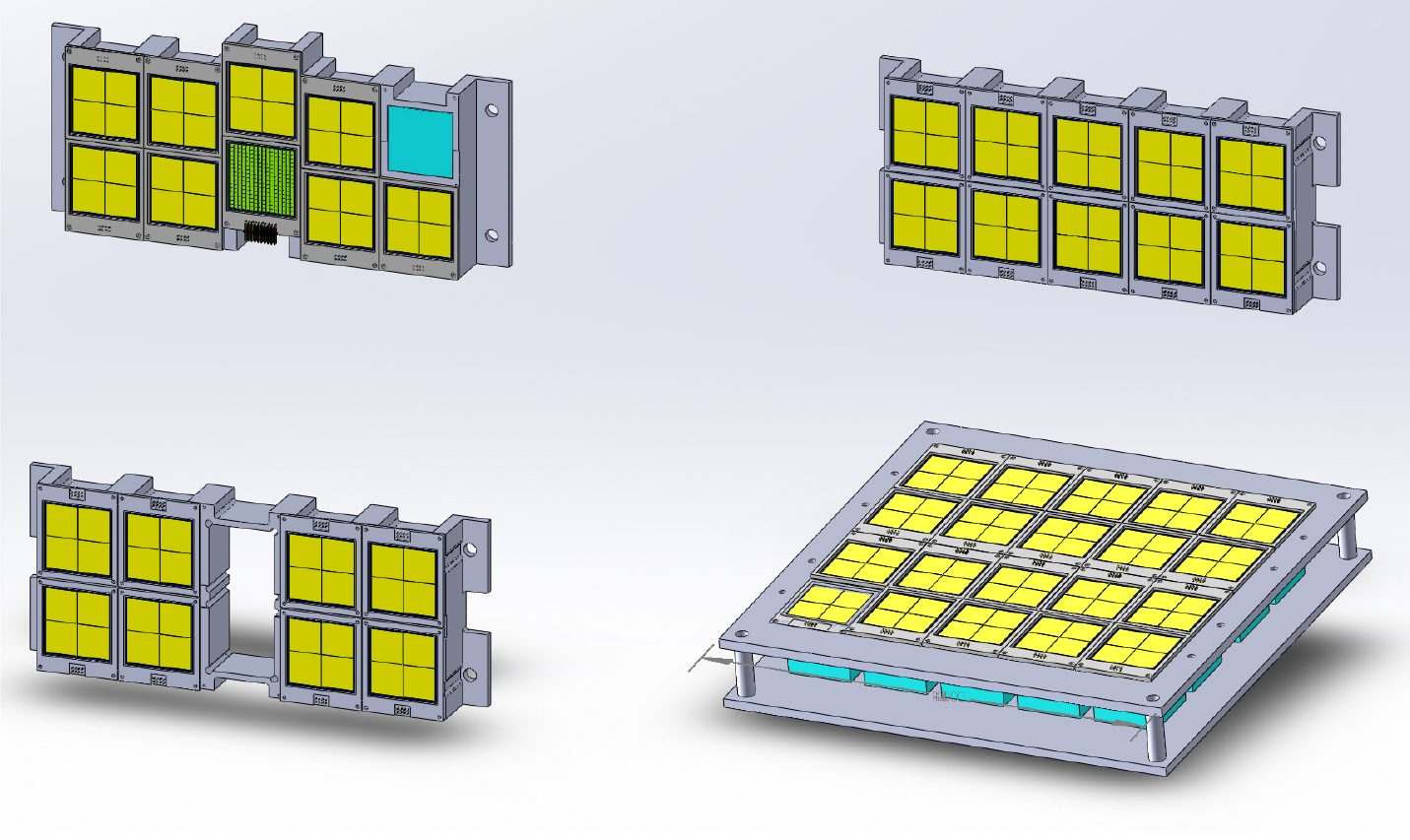}
    \caption{Rendering of the solid state telescopes of TexAT array: Forward array (top left); Side array (top right); Backward array (bottom left); 
    Bottom array (bottom right). 
    One of forward Si detector is removed to show an arrangement of CsI(Tl) scintillation detectors.} 
    \label{fig:WallsAll}
\end{figure}

Solid state telescopes are arranged in 4 different configurations matching the TPC geometry (shown in Fig. ~\ref{fig:WallsAll}). All side's arrays 
are mounted at independent doors providing an easy access to the scattering chamber and handy replacement of components. The bottom array 
is placed onto the main body of the TexAT scattering chamber.

Auxiliary detector systems, such as a windowless ionization chamber located at the entrance of the TexAT scattering chamber, right after 
the entrance window, and a scintillator located before the entrance window in vacuum have been used for some (but not all) TexAT experiments 
for normalization and beam diagnostics. These systems are not part of TexAT and will not be discussed further in this paper.

\subsection{Readout electronics \label{GET}}

The electronics channel count for TexAT is 1,024 channels for TPC readout, 232 channels for Si detectors and 58 channels for
 CsI(Tl) scintillation detectors, for a total of 1,314 channels. The readout electronics for all of these detectors is based on General Electronics 
 for TPCs (GET). The details of the GET system are published in Ref.~\cite{Pollacco}. Briefly, the TexAT readout system consists of 24 AGET chips. 
 The AGET chip amplifies and shapes the signals while performing pole-zero corrections. For each channel, the signal is stored in 512 time
 buckets (using switch capacitor arrays) with a frequency that can vary between 1 to 100 MHz (from 1$\mu$s to 10 ns per time bucket). The 
 signals are compared to a threshold to provide a channel-level trigger. Each AGET chip has 64 independent channels and includes four extra 
 channels to measure the noise. These ``noise'' channels are structurally identical to the 64 input channels and provide a way to subtract the 
 electronic noise in the system.

The 24 AGET chips are installed onto 6 AsAd boards (4 chips per boards). The main role of the AsAd board is to digitize the signals from 
each AGET chip when a trigger is issued. The digitizer of the AsAd board is a four channel 12-bit ADC \cite{Pollacco}. Up to four AsAd boards 
can be connected to the top level of the GET electronics, the CoBo (\textbf{Co}ncentration \textbf{Bo}ard). When a trigger is sent to the CoBo, 
it collects and time stamps the data and sends the data to be stored. An additional board called the 
MuTAnT (\textbf{Mu}ltiplicity \textbf{T}rigger \textbf{An}d \textbf{T}ime) is used to synchronize all of the CoBos. The MuTAnt board 
also collects the triggers from all of the CoBos and generates a global trigger. There are three different types of triggers in the GET 
system: Level 0, Level 1 and Level 2. The Level 0 trigger is an external trigger. A Level 1 trigger is obtained by summing the multiplicity 
triggers from the CoBos to generate a global trigger. The last trigger type, Level 2, can trigger on complex predefined pattern of channels 
that fired. The Level 2 trigger has not been tested with TexAT yet.

\subsection{Transition ZAP- boards}

To connect detectors to the AsAd cards, a special readout adapter board (referred as ZAP) has been designed for each kind of detector 
(MM, Si and Csi(Tl)). The ZAP-board allows the readout of signals from the detectors, to bias them (individually or in groups) and to 
protect the sensitive electronics of the AsAd cards and AGET chips from the electrical breakdown caused by potential sparks in Micromegas detector.

The PCB design was made using prototyping software Design Spark PCB 7.1~\cite{DesignSpark}. The boards have 4 layers and all elements 
are surface mounted. The electronic components are placed outside of TexAT scattering chamber to simplify the replacement in case of 
electronic failures. All transition ZAP- boards have a similar form- factor. They are epoxy- sealed to the custom vacuum flange, making 
a signal feed-through to the TexAT scattering chamber.  



\subsubsection{Micromegas ZAP- board design}
The basic circuitry for the signal readout allowing protection, suggested by developers of GET electronics 
 (\cite{GET}). The protection circuitry is based on a diode bridge. A ``common mesh" circuit is used for TexAT's Micromegas detector:
  pads/strips/chains were set at positive DC potential (200 to 600 V) to create an avalanche in Micromegas detector. Such an approach allows the
   tension to be set individually at different zones of the TexAT Micromegas detector (see Sec. \ref{lowANDhigh}) to create avalanche areas
    with different gas gain. The board-to-board connection concept was used to reduce noise and to avoid eventual connection problems with
     high density cables between electronic components. Every Micromegas ZAP card has 4 individual biasing lines through the SHV-  connectors 
     to ensure smooth functioning at the higher tension ($>400$ V). Micromegas ZAP ``input" interface matches to
        MM readout board interface (SAMTEC male/female) and the ``output'' ZAP- interface matches to AsAD card interface (SAMTEC female/male). 

The micromesh is either grounded or fed to the charge sensitive preamplifier (Canberra). In the latter case, the signal from the mesh can 
be monitored with an oscilloscope and/or to be used to generate a trigger.

\subsubsection{Si detectors ZAP-board design}

	A ZAP card consists of four identical groups of channels, each one maintains up to 16$\times$4 quadrant Si detectors ( MSQ25-1000 
	or KDP-1K, 64 total individual channels). The Si detectors connected to ZAP board inside scattering chamber by flat cables combining 
	8 detector outputs to 32 ribbon cable, plugged to standard 34 pin header.
10M$\Omega$) resistors is used instead of (10M$\Omega$ - 100M$\Omega$). 
Detectors are biased through the 16 position connector header with a single bias line supporting 4 Si detector front channels.
In the full configuration, TexAT consists of 58 Si detectors. The current ZAP-board configuration allows to read out signals from the 
``front"- side of all detectors provided that ``backs" of all detectors are common. The signals from the back side of Si detectors can 
be also readout, but in this case the total number of detectors is limited to 48 (a single AGET is not able to work with ``mixed" polarity 
signals). All Si detectors are biased from MPOD Power supply unit.


Signals from the detectors go through the SAMTEC connectors to the internal AGET preamplifier. 

\subsubsection{CsI(Tl) scintillation detectors ZAP- board design \label{CsIZAP}}

\begin{figure}[hbt!]
	\centering
	\includegraphics[width=1.0\columnwidth]{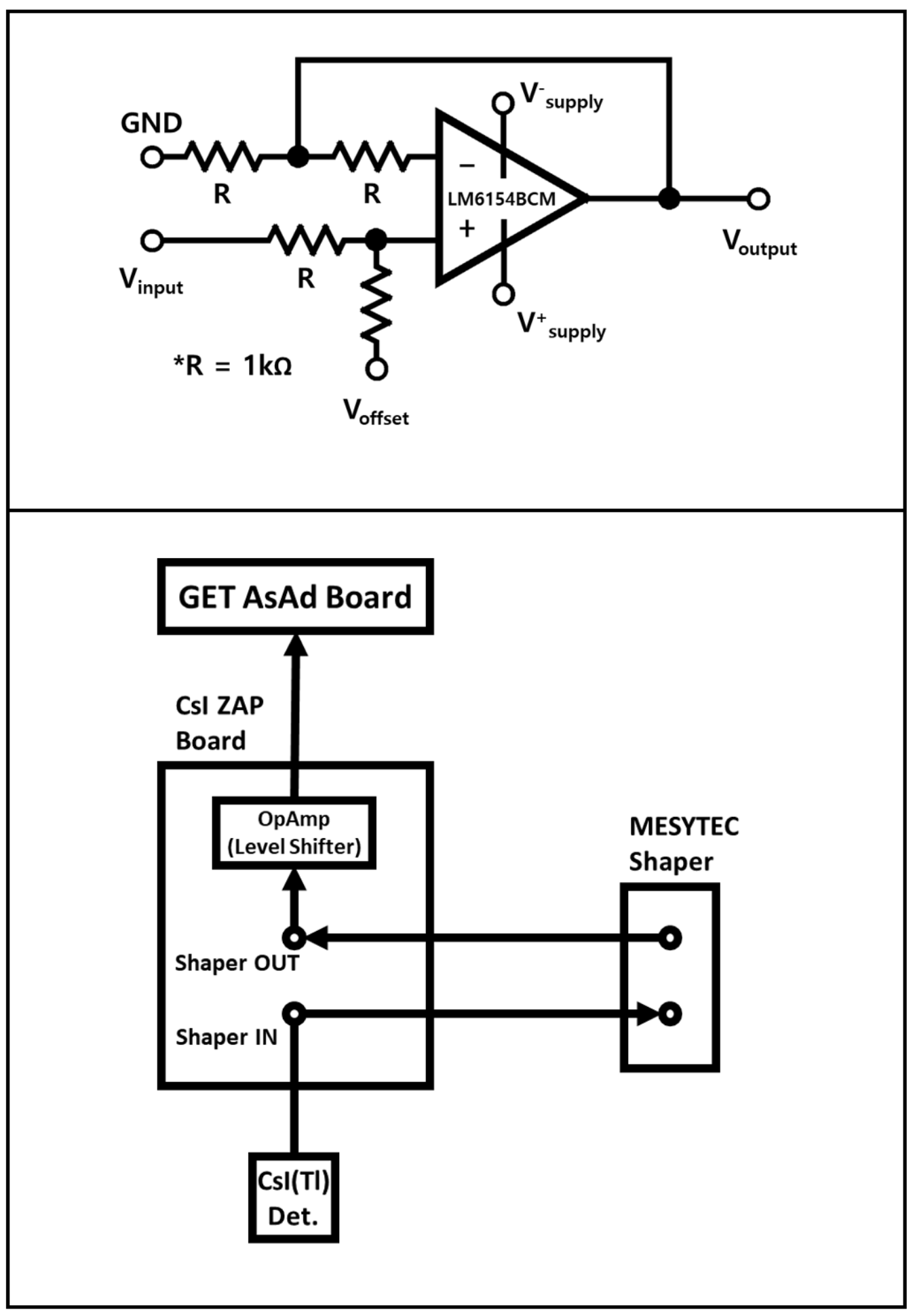}
	\caption{Top: A schematic of the non-inverting level shifter circuit using the LM6154BCM OpAmp. The supply voltage, V$^{+}_{supply}$ 
	or V$^{-}_{supply}$, was adjusted to match the impedance of the circuit and GET electronics input. Bottom: A signal processing diagram 
	using the external shaper for CsI(Tl) scintillation detector.}
	\label{fig:CsIZAP_circuit}
\end{figure}

Signals from the CsI(Tl) scintillation detector were output of its own custom preamplifier, located right on the PIN diode to reduce the electronic noise.
 The Charge Sensitive pre-Amplifier (CSA) function and filter stages of the AGET chip are bypassed. In addition, the ``slow'' output of the detector
  required an optimal peaking time of 4 to 6~$\mu$s, while the available range of peaking time for the AGET chip extends from 50 ns to 1 $\mu$s 
  (sixteen values total). In order to optimize the signal processing of the detector, we use a 16-channel MESYTEC shaper (MSCF-16)~\cite{MESYTEC} 
  and feed the shaped output signals into an Inverting x2 Gain (G-2) function of the AGET chips (See \cite{GET} for details). The G-2 stage provides
   an extra inverting voltage gain and a buffering for signal sampling in a Switch Capacitor Array (SCA) in the AGET chip. The offset voltage of the 
   Gain-2 input signal was adjusted by the non-inverting level shifter circuit with the 75~MHz gain bandwidth LM6154BCM Operational Amplifier 
   (OpAmp)~\cite{OpAmp} (Fig. ~\ref{fig:CsIZAP_circuit}), depending on the polarity of input signals: +2.2~V for positive polarity or 0.7~V for
    negative polarity.

A custom ZAP board has also been designed. The sixteen total fast OpAmp integration circuits were implemented on the board. 
While one offset voltage and one supply input can be set for all the OpAmp circuits, one OpAmp handles 4 independent channels 
resulting in 16 outputs for each AGET chip input corresponding to the full channel numbers of one MESYTEC module. One ZAP 
board allows us to process signals from 64 total CsI detectors, placed inside the TexAT chamber.

\subsection{Scattering Chamber}

\begin{figure}[hbt!]
	\centering
	\includegraphics[width=1.0\columnwidth]{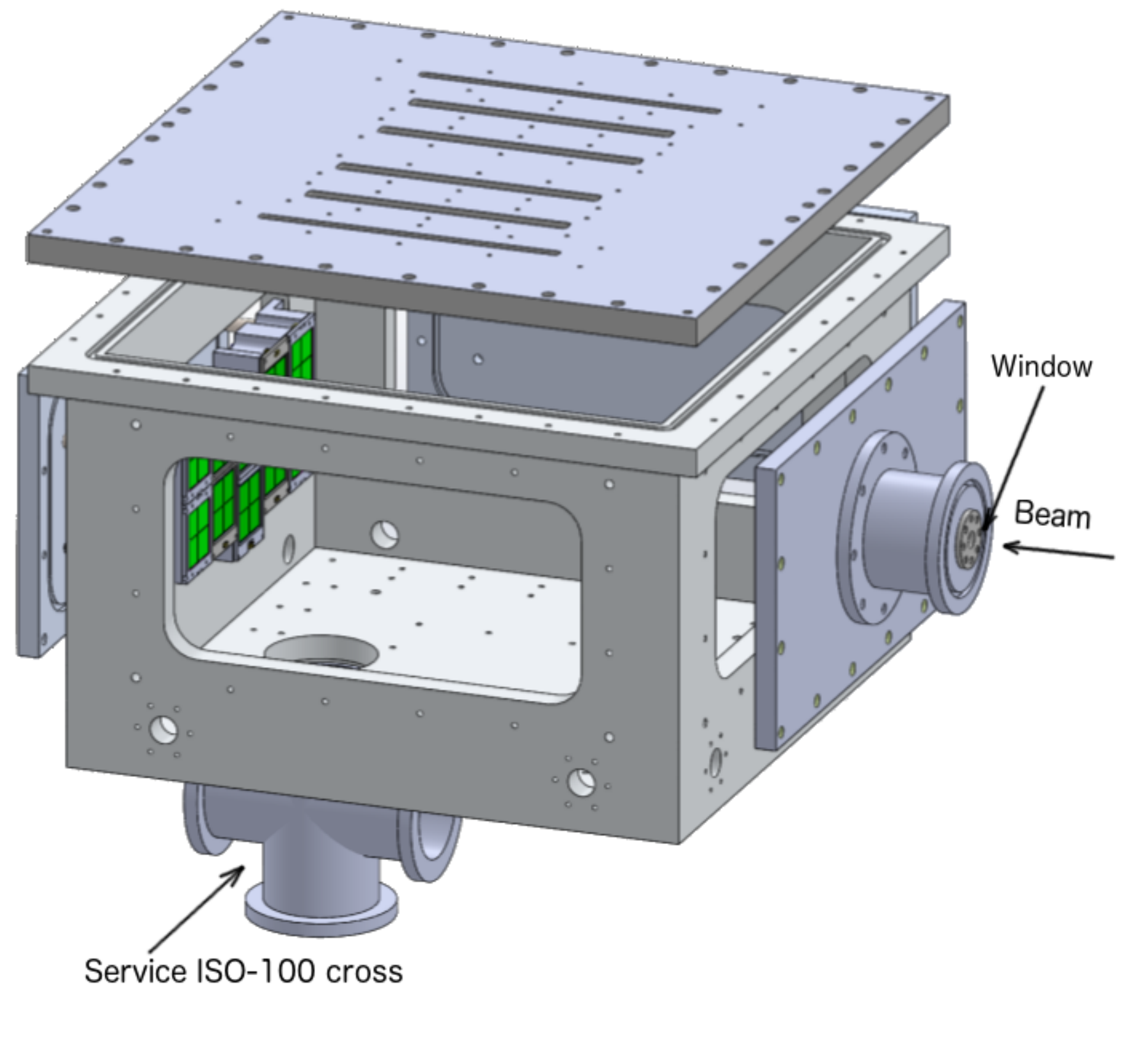}
	\caption{3D- drawing of TexAT scattering chamber. One of the Si/CsI(Tl) detector array is shown attached to the ``forward"- door}
	\label{fig:TexATChamber}
\end{figure}

A custom scattering chamber has been designed to hold the detector array and to maintain the workable conditions for both gaseous 
(Micromegas and ion chamber) and solid state (silicon and scintillation) detectors. It has an aluminum body/shell of cuboid shape: 
508mm x 508mm x 343mm, made from 25.4 mm thick aluminum sheet. The top cover is designed to secure a Micromegas based 
time-projection chamber (TPC). It also has six slots for vacuum feedthroughs (ZAP-boards) to read out the anode signals from 
Micromegas and solid state (semiconductor and scintillation) detectors. The general layout of scattering chamber parts is shown in
 Fig. ~\ref{fig:TexATChamber}.
 All readout electronics are coupled directly to the feedthroughs, eliminating cables. (Fig. ~\ref{fig:TexATChamber}

Each of the four chamber's sides has a square 355.6 mm  by 203.2 mm) port, plugged with a O-ring's vacuum sealed doors. Such design 
allows one to mount solid state detectors, surrounding the TPC at each flange independently. It provides an easy access to all components 
of detector array and versatile installation at any beam line (either at in-flight exotic secondary beam line using MARS- separator, or at future 
project of re-accelerated exotic beam line) through the standard ISO-100 upstream vacuum flange. The scattering chamber's gas volume is 
separated from the beam line with a thin vacuum tied Havar/Kapton/Mylar entrance window with a diameter of 12.5 mm, mounted at the 
forward wall, connecting TexAT to the beam-pipe (shown in (Fig. ~\ref{fig:TexATChamber}).
	
A small planar windowless Ion Chamber (IC) with a Frisch grid can be mounted at upstream flange. The IC works in the same target gas 
media to provide a particle ID and count ions of incoming beam and also provides additional normalization and beam-tracking capabilities.
	
The bottom side of the chamber is equipped with ISO-100 compatible style port for vacuum pumps connection. 
Another set of KF40 ports are placed on the side walls (2 at each wall, 8 total). These ports are used for gas inlet/outlet, 
vacuum/pressure/temperature sensors, IC voltage/readout and high voltage feedthroughs, supplying different circuits. 

The chamber finish is mechanically polished and de-ionizied to meet the pretty narrow requirement for steady operation of the Micromegas 
detector which is sensitive to impurities and dust. The tested chamber leak rate was in the range of 10$^{-8}$ atm-cc/sec. 
All major operations with TexAT (e.g., mounting of MM-, Si- and scintillation- detectors, etc.) during the experiment's preparation are 
conducted in a clean  room  environment. 

\subsection{Gas handling system}
	
The multipurpose vacuum and gas handling system
has been developed to provide a possibility of running experiments with
exotic beams in both ``detector" (for example, decay-experiments) and ``active target" (resonance scattering,
transfer reactions, etc.) modes. It is similar to those built for the ANASEN detector \cite{ANASEN}.
The gas pressure and flow is controlled by an integrated pressure
controller with mass flow meter $\pi$PC-99 from MKS Instruments.
It can control gas pressure through TCP/IP protocol at the range of 15 Torr to 1250 Torr with an accuracy of $\pm$1\%.
The system was tested for gases like helium/CO$_2$ mixture, carbon dioxide, argon and mixtures (P5/P10).
It also meets safety specifications for use of flammable gases (hydrogen/deuterium, methane,
iso-butane, etc.) and was tested with those gases as well.

The reliability of the system has been demonstrated during several 3 - 4 week long runs.
	
\section{TexAT Monte Carlo Simulation Package}

\begin{figure}[hbt!]
    \centering
     \includegraphics[width=1.0\columnwidth]{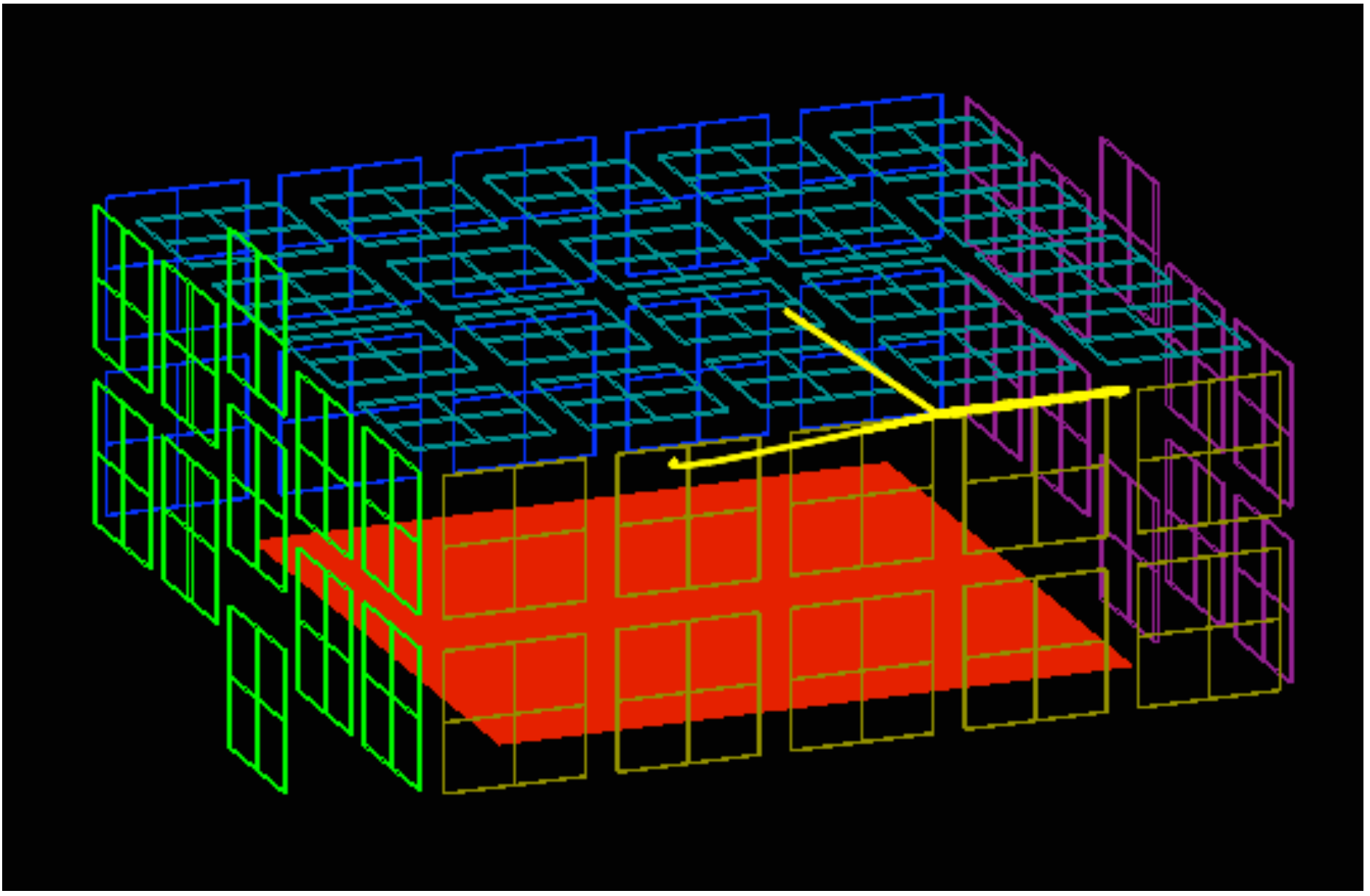}
     \caption{Example of simulated track for the $^4$He($^{18}$Ne,p)$^{21}$Na reaction in the TexAT detector as visualized in Geant4. 
     The light recoil (proton) hits the Si detector on the top plate and the heavy recoil ($^{21}$Na) stops in the gas volume.}
     \label{fig:Simulation_1}
   \end{figure}

A complete TexAT Monte Carlo simulation package was developed utilizing the Geant4 \cite{Geant4} and Garfield++\cite{GARFIELD} 
libraries. Two reactions were studied as the test cases: $^{12}$C(p,p)$^{12}$C and $^{18}$Ne($\alpha$,p)$^{21}$Na. In the latter, 
residual $^{21}$Na nuclei in both the ground state and first excited state were populated. In the simulation, the incoming ions were 
forced to undergo the reaction of interest at a random point within the TPC using binary reaction kinematics. At each step of an incoming 
or product track, the total ionization energy was converted to a number of electron and ion pairs, and each electron was drifted under a 
constant electric field to the Micromegas board. Time-dependent histograms of the number of detected electrons for each pad of the 
Micromegas board were generated and stored for offline processing incorporating the properties of the gas being simulated. Additionally, 
the energy collected in each detector of the silicon array was also recorded per event. Fig. ~\ref{fig:Simulation_1} shows an example event 
visualized in Geant4.
	
Custom track reconstruction routines have been developed for the TexAT detector. For
particles traversing the left and right regions, X-Y coordinates are determined by matching strips
to chains (or vice versa) using the average recorded drift time. In the central region, the X-Y
coordinates of a track are determined solely from the position of the rectangular pad with the
highest energy deposition per row. Such a procedure yields two tracks: one for the light product and
one combined track for the incoming ion and heavy recoil. By finding the point of closest
approach of the fitted light product and the heavy ion tracks, the heavy ion track in the central
region can be split into separate incoming and recoil tracks. Finally, the three tracks are re-fit
simultaneously to find the vertex of the interaction and the relative angles of the reaction products extracted. 
Fig.~\ref{fig:Simulation_1} shows an example of this track reconstruction.

\begin{figure}[hbt!]
    \centering
     \includegraphics[width=1.0\columnwidth]{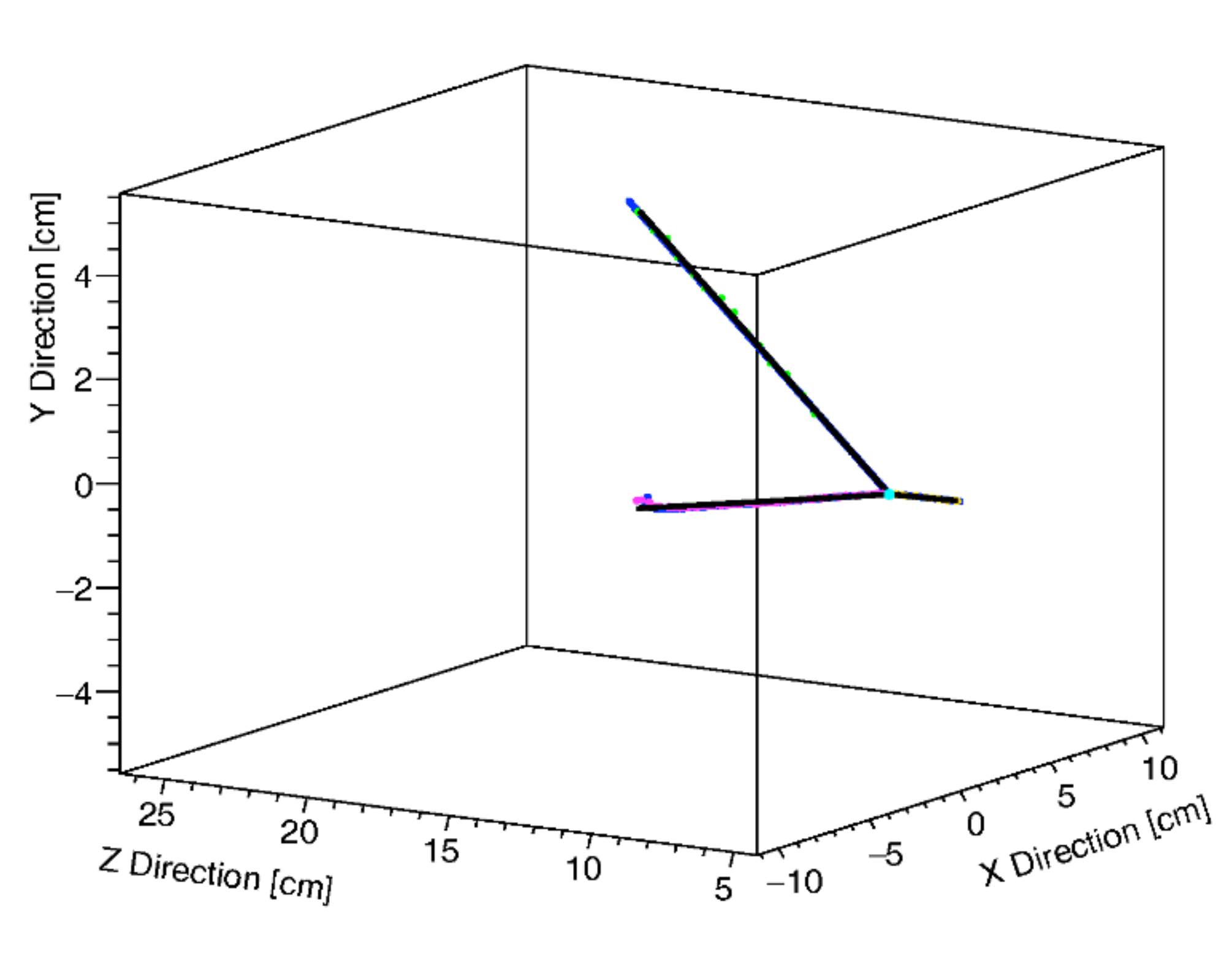}
     \caption{Same track as in Fig.~\ref{fig:Simulation_1}, but reconstructed from drifted electrons
detected by the Micromegas pads.}
     \label{fig:Simulation_2}
   \end{figure}

\begin{figure}[hbt!]
    \centering
     \includegraphics[width=1.0\columnwidth]{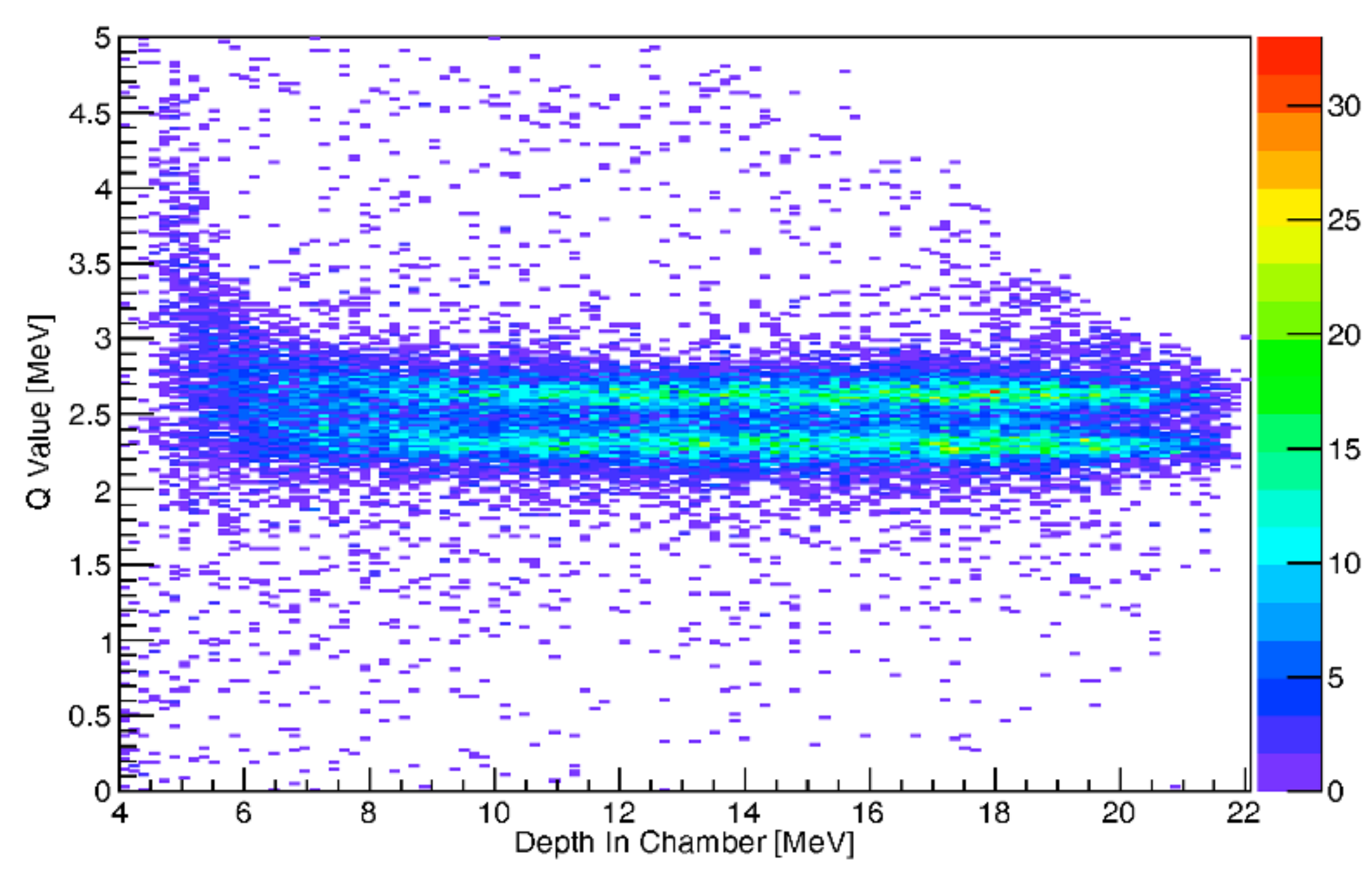}
     \caption{Reconstructed Q-value vs. chamber depth as determined by the energy of the $^{18}\mathrm{Ne}$ for the simulated 
     $^{18}$Ne($\alpha$,p)$^{21}$Na reaction}
     \label{fig:Reconstruction_1}
   \end{figure}

The blue points are the simulated interaction points, while the yellow, purple and green points are reconstructed from the
information recorded from the simulation for the Micromegas board. The black lines are the
fitted tracks, and the cyan point indicates the fitted reaction vertex. For the reactions of interest,
the vertex reconstruction error was found to be approximately 1.5 mm on average, with a
distribution that peaked below 1 mm. The angular resolution determined from the simulation was
approximately 3$^{\circ}$ at FWHM. If the Q-value of the reaction is unknown, the center of mass
energy of the interaction can only be reconstructed from the vertex position and the average
energy loss in the gas, and is limited by energy straggling effects. For the incoming $^{18}$Ne ions,
this yielded a center of mass energy resolution of 190 keV at FWHM. Using the interaction
energy as provided by the vertex, in addition to the angle and energy (measured in Si detectors) of
the light product, the Q-value can be reconstructed. In the case of the $^{18}$Ne($\alpha$,p)$^{21}$Na reaction,
transitions to the ground and first excited states of $^{21}$Na (separated by only 330 keV) could be
well resolved (see Fig.~\ref{fig:Reconstruction_1}). (Assuming that energy resolution of the incoming $^{18}$Ne 
beam corresponds to the ``typical'' energy resolution of reaccelerated radioactive beams.)

\begin{figure}[hbt!]
    \centering
     \includegraphics[width=1.0\columnwidth]{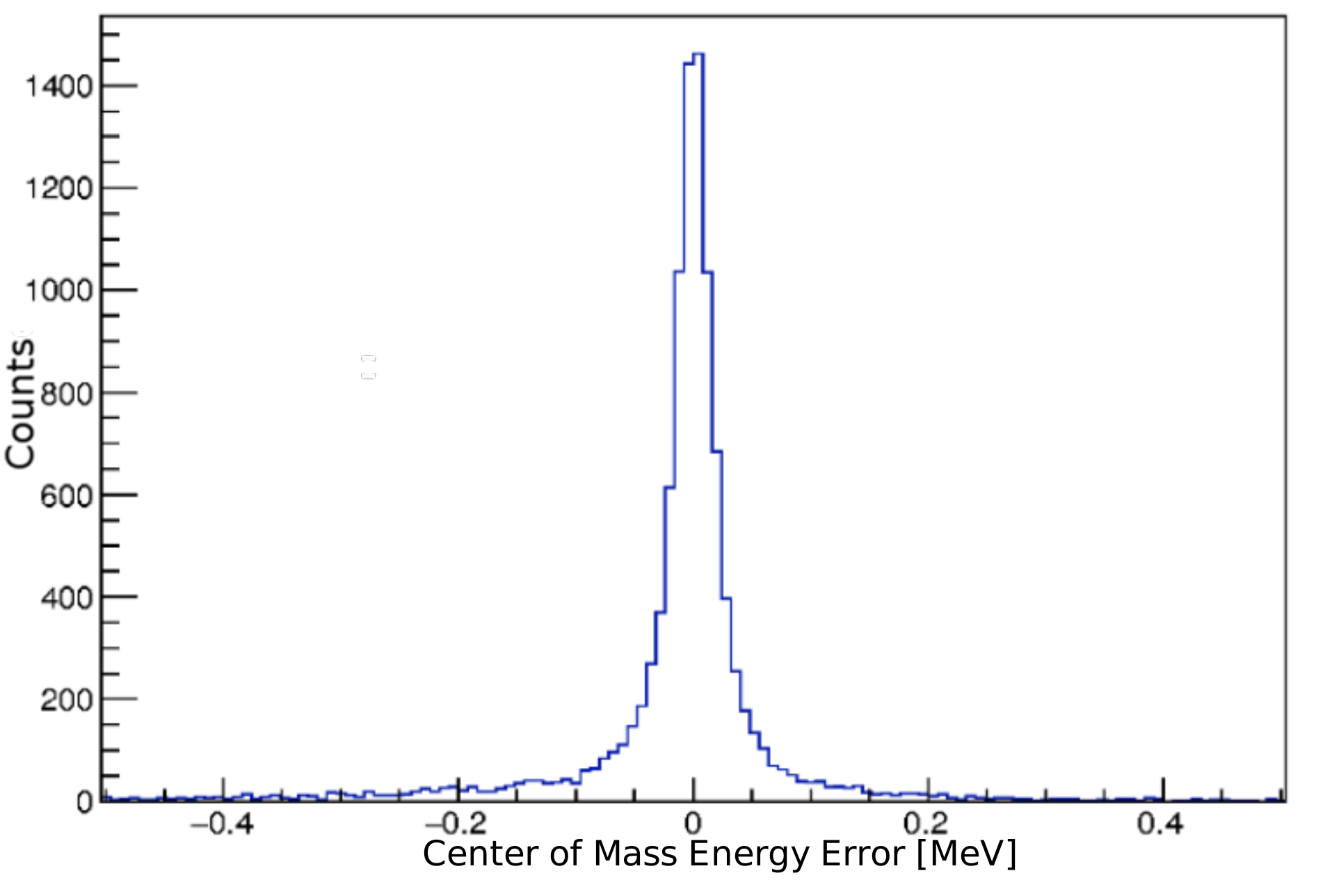}
     \caption{Center of mass energy resolution for the simulated $^{12}$C(p,p) $^{12}$C reaction.}
     \label{fig:Reconstruction_2}
   \end{figure}

Once the reaction mechanism is identified (e.g. elastic or inelastic scattering), the center of mass energy resolution can be improved 
by recalculating using the Q-value, the detected energy in the Si detector, and emitted light product angle. In the present study, the 
center of mass energy resolution for these events was found to be 40 keV at FWHM (see Fig. ~\ref{fig:Reconstruction_2}).
		
Additionally, these energy and angle resolutions were found to be mostly insensitive to the depth within the scattering chamber and 
therefore interaction energy and scattering angle. The simulations demonstrate that the segmentation implemented in TexAT is 
sufficient to reconstruct the observables with high enough precision to identify the events due to population of two closely spaced 
excited states ($\sim$300 keV) in the ($\alpha$,p) reaction. It was shown that track reconstruction allows for energy resolution of 
40 keV and angular resolution of $\sim$3$^{\circ}$ over wide range of excitation energies and scattering angles.

\section{Data Analysis}

The analysis of the data recorded using the GET DAQ with the TexAT detector involves a multi-step process. It involves subtraction 
of the baseline for all signals, the wave form fitting to determine waveform maximum amplitude values and peak times, matching 
chains and strips in the side region to create three-dimensional points for track reconstruction and fitting tracks using various 
methods for noise reduction.

\subsection{Baseline Corrections and Obtaining the Energy and Timing}

\begin{figure}[hbt!]
  	\includegraphics[width=1.0\columnwidth]{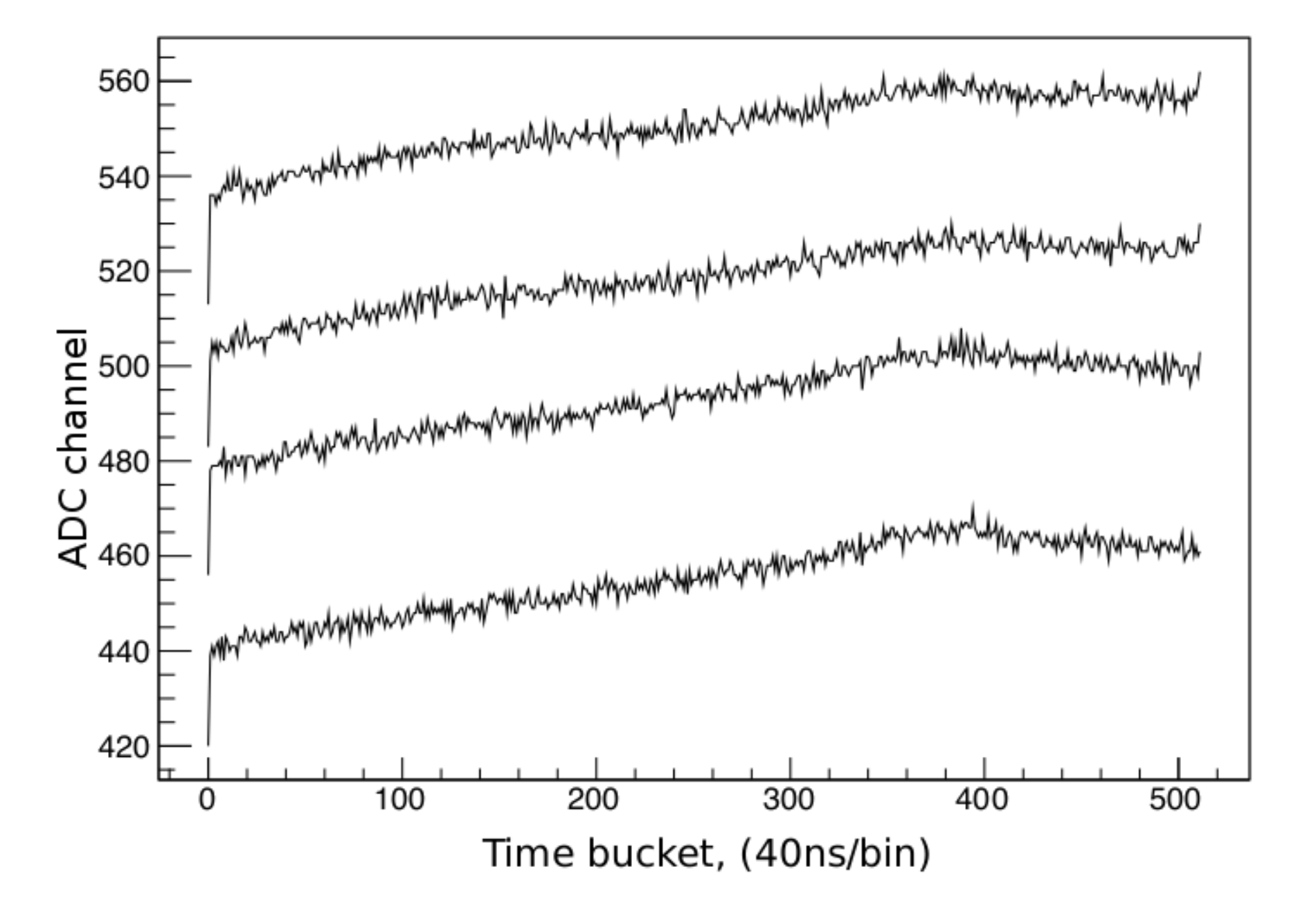}
         \includegraphics[width=1.0\columnwidth]{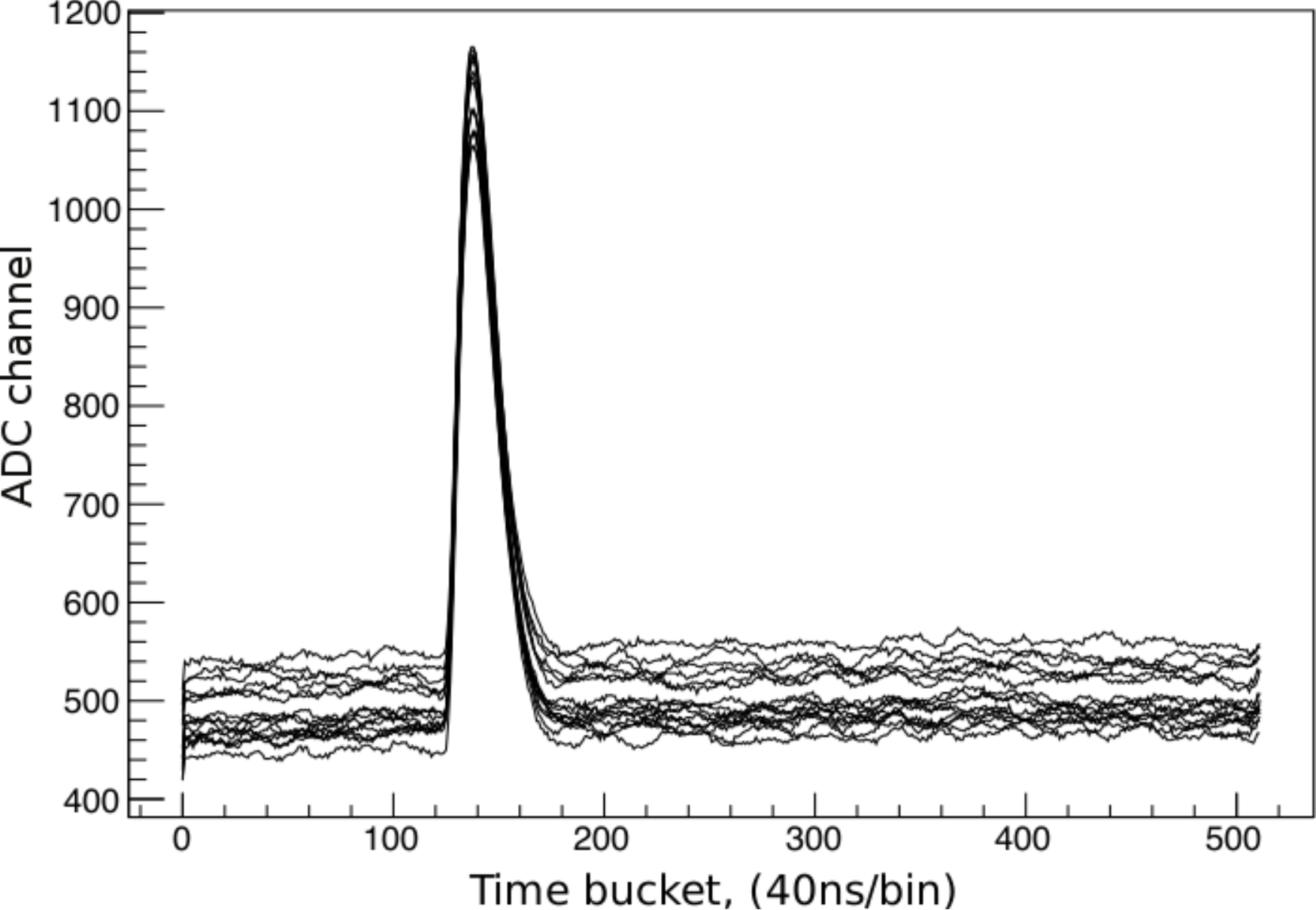}
	\caption{(Top) Example waveforms of the four FPN channels. (Bottom) Raw Waveforms without any FPN or background subtraction.}
	\label{fig:FPN_Raw}
\end{figure}

The AGET chips have 64 input channels but provide 68 output channels. Four of these channels are called fixed-pattern noise (FPN) 
channels that are not connected to any detector and record the intrinsic noise and electronic baseline \cite{Pollacco}. These four 
channels are averaged together and subtracted from each waveform of the remaining 64 channels to remove the intrinsic noise in 
the electronics. Examples of the FPN channels and raw waveforms are shown in Figure \ref{fig:FPN_Raw}.

\begin{figure}[hbt!]
    \centering
    \includegraphics[width=1.0\columnwidth]{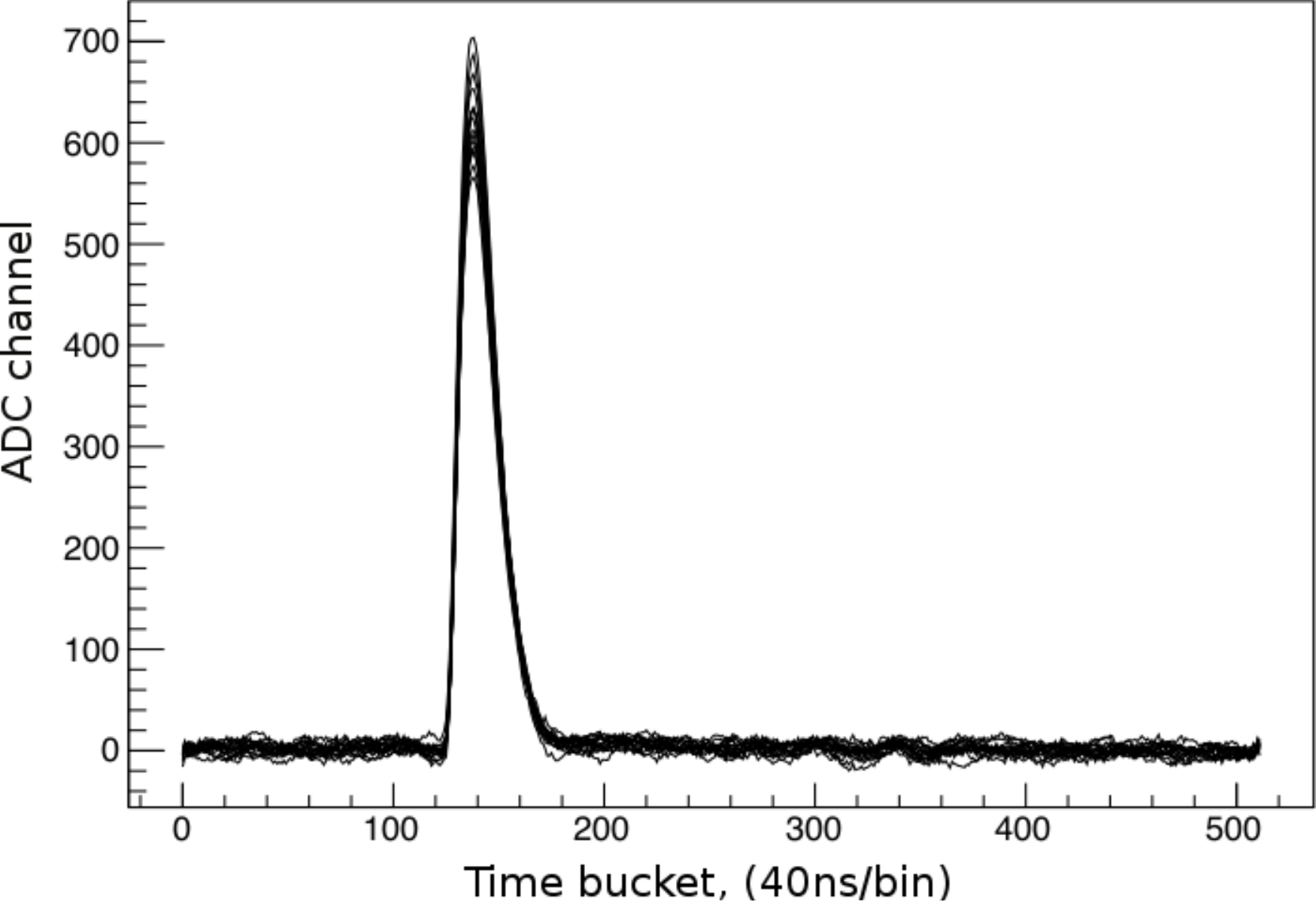}
    \caption{Waveforms corrected with FPN and background subtraction.}
    \label{fig:Waveform_Corrected}
\end{figure}

Further correction is achieved by subtracting the overall baseline, found by averaging the first and last 10 time buckets, from the 
FPN corrected signal. Examples of the waveforms corrected by the FPN and baseline subtraction are shown in Fig. \ref{fig:Waveform_Corrected}.

\subsection{Chain and Strip Matching \label{CSmatching}}

\begin{figure}[hbt!]
	\centering
    \includegraphics[width=0.5\textwidth]{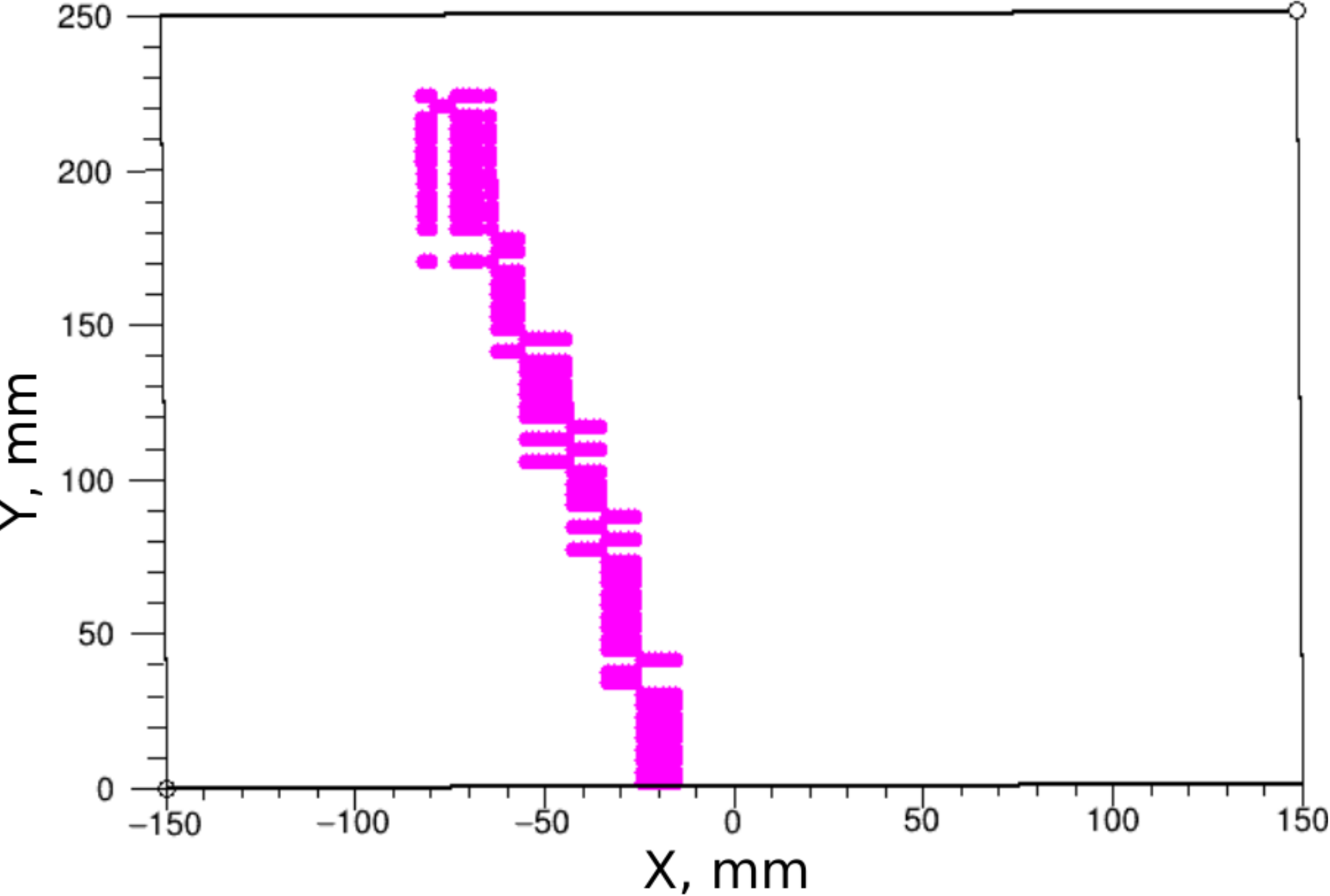}
    \caption{An example of matched chains and strips for a single event that can then be used to reconstruct a track.}
    \label{fig:ChainStripExample}
\end{figure}

To perform track reconstruction in the side regions of the Micromegas detector, chains and strips must be converted to matched X-Y pixels since 
they only contain either X-pixel of the track (strips) or Y-pixel of the track (chains), respectively, as shown in the top panel of
 Fig. ~\ref{fig:MMBoardLayout}. This matching process can be done by using the time of the signal recorded for each channel. When the
  charged particle travels over the side region, any X-pixel from strips and Y-pixel from chains that correspond to the proton track position 
  should have the same drift time. If the track is not parallel to the Micromegas plane (time among strips or chains $>$40~ns), the matched 
  X-Y pixels based on the same signal time within 40~ns should form a well defined track. An example of such event is shown in 
  Figure \ref{fig:ChainStripExample}. When the track is parallel to the Micromegas detector (time among strips or chains $<$ 40~ns), 
  the timing 
  for the chains and strips will be the same and the matched X-Y pixels form a square instead of a defined track like that in
   Figure \ref{fig:ChainStripExample}. For cases like this, the first point in the matched square closest to the origin and the point furthest away 
  from the origin are chosen to form the track.

\subsection{Track Reconstruction - Hough Transform}

The signal in the TPC is often contaminated by ``noise'' - random pads firing in coincidence with the track but seemingly unrelated to it. 
We have used the two dimensional Hough transform to reduce the influence of this ``noise'' on the track reconstruction.

\begin{figure}[hbt!]
	\centering
    \includegraphics[width=0.5\textwidth]{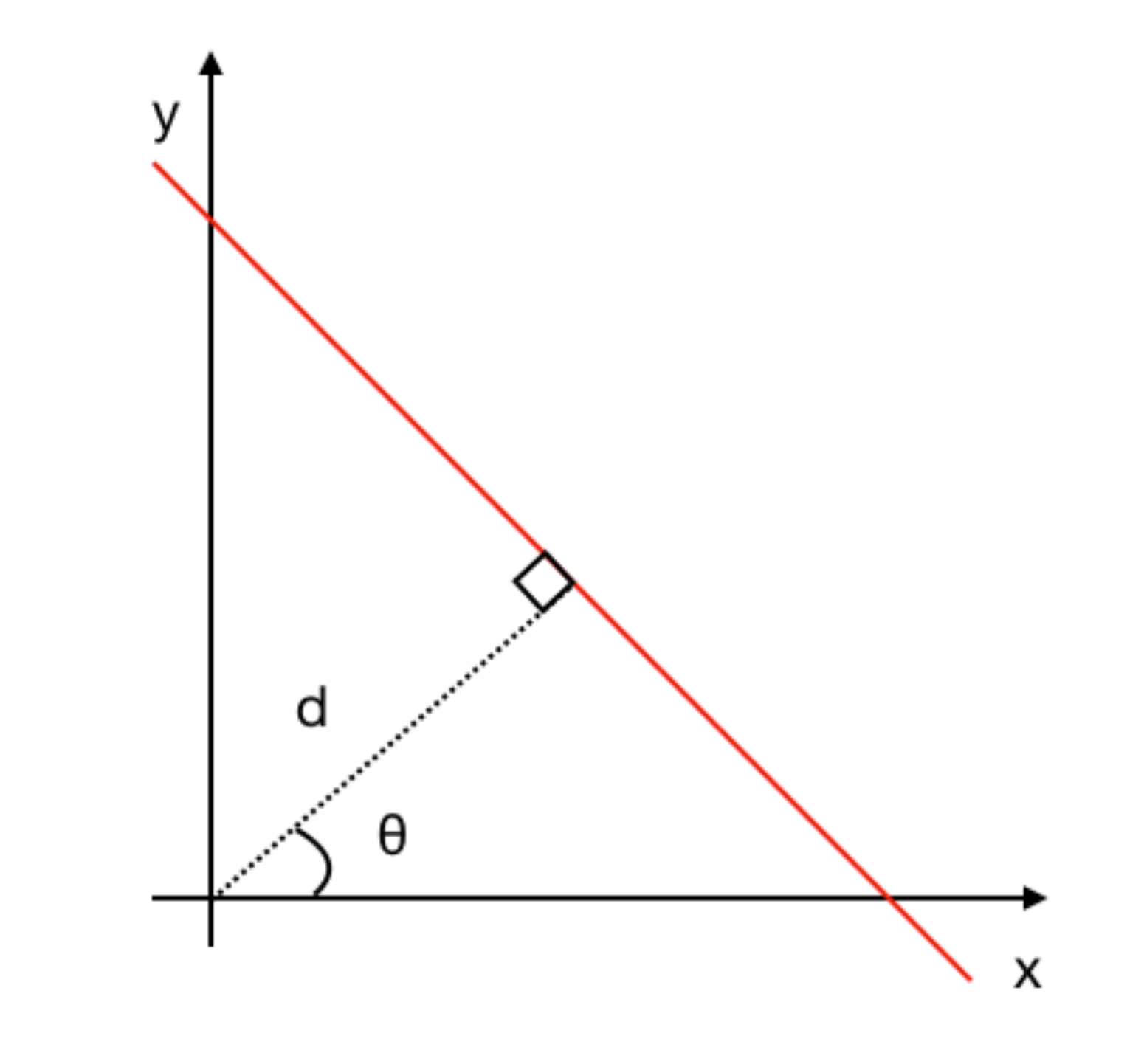}
    \caption{Diagram of the $d, \theta$ parameters in two dimensions.}
    \label{fig:HoughDiagram}
\end{figure}

The Hough transform is a feature extraction technique that has been very successful in image analysis and image processing \cite{Duda:1972}.
 The Hough transform was originally used to identify lines in an image and in this analysis, to find tracks (lines) through data points. This method 
 works by transforming points into a parameter space and a voting procedure is used to find peaks in this parameter space. 
 For each point $(x, y)$ the Hesse normal form is calculated.

\begin{equation}
	d = x \cos \theta + y \sin \theta,
	\label{eq:Hesse}
\end{equation}

where $d$ is the distance of closest approach to the origin and $\theta$ is the angle between the x-axis and the line connecting 
the origin to the closest point as shown in Figure \ref{fig:HoughDiagram}. The $(d, \theta)$ parameter space is called the Hough 
space. For every point in the data, $\theta$ is varied from $0$ to $\pi$ and $d$ is calculated.

\begin{figure}[hbt!]
	\centering
	\begin{subfigure}{.5\textwidth}
  		\centering
  		\includegraphics[width=\linewidth]{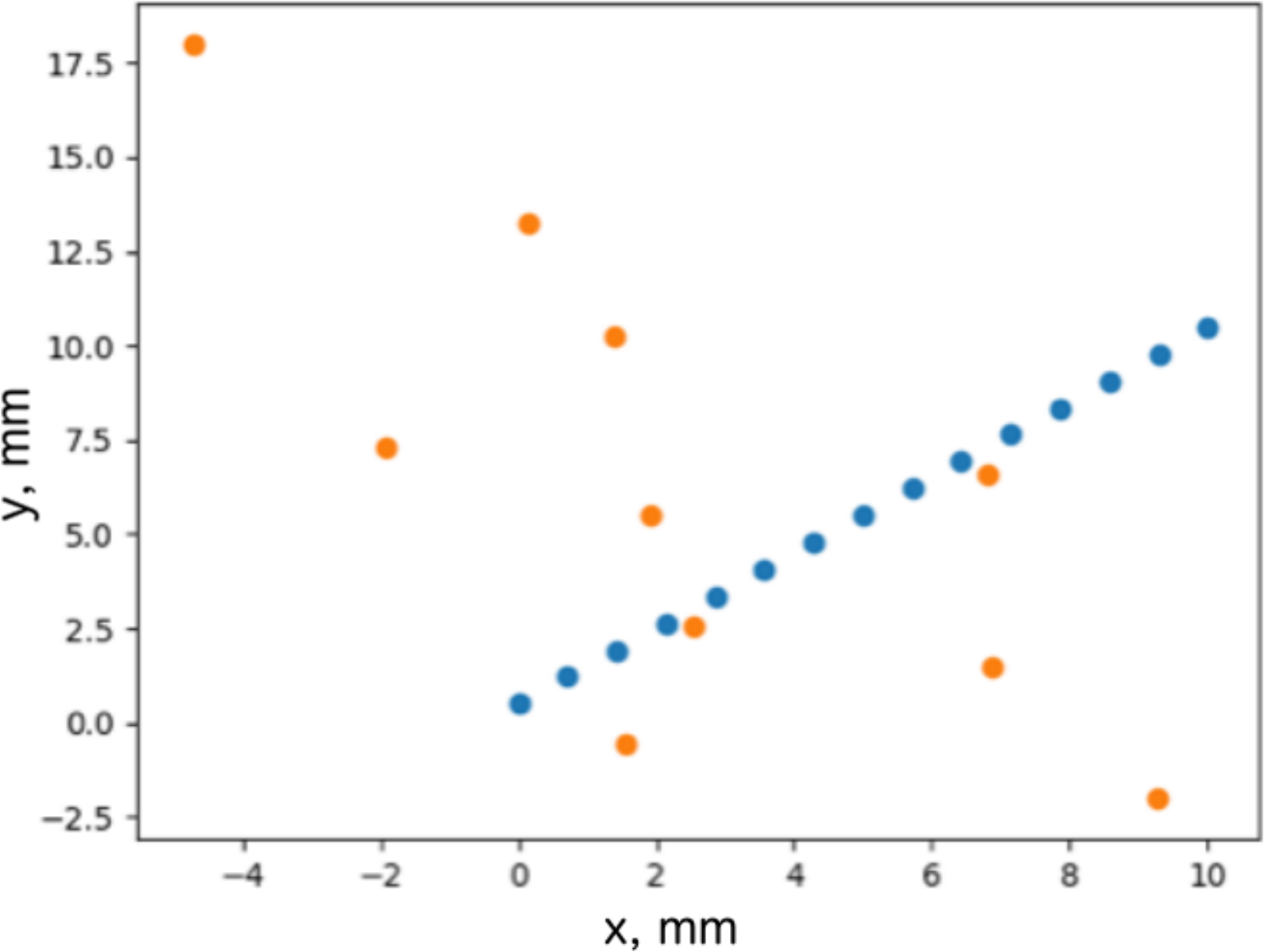}
	\end{subfigure}%
	\\
	\begin{subfigure}{.5\textwidth}
  		\centering
  		\includegraphics[width=\linewidth]{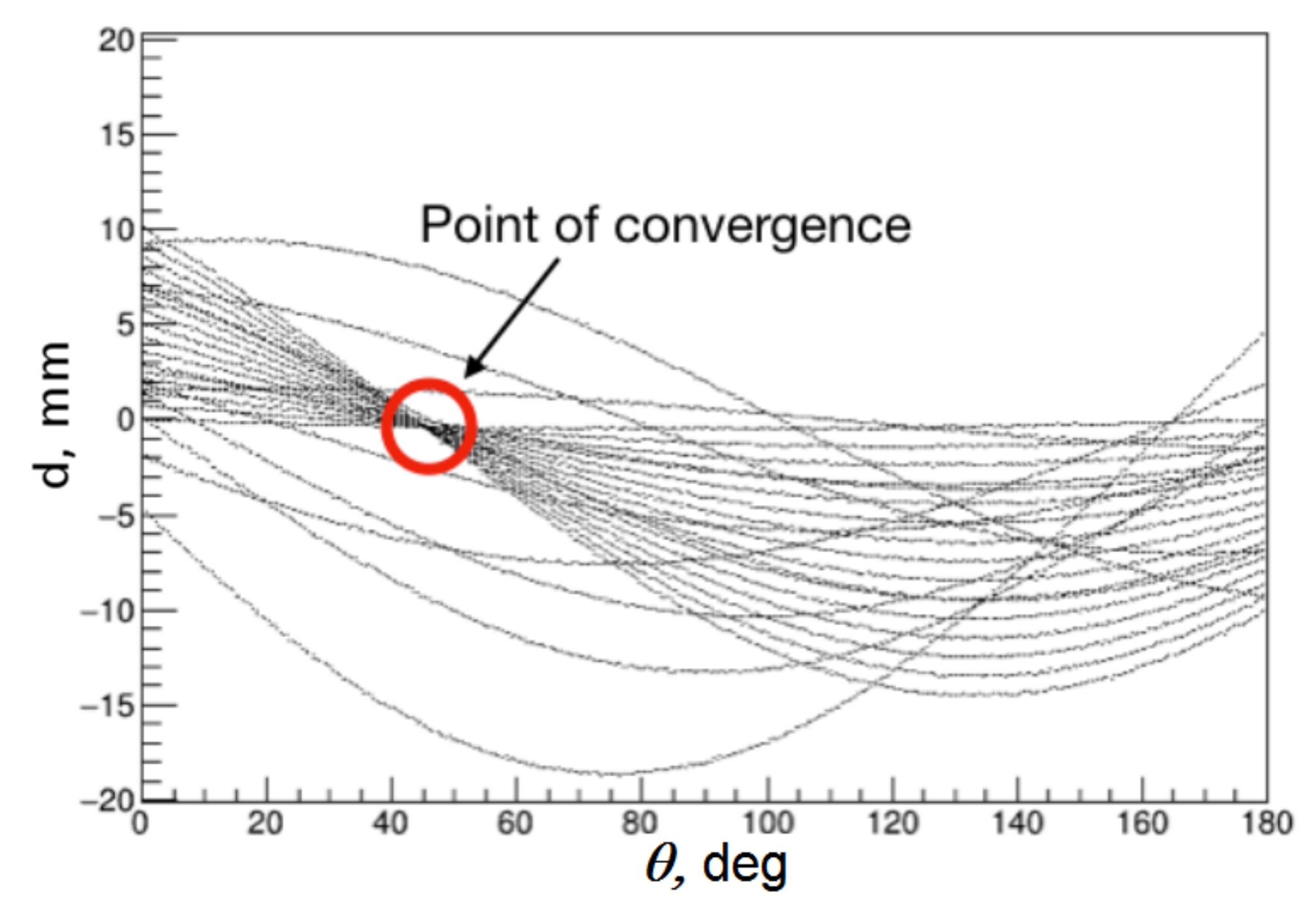}
	\end{subfigure}
	\caption{(Top) A straight line (blue points) with scattering noise (orange). (Bottom) The Hough space of the top points. Outlined 
	in the red is the minimal standard deviation of $d$ corresponding to the straight line of blue points.}
	\label{fig:HoughExample}
\end{figure}

The algorithm used to find the optimal $(d, \theta)$ was to search through all angles and find the lowest standard deviation in $d$. A simple 
example with noise is shown in Figure \ref{fig:HoughExample}. In the top of Figure \ref{fig:HoughExample}, there is a straight line formed by 
the blue colored points while the orange colored points are noise. The Hough space is shown at the bottom of Figure \ref{fig:HoughExample} 
illustrating the point where the standard deviation is minimal. By finding the optimal parameters $(d, \theta)$, we find the best fit of the data 
that is not affected by noise.

 \subsection{Alpha Source Test in Gas}
 
\begin{figure}[hbt!]
	\centering
	\includegraphics[width=0.5\textwidth]{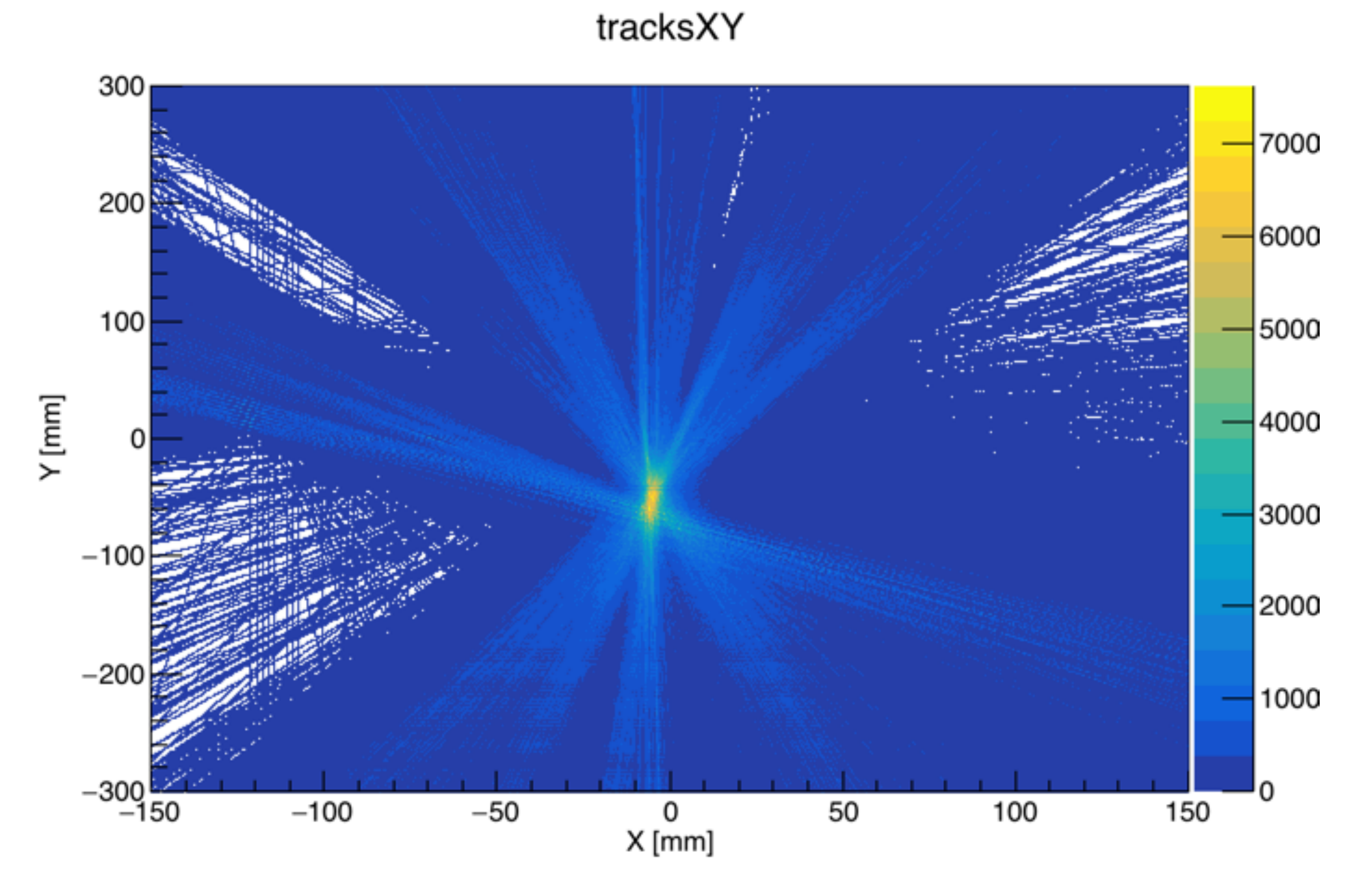}
    \caption{An accumulation of the alpha source tracks in the XY-plan in the forward and side walls. 0 mm on y-axis corresponds 
    to the beginning of the Micromegas plate. All of the tracks converge to $\sim -60$ mm where the source was located.}
    \label{fig:TexATAlphaTracksXY}
\end{figure}

The chains and strips matching technique and a technique to fit tracks with noise using the Hough transform using an $\alpha$ source in gas
 were tested. The gas chosen for this measurement was methane at $50$ Torr so that the 5 MeV $\alpha$ particles propagate through the entire
  active volume of the scattering chamber and make it to the Si detector while also depositing enough energy in the gas to make tracks. 
  The $\alpha$ source was placed approximately 60 mm behind the Micromegas plate facing the Si detectors. An accumulation of the tracks in 
  the XY-plane are shown in Fig. \ref{fig:TexATAlphaTracksXY}. Zero on the y-axis in Figure \ref{fig:TexATAlphaTracksXY} corresponds to the 
  start of the Micromegas plate and shown are all the tracks that end up in the forward and side wall Si detectors. The tracks are converging at
   the (0 mm,-60 mm) point that has about the size of the source ($\sim 5 mm$ along the x-axis).

\begin{figure}[hbt!]
	\centering
	\includegraphics[width=0.5\textwidth]{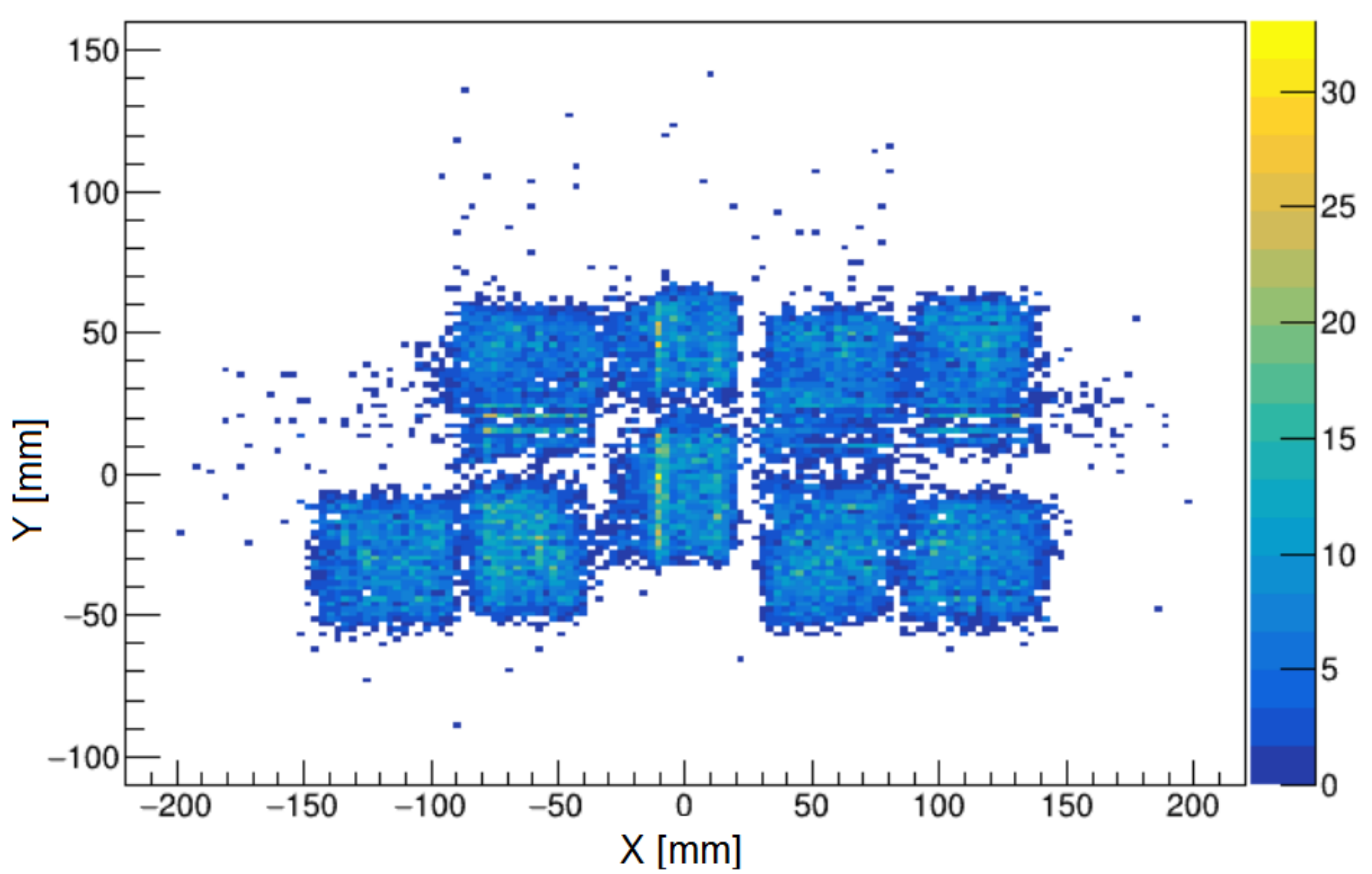}
    \caption{The XZ projection of the forward Si wall.}
    \label{fig:TexATTracksForwardSi}
\end{figure}

\begin{figure}[hbt!]
	\centering
	\begin{subfigure}{.5\textwidth}
  		\centering
  		\includegraphics[width=\linewidth]{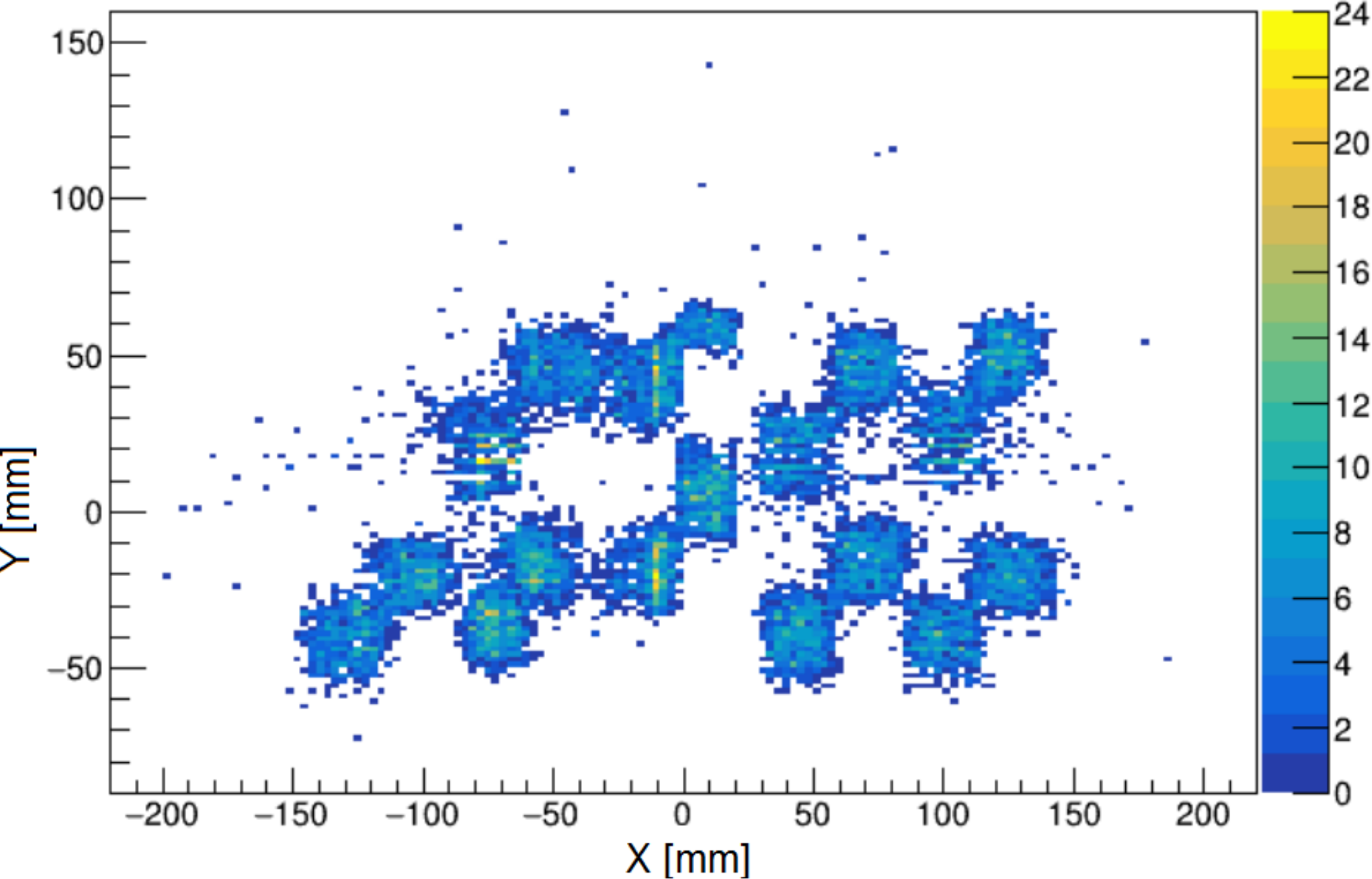}
	\end{subfigure}%
	\\
	\begin{subfigure}{.5\textwidth}
  		\centering
  		\includegraphics[width=\linewidth]{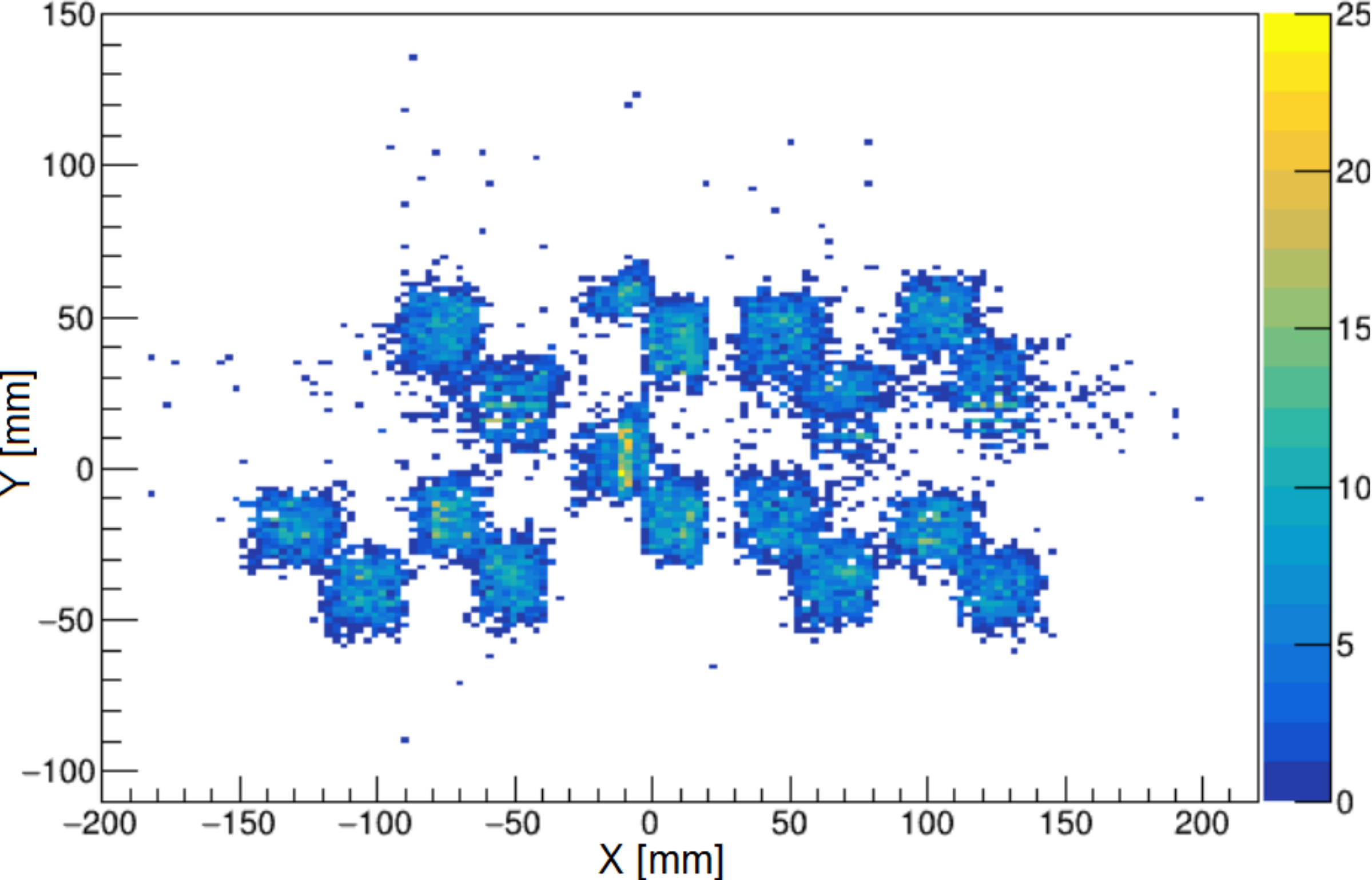}
	\end{subfigure}
	\caption{(Top) XZ projection of the Si wall choosing the lower left and upper right quadrants of each Si detector. (Bottom) 
	XZ projection of the Si wall choosing the upper left and lower right quadrants of each Si detector.}
	\label{fig:TexATTracksForwardSi_Corners}
\end{figure}

Beyond plotting the XY-plane of all the tracks, we also judge how well our track reconstruction performs by plotting the projected end point 
of the track at the location of the Si detectors plane. When fitting the tracks, none of the Si information is used to constrain the track
 reconstruction. The XZ-plan reconstruction of track end point at the location of the Si detectors for the forward wall are shown in 
 Figure \ref{fig:TexATTracksForwardSi}. We clearly see the definitions of all nine Si detectors on the forward wall and also the gap 
 corresponding to the missing Si detector in the top left corner. Although no Si information is used in the track reconstruction, it is
 recorded which Si quadrant fired. By using this information, we can get a better idea to how well the track reconstruction is by only
  plotting the XZ projection if certain quadrants have fired. By plotting the reconstructed track endpoints only for opposite corners of
   the Si detectors a checker-board pattern is achieved, just as expected \ref{fig:TexATTracksForwardSi_Corners}. Very good position 
   reconstruction for each of the quadrants is evident.

\section{Commissioning experiments}

To date we have conducted four types of experiments with TexAT:

\begin{itemize}
	\item Resonance elastic scattering of protons with stable and rare isotope beams - $^{12}$C+p and $^8$B+p
	\item Resonance elastic scattering of $\alpha$-particles with rare isotope beams - $^{10}$C+$\alpha$ and $^{14}$O+$\alpha$
	\item Direct measurements of fusion reaction cross section - $^8$B+$^{40}$Ar
	\item $\beta$-delayed charged particle emission - $^{12}$N $\beta$ decay to $^{12}$C Hoyle state with subsequent emission of 
	three $\alpha$-particles.
\end{itemize}

In this paper we will only focus on describing the experimental procedures for the proton elastic scattering measurements, the 
experiments that were used as commissioning runs for TexAT and are at the most advanced stages of analysis at this time.

For these experiments, stable beams of $^{12}$C and $^6$Li were delivered by the K150 cyclotron accelerator at the Texas A\&M 
University Cyclotron Institute.

Tests with stable beam of $^{12}$C ions have shown that TexAT can handle up to the 2 x 10$^5$ pps with the ``fast" gas (methane). 
For the commissioning experiment the intensity of the stable $^{12}$C beam have been reduced to the ``safe" level of 10$^5$ pps .

 The $^8$B beam was produced using Magnetic Achromat Recoil Separator (MARS) \cite{MARS} with 
the reaction $^6$Li($^3$He,n)$^8$B. The energy of the $^8$B beam was 60 MeV and the intensity was about 1,000 pps.
In addition to the regular TexAT detector components/arrays, we installed a thin scintillator, read out by two PMTs, upstream from the TexAT
 chamber. This detector was used during the run with $^8$B beam to count the beam particles before the TexAT setup and help removing the
  beam contaminants. The scintillator was removed for the run with $^{12}$C (the intensity was about 10$^5$ pps). The PMT signals from the 
  sides of the scintillator were fed to a CFDs with a threshold high enough to trigger only on the $^{8}$B beam particles leaving the beam
   contaminants, primarily $^3$He, below the threshold. The two outputs from the CFD were sent to a logic unit that triggered when both PMT
    signals were above the threshold. The trigger was then sent to a CAEN VME Scalar unit to count the number of beam ions for overall 
    normalization. The CAEN Scalar was placed on its own data acquisition system separate from the GET electronics. A windowless Ionization 
    Chamber (IC) was also placed inside the TexAT chamber, just after the entrance window to count and identify the beam particles. To trigger
     the GET electronics, a coincidence between the external signal from the IC and at least one Si detector channel was required (this is the so-
     called L0L1 mode in GET electronics). This was important to avoid triggering on events associated with $^3$He ions in the beam. Unlike $^8$B, 
     these light ions do not stop in the gas and trigger the Si detector located on the beam axis, however they do not leave enough energy to trigger
      the IC, therefore they are vetoed. The digitizer was set to 25 MHz meaning that the waveforms were recorded every 40 ns and the total 
      timeframe window was 20.48 $\mu$s (512$\times$40 ns). The system was set to zero suppression mode so that only the channels 
      that fired above the threshold were recorded and written to disk, all the other channels were ignored.

\subsection{Beam Particle Identification}

\begin{figure}[hbt!]
	\centering
	\begin{subfigure}{.5\textwidth}
  		\centering
  		\includegraphics[width=\linewidth]{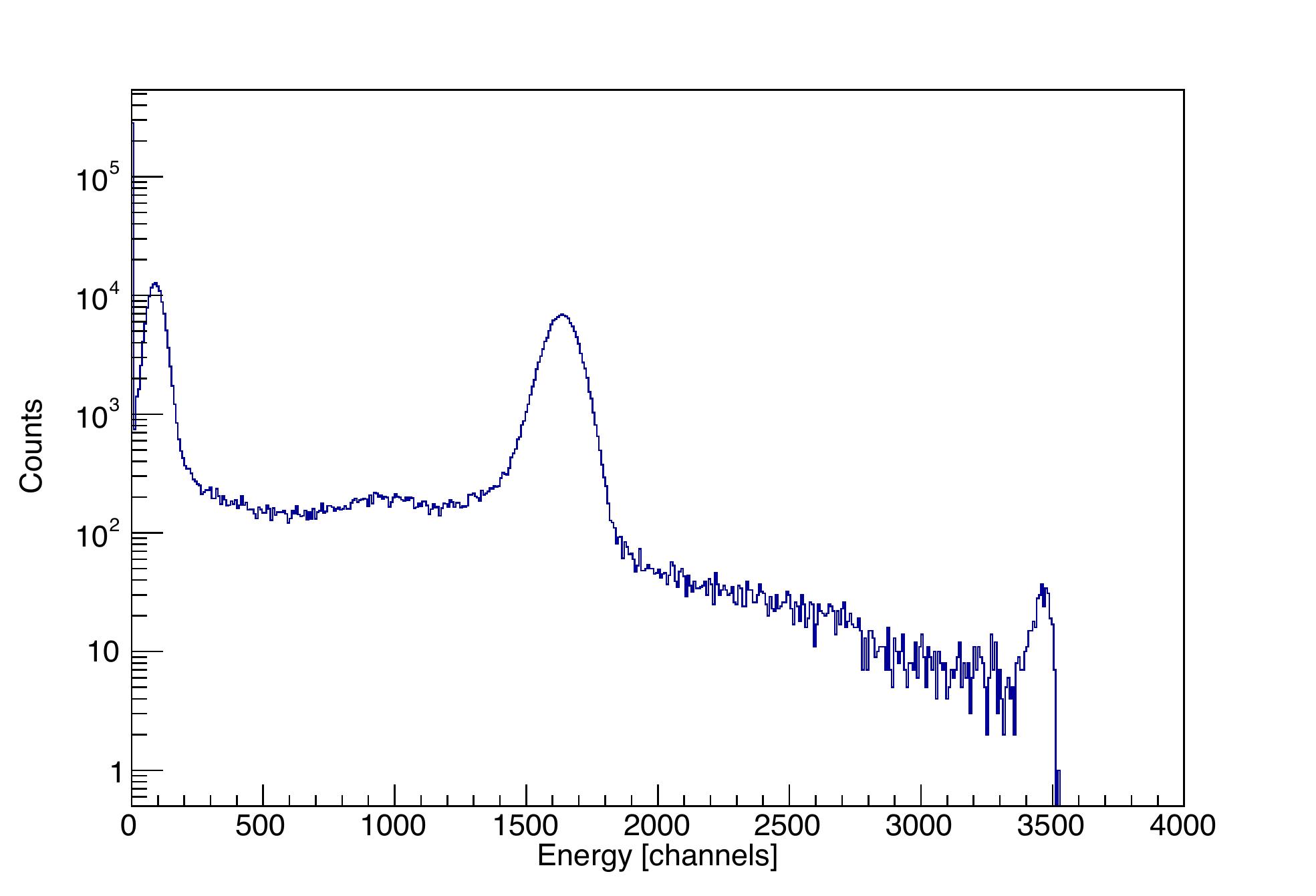}
	\end{subfigure}%
	\\
	\begin{subfigure}{.5\textwidth}
  		\centering
  		\includegraphics[width=\linewidth]{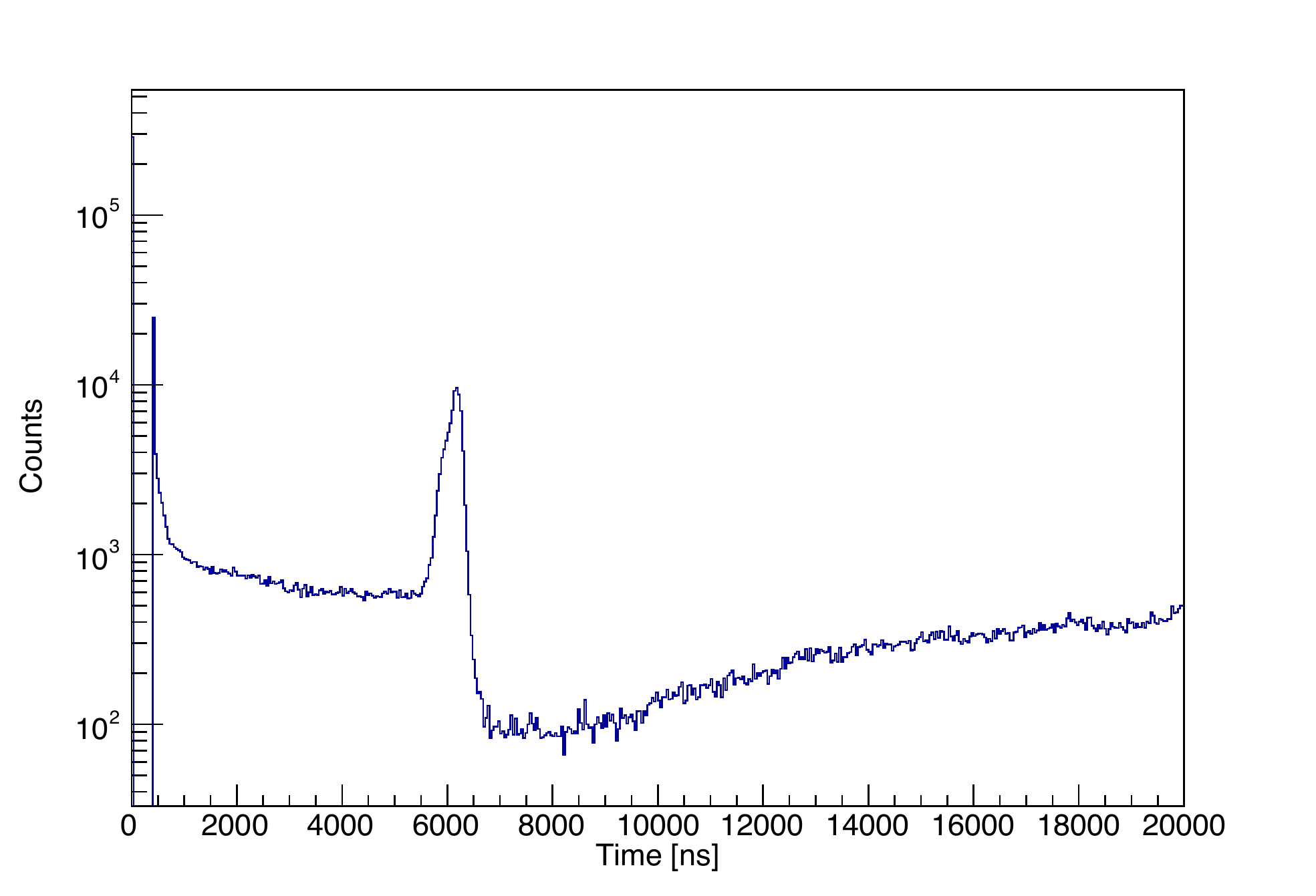}
	\end{subfigure}
	\caption{(Top) The energy deposited in the ionization chamber. The $^{8}$B peak is located between channels 1300-1900. (Bottom) The time of the maximum of the ionization chamber. The beam particles corresponding to the Si hit fall between 5400 ns to 6500 ns.}
	\label{fig:9CIonization}
\end{figure}

In the experiment with the stable beam ($^{12}$C), there were no beam contaminants. In the rare isotope beam experiment the separation of
 the $^{8}$B ions from the small contamination, mostly of $^3$He, was done using the energy and timing from the ionization chamber. By
  measuring the energy deposited in the IC and the time relative to the signal in the Si detector, it was possible to cut out the small amount
   of contamination. Note that the CH$_4$ gas pressure was set to (435 Torr) so that $^8$B ions were stopped long before they could hit the 
   Si detector at zero degrees. Therefore, the correlation between the time of a hit in a Si detector and that of a $^8$B ion in the ionization
    chamber is a signature that a nuclear reaction with $^8$B produced a light recoil. Fig. \ref{fig:9CIonization} shows the energy deposited 
    in the IC and the time relative to the Si detector. For all ``good'' events, the ionization chamber time is mostly determined by the shaping 
    time delay between the ionization chamber signal and that of the Si detectors. We used an external MESYTEC shaper for the ionization chamber
     signal that was then fed directly into the Gain-2 stage of GET electronics, bypassing the internal AGET preamp, shaper and filter, as described
      in section \ref{CsIZAP}. The energy and timing peaks of the $^{8}$B beam in the ionization chamber is clearly seen in Fig. \ref{fig:9CIonization}.
       These two quantities were used to make a cut on the events related to $^{8}$B beam.

\subsection{Tuning of the beam}

\begin{figure}[hbt!]
	\centering
    \includegraphics[width=0.5\textwidth]{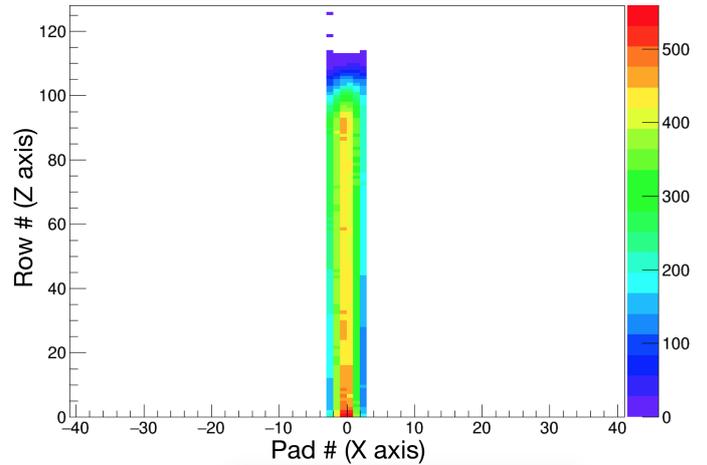}
    \caption{The cumulative counts in the central region of the Micromegas for the $^8$B beam. The pressure was adjusted to stop the 
    beam about 1/8th from the end of the Micromegas.}
    \label{fig:9CBeamCumulative}
\end{figure}

\begin{figure}[hbt!]
	\centering
    \includegraphics[width=0.5\textwidth]{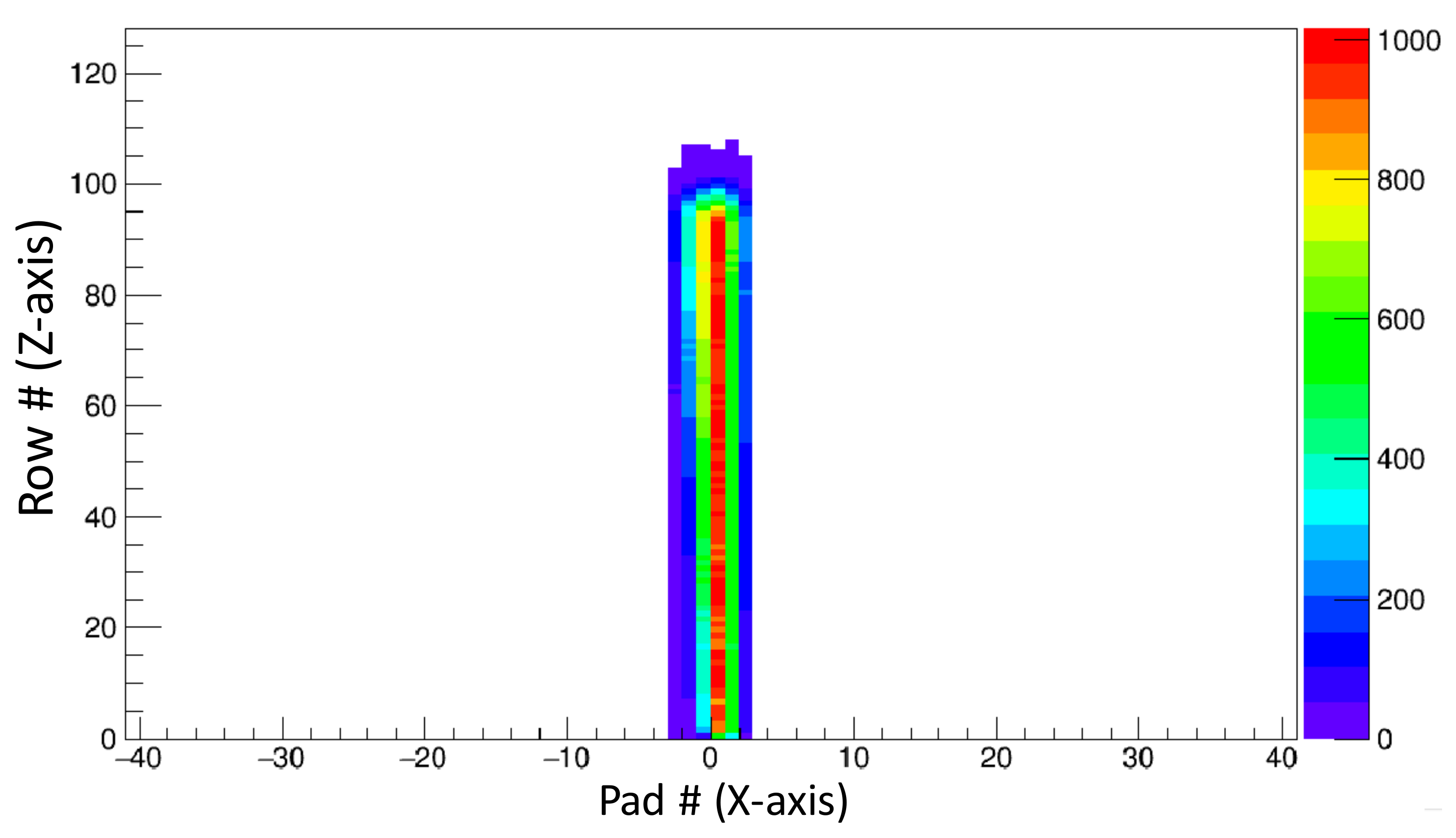}
    \caption{The cumulative counts in the central region of the Micromegas for the $^{12}$C beam. The pressure was adjusted to 
    stop the beam about 1/8th from the end of the Micromegas.}
    \label{fig:12CBeamCumulative}
\end{figure}

\begin{figure}[hbt!]
	\centering
    \includegraphics[width=0.5\textwidth]{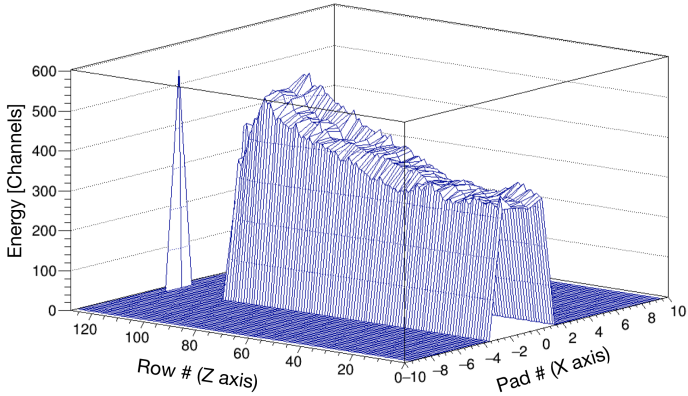}
    \caption{The energy deposited in each of the central region pads. The Bragg peak occurs around row number 90.}
    \label{fig:9CBeamEnergy}
\end{figure}

An important benefit of active target detector is the ability to measure the tracks of the incoming beam ions. This is especially useful when 
tuning the RIB into the chamber to make sure that it is coming in at the correct angle and stopping in the correct location. Shown in 
Figure \ref{fig:9CBeamCumulative}  and Figure \ref{fig:12CBeamCumulative}  are the cumulative counts in the central pads during the tuning process. 
One can see, that  the beam stops 
before the last 1/8th of the pads. Figure \ref{fig:9CBeamEnergy} is the average specific energy loss in the pads. The figure shows the Bragg 
curve which occurs as the beam travels further into the chamber and loses more and more energy, the specific energy loss over each pad 
becomes larger until it reaches the Bragg Peak around row number 90 (out of 128 total). Note that the radioactive beam of $^8$B has much 
wider profile in the scattering chamber, which is defined by the size of the entrance window, while the profile for the stable $^{12}$C beam 
is rather narrow (see Fig. \ref{fig:12CBeamCumulative}).

Figure \ref{fig:9CBeamEnergy} is the average specific energy loss in the pads. The figure shows the Bragg curve which occurs as the 
beam travels further into the chamber and loses more and more energy, the specific energy loss over each pad becomes larger until 
it reaches the Bragg Peak around row number 90 (out of 128 total).

\subsection{Low and High Gain Areas in the Micromegas Detector \label{lowANDhigh}}

The Micromegas detector was split into effectively two regions: one with low gain and another with high gain. The low gain region consists 
of the first 7/8th of the central pads in the beam direction. This low gain region is used to measure the incoming beam particle and the 
energy deposition along the pads. It had a Micromegas bias of 400 V. The last 1/8th of the central pad region closest to the Si detectors
 was biased at 600 V and is a high gain region which was used to measure the protons as the beam is stopped before this region. Both 
 side regions were high gain regions biased at 570 V to obtain the proton tracks.

\subsection{Selecting Proton Events}

\begin{figure}[hbt!]
	\centering
    \includegraphics[width=0.5\textwidth]{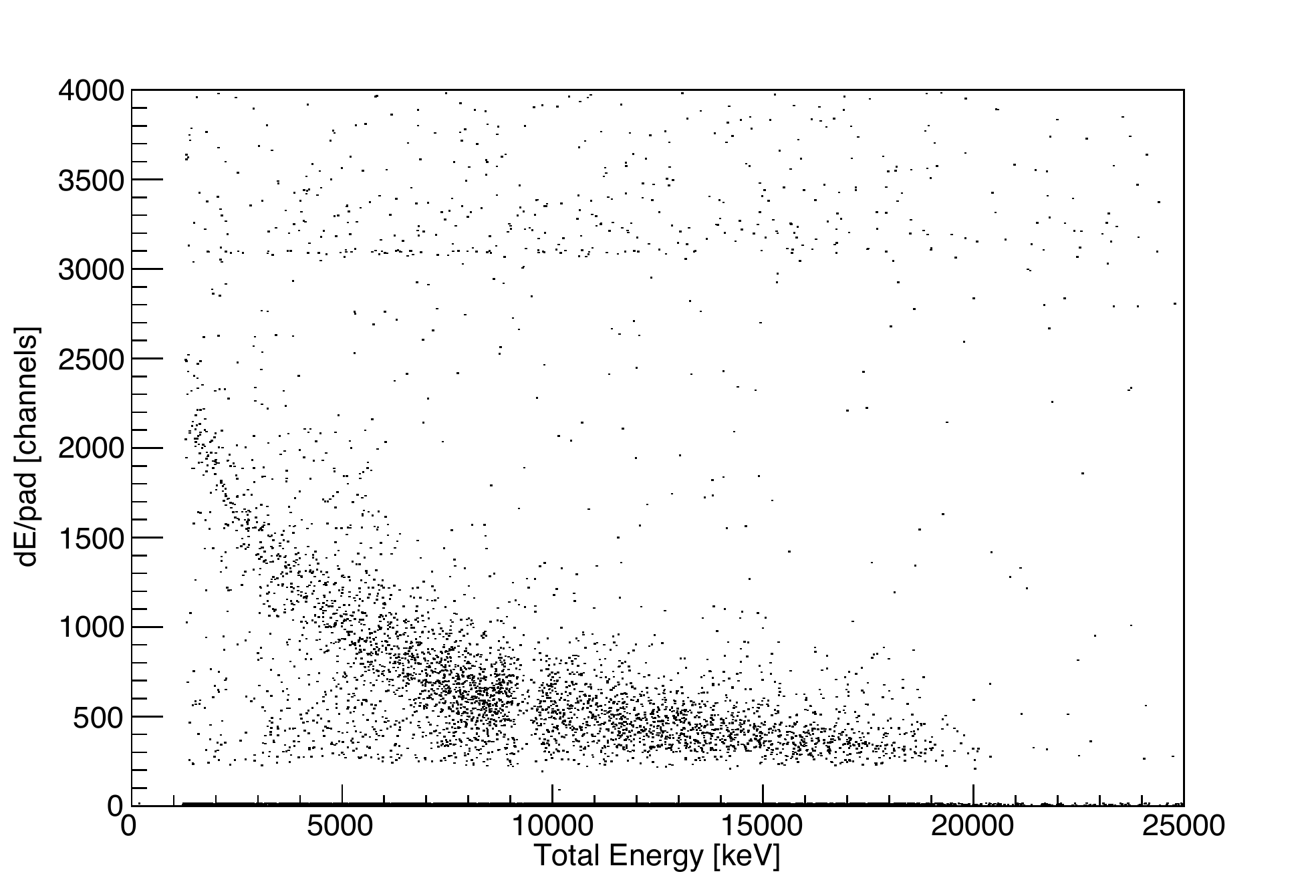}
    \caption{The specific energy loss per unit pad in the central region plotted against the total energy measured in the Si and CsI detector. 
    The energy loss is measured in the last 1/8th in the central pads.}
    \label{fig:9C_dEE5}
\end{figure}

\begin{figure}[hbt!]
	\centering
    \includegraphics[width=0.5\textwidth]{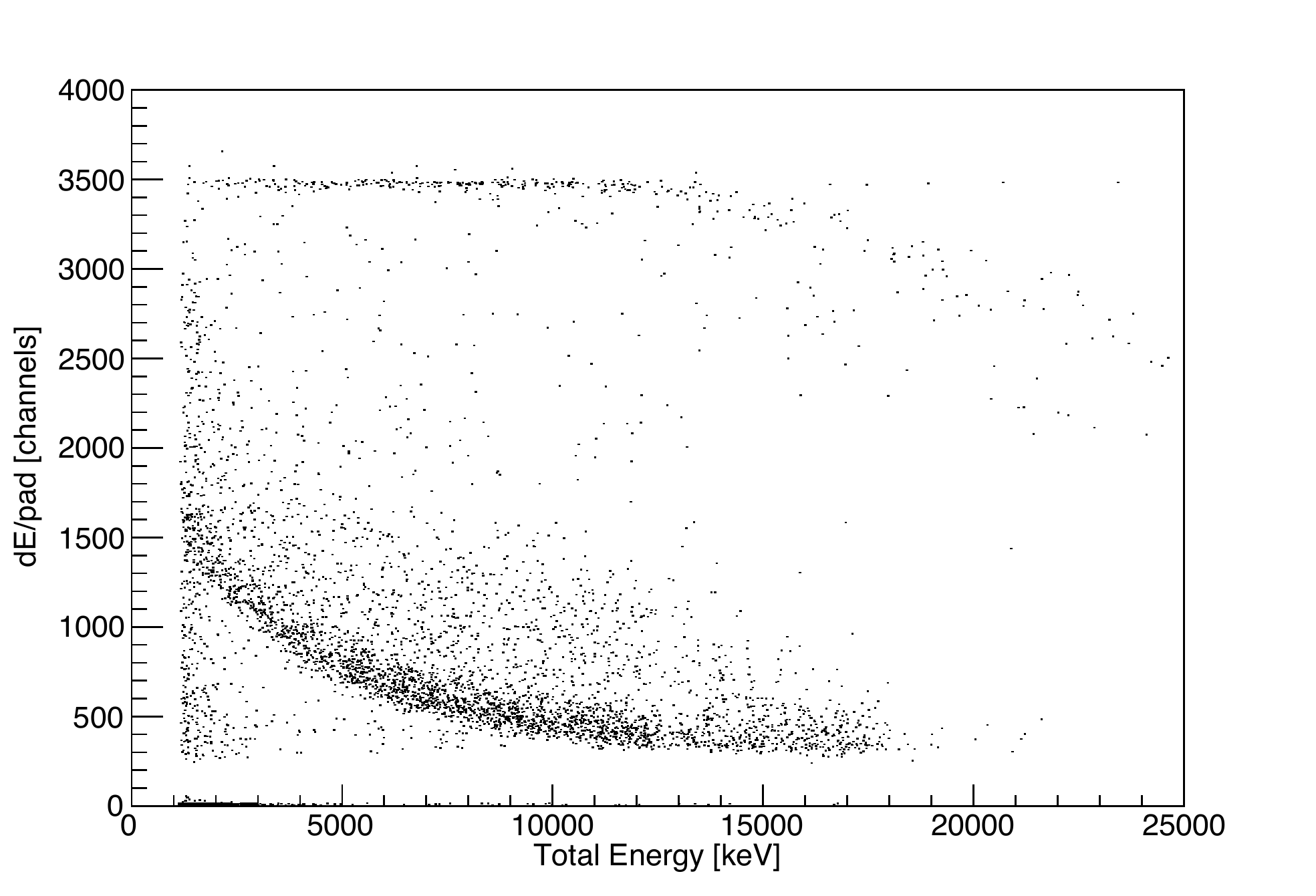}
    \caption{The specific energy loss per unit pad of the strips in the side region plotted against the total energy measured in the Si and CsI detector.}
    \label{fig:9C_dEE9}
\end{figure}

\begin{figure}[hbt!]
	\centering
    \includegraphics[width=0.5\textwidth]{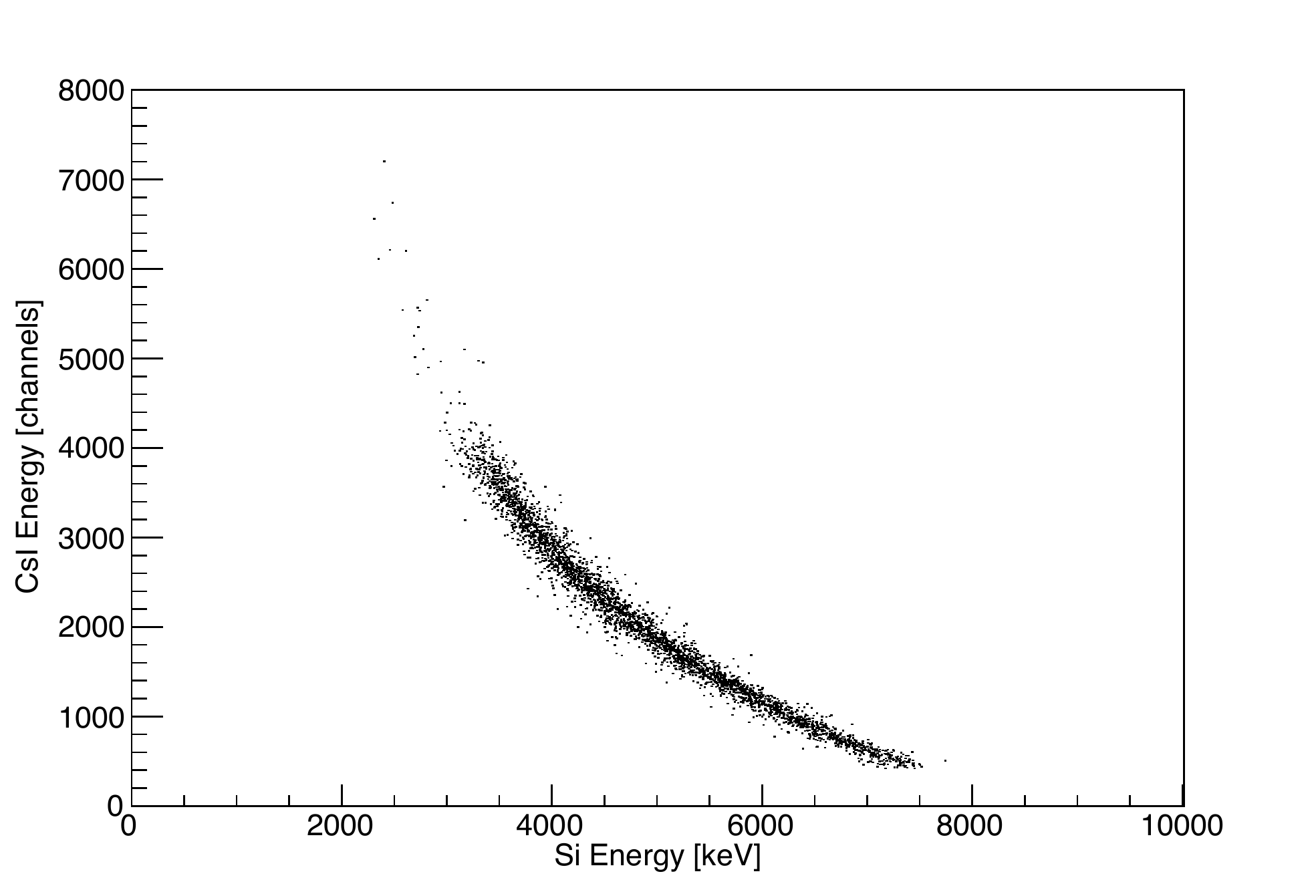}
    \caption{The energy recorded in the CsI detector in channel number plotted against the measured energy(in keV) in the Si detector.}
    \label{fig:TexAT_SiECsIE}
\end{figure}

To identify proton events, we used the specific energy loss in the high gain regions vs the total energy in the Si and CsI detectors. Since 
trajectories have variable path length over the active region of Micromegas detector, we use specific energy loss per unit pad for particle 
ID. The specific energy loss per unit pad for the central detectors plotted against the total energy is shown in Fig. \ref{fig:9C_dEE5}. For the 
side regions, only the energy deposited in the strips are used to get the specific energy loss since the strips run perpendicular to the beam 
axis. This means that protons traveling in the forward direction will not skip over any strips nor will the geometry of the pad prevent the 
full measurement of the energy. The specific energy loss per unit pad plotted against total energy in the side region is shown in 
Fig. \ref{fig:9C_dEE9}. As shown in both of these figures, protons are easily identified. The protons that punch-through the Si 
detectors ($>9$ MeV for 700 $\mu$m of Si) produce a signal in the CsI(Tl) providing additional identification shown in Fig. \ref{fig:TexAT_SiECsIE}.

The gap in Fig. \ref{fig:9C_dEE5} between $\sim 9$ and $\sim 10$ MeV corresponds to the threshold in the CsI detectors. The protons 
in this energy region do not deposit enough energy in the CsI to be recorded due to the threshold for the CsI(Tl) channels in GET electronics. 
This threshold can actually be eliminated if complete readout mode is used in GET, in which case the ``threshold'' would be defined by the 
CsI(Tl) noise level. However, it was not done in the commissioning run. Shown in Fig. \ref{fig:TexAT_SiECsIE} is the Si energy plotted against 
the energy deposition in the corresponding CsI(Tl). As shown, we start recording a signal in the CsI detector when the punch-through 
protons are only depositing less than 7.5 MeV in the Si detector.

\subsection{Identification of inelastic scattering events}

The main focus of the commissioning run was measuring the excitation function for $^8$B+p resonance elastic scattering. However, inelastic scattering 
events were clearly observed. There are no proton-bound excited states in $^8$B because its protons decay threshold is located at 137 keV 
and the first excited states (1$^+$) is at 770 keV. As a result, any inelastic scattering event will produce two protons and a $^7$Be recoil 
(which itself may be in its ground or excited state). Investigating  the events with two proton tracks, we can identify these inelastic scattering 
events.

\begin{figure}[hbt!]
	
  		\includegraphics[width=1.0\linewidth]{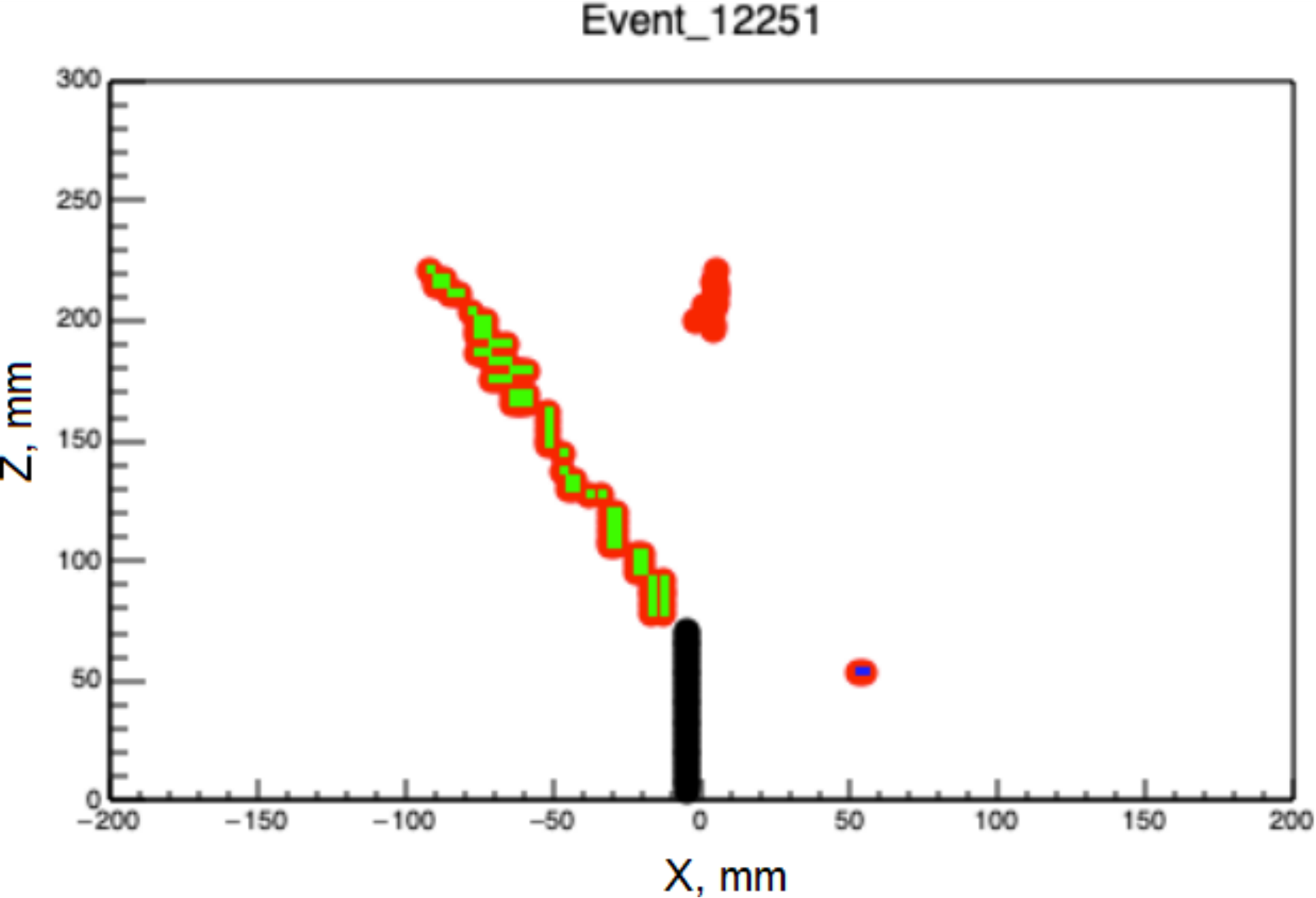}
  		\includegraphics[width=1.0\linewidth]{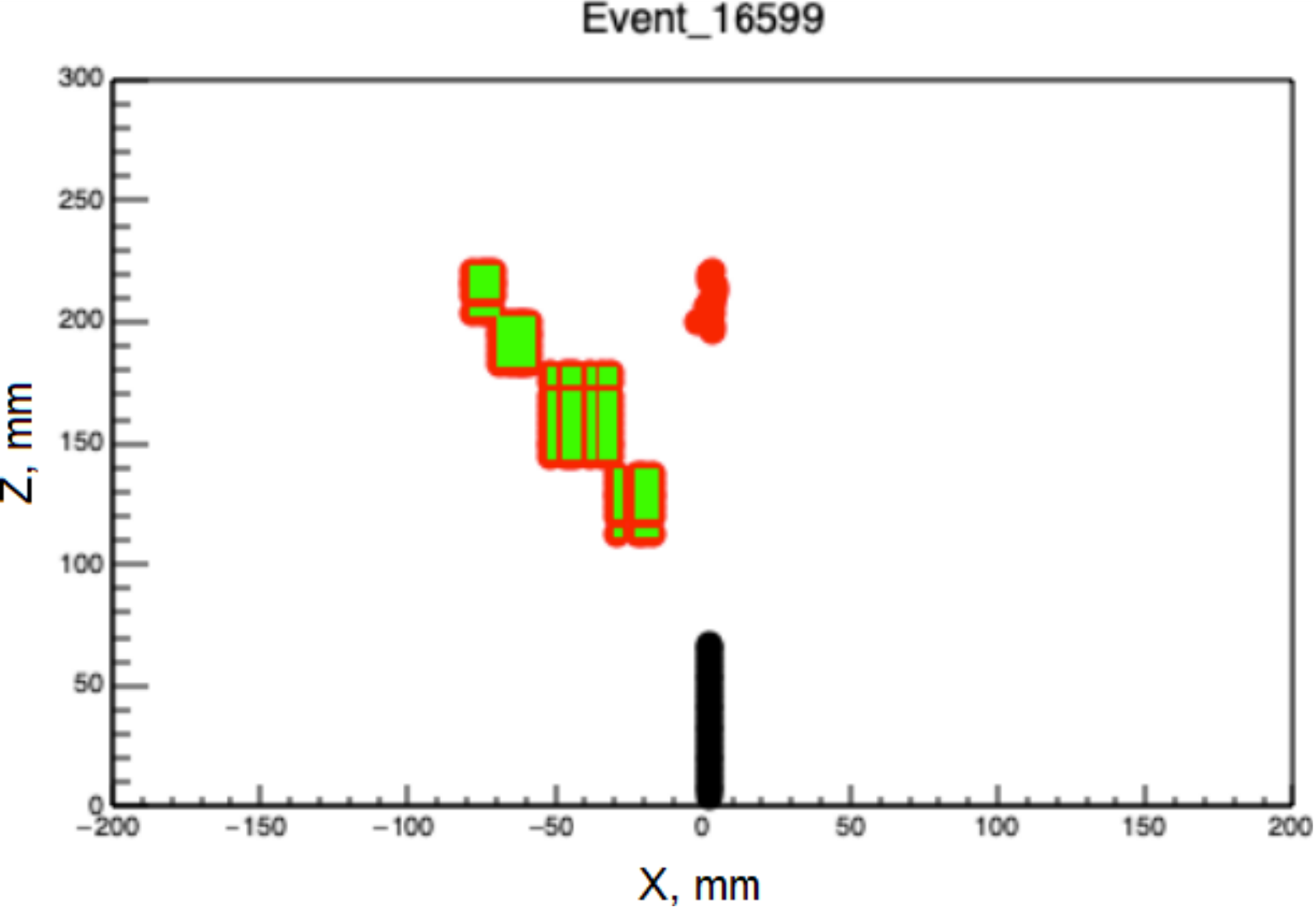}
	\caption{Inelastic events where a proton is measured one side of the Micromegas plate and a proton in the central region.}
	\label{fig:9C_Inelastic_Center}
\end{figure}

\begin{figure}[hbt!]  
	  \includegraphics[width=\linewidth]{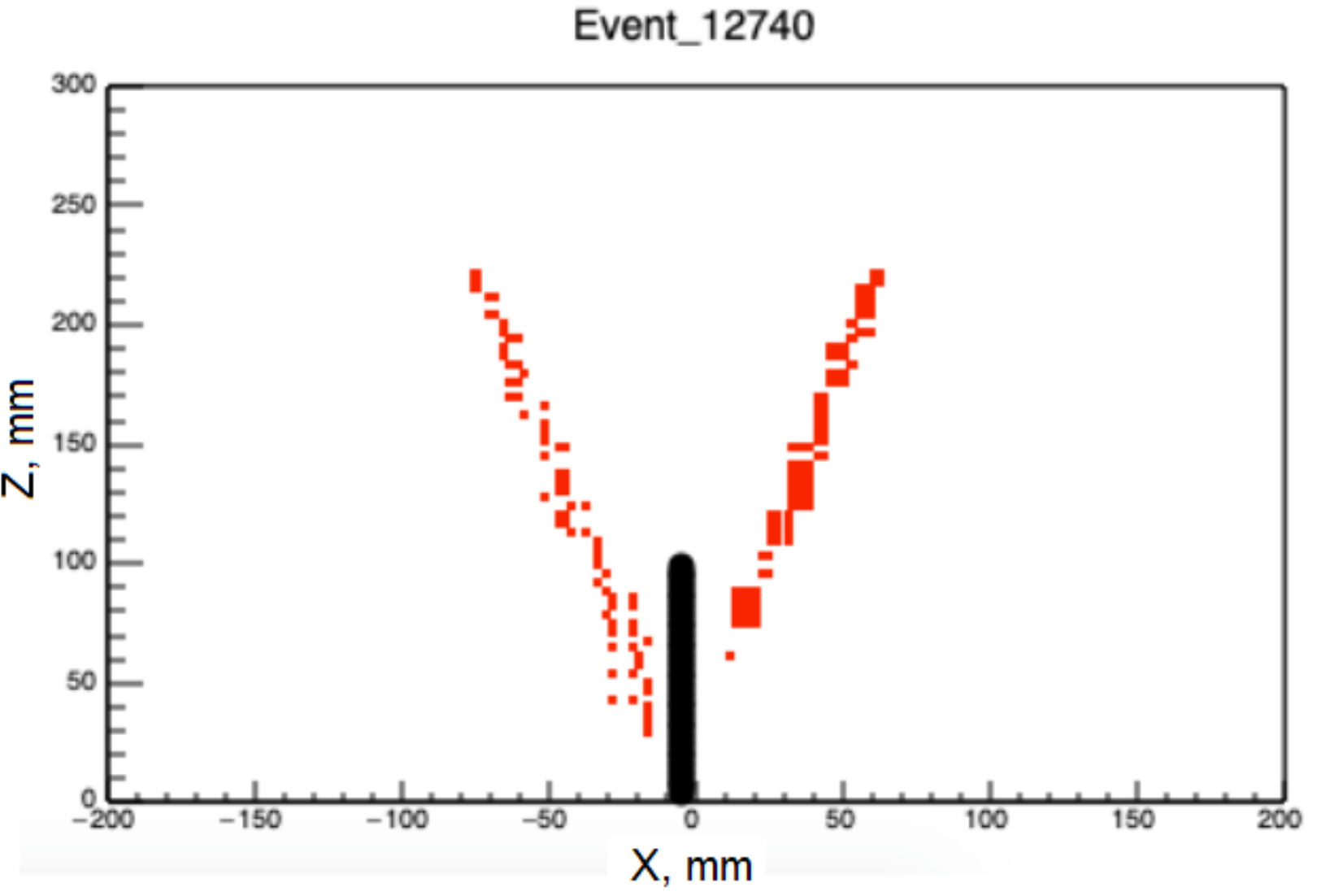}
   	  \includegraphics[width=\linewidth]{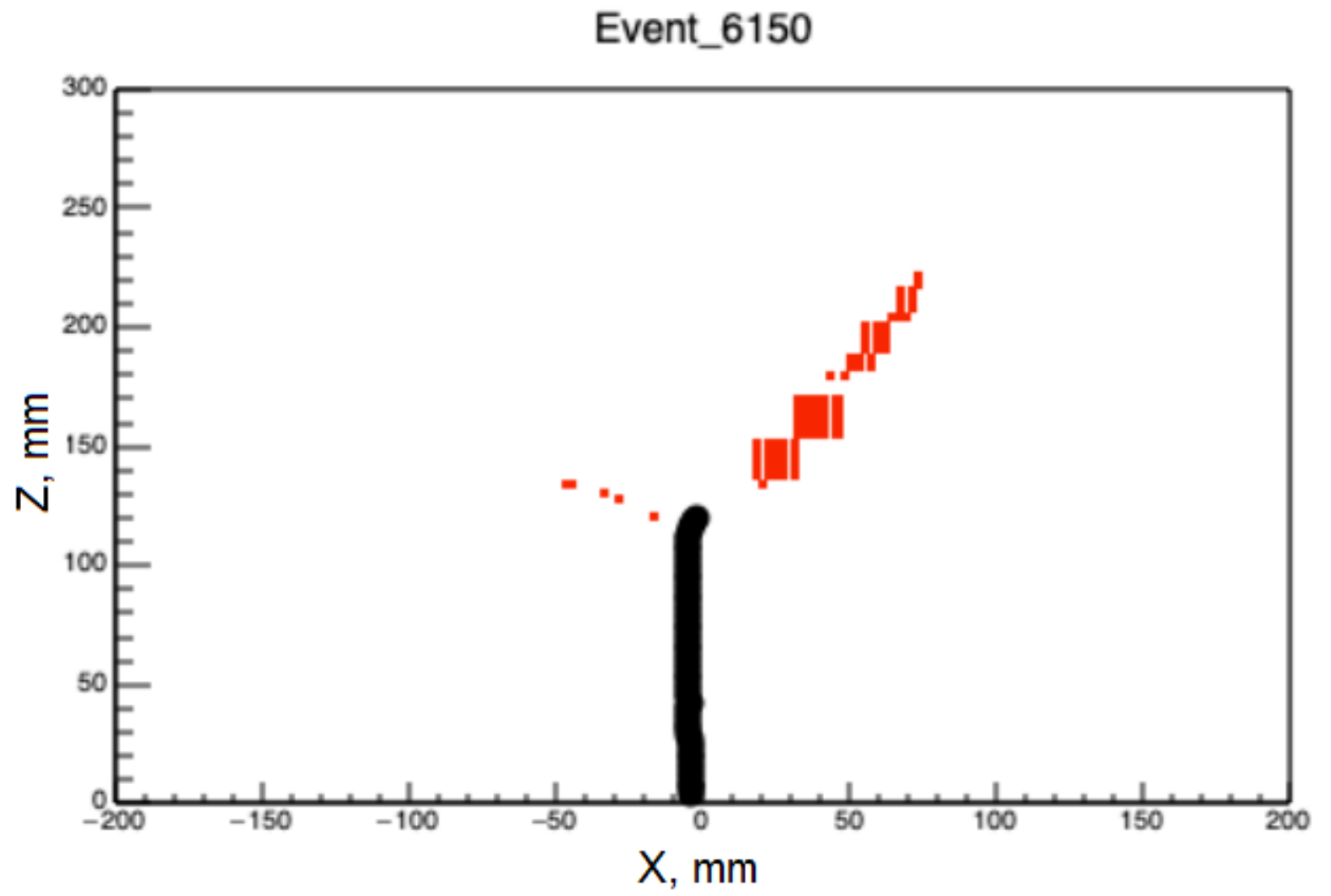}
	\caption{Inelastic events where protons are measured in both sides of the Micromegas plate.}

	\label{fig:9C_Inelastic_Side}
\end{figure}

Using the total energy in the Si and CsI detectors and the Micromegas, we can clearly identify protons that hit one of the Si detectors and 
triggered the DAQ, but ``second'' protons produced by an inelastic scattering event do not always hit a  Si detector. Therefore, we cannot 
rely on specific energy loss vs the total measured energy and need to identify protons by only using their tracks in the gas. The energy loss 
of Z=1 nuclei is very different from that of heavier recoils (e.g. $^7$Be) and there appears to be no or very few deuterons or tritons produced 
in the interaction of $^8$B with the methane gas (see Fig. \ref{fig:9C_dEE5} and \ref{fig:9C_dEE9}). So, by comparing specific energy loss per 
unit pad we can determine if the particle is a heavy recoil or a proton. Examples of protons measured in the side regions and the second 
proton is measured in the central region (Figure \ref{fig:9C_Inelastic_Center}) and opposite side region (Figure \ref{fig:9C_Inelastic_Side}) 
are shown. By identifying these events, we are able to exclude them from the excitation function for resonance elastic scattering.

\subsection{Vertex Reconstruction in the Central Region}

\begin{figure}[hbt!]
  		\includegraphics[width=\linewidth]{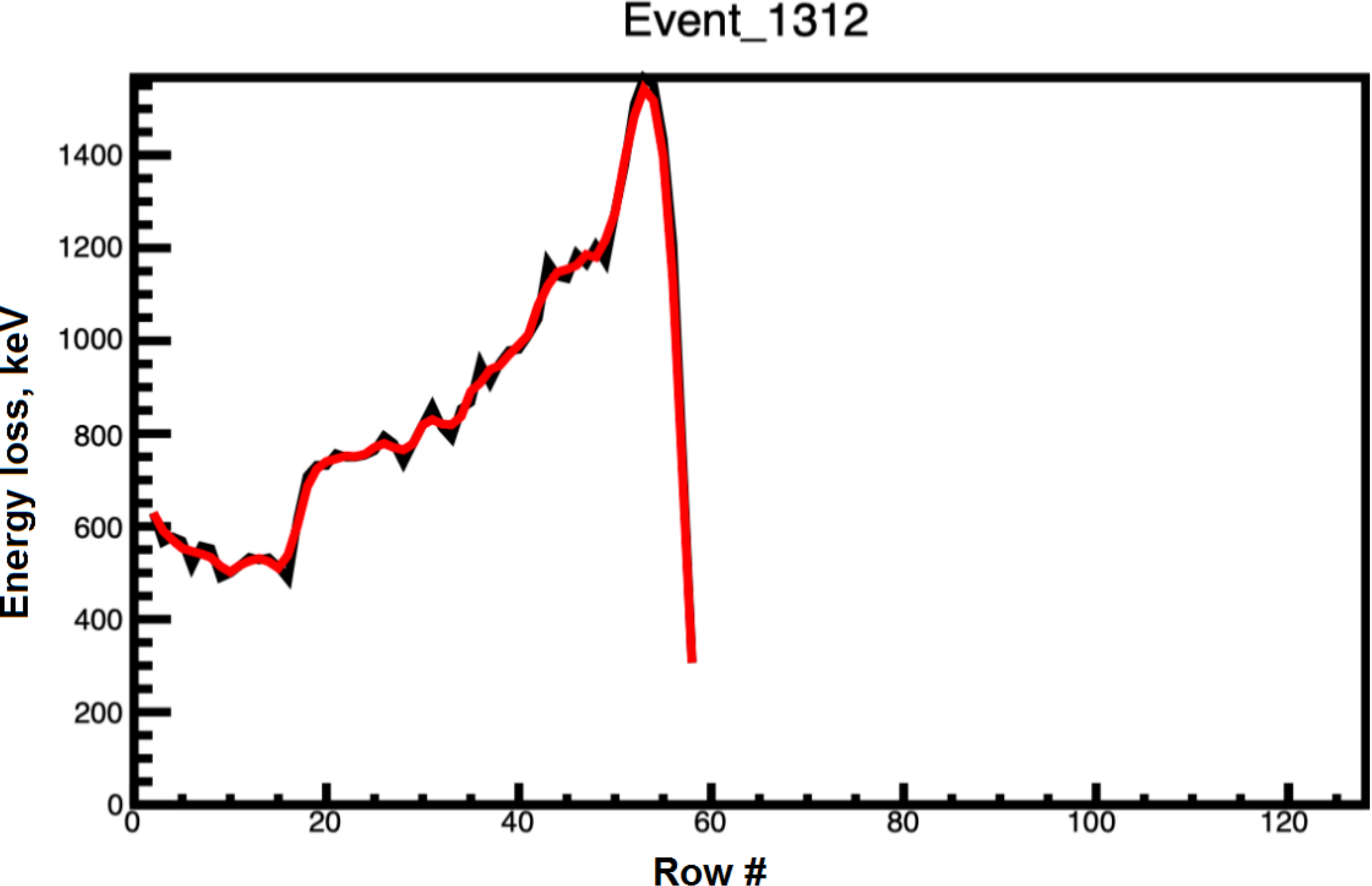}
  		\includegraphics[width=\linewidth]{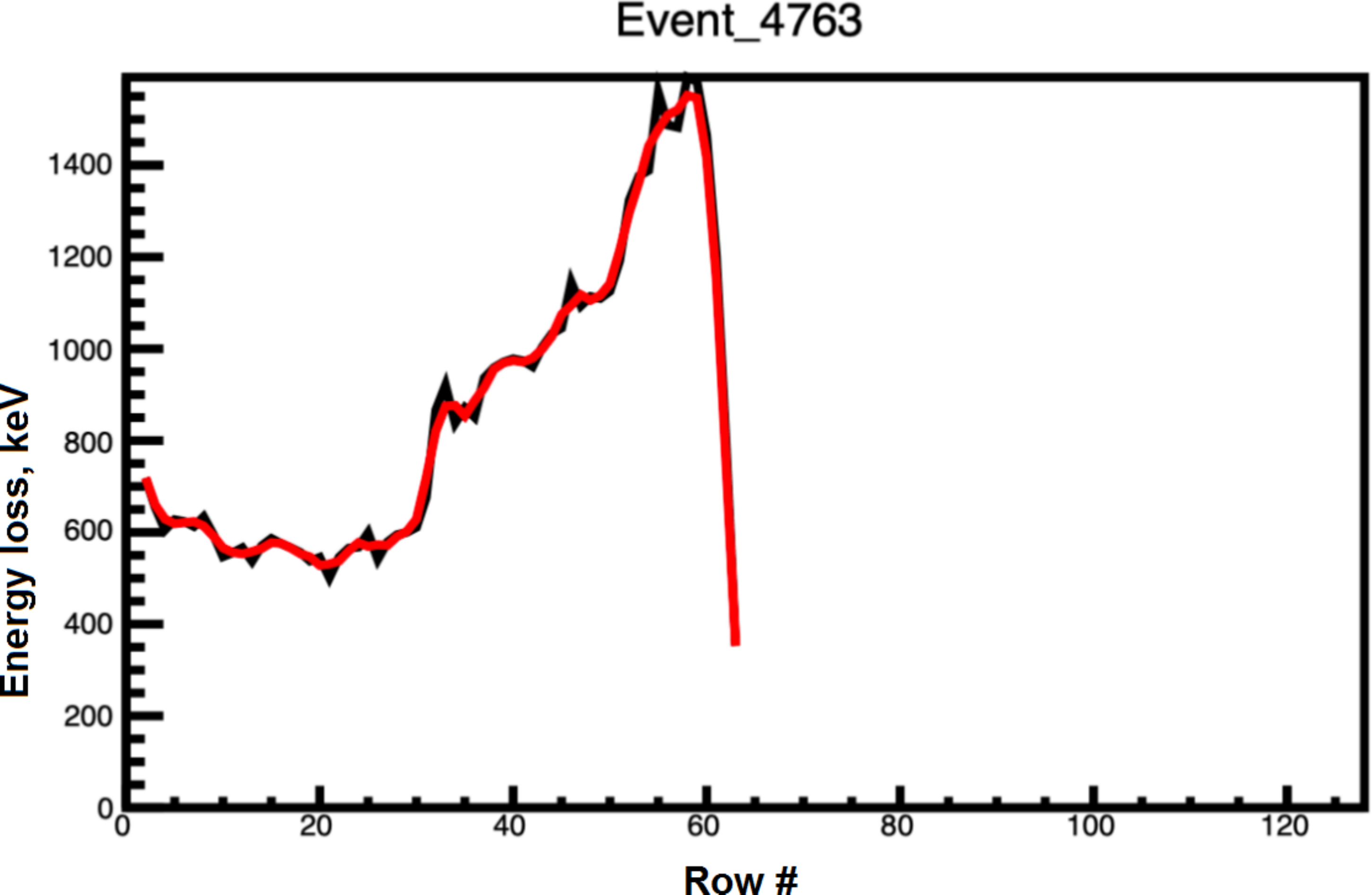}
	\caption{Specific energy loss of the beam and heavy scattered recoil vs row number in Micromegas along the beam axis. Black lines are the 
	raw energy values while the red curve is the running average. (Top) The vertex location is around row 20 while the maximum specific 
	energy loss is around row 50. (Bottom) The vertex location is around row 30 while the maximum specific energy loss is around row 60.}
	\label{fig:9C_Center_Energy_Deposition}
\end{figure}

\begin{figure}[hbt!]
	\centering
    \includegraphics[width=0.5\textwidth]{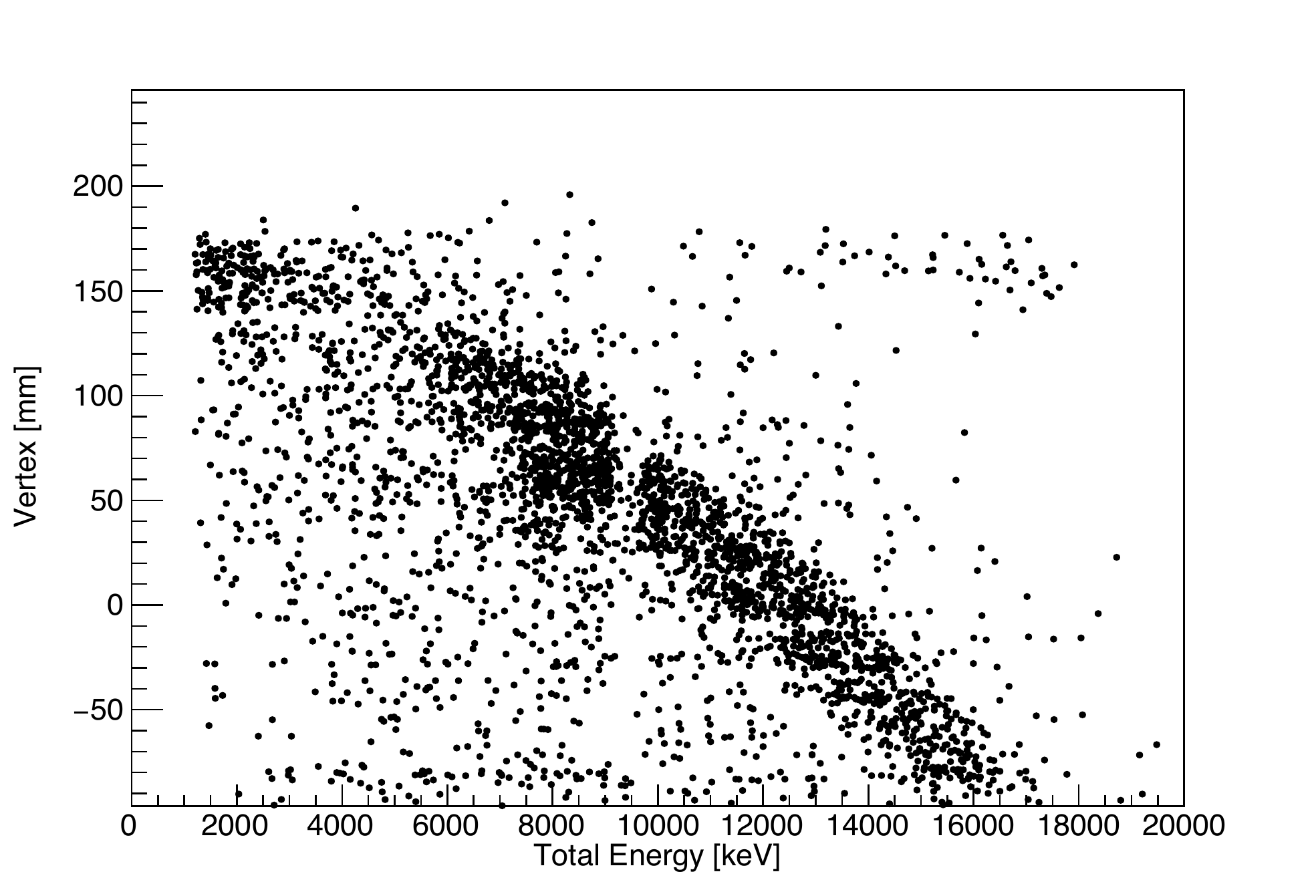}
    \caption{Vertex position vs total energy measured in the Si and CsI detectors for the central forward detectors.}
    \label{fig:9C_VertexReconstructionRegion1}
\end{figure}

The following procedure is used for reconstruction of the event kinematics in the central region.
As the $^{8}$B beam propagates over the active region of Micromegas, it deposits energy by ionizing gas in the region of central pads. 
When a reaction occurs over the Micromegas, such as elastic proton scattering, the energy deposition changes. At this reaction point, 
there is a jump in specific energy loss because the $^{8}$B ion transfers a fraction of its energy to the target proton and therefore specific 
energy loss changes instantaneously at the interaction point. This sudden energy change can be directly observed in the pads as long as 
the vertex is over the Micromegas. In this measurement, only events below $E_{c.m.} = 3.2$ MeV will have a vertex position over the 
Micromegas plate. An alternative way to determine the vertex location for central events and for events produced at c.m. energies above 
$3.2$ MeV is to identify the location of the Bragg peak for a heavy recoil. Since the elastically scattered proton events measured in the 
central detectors travel at angles close to zero degrees, the heavy recoil also travels at angles close to zero (due to momentum conservation). 
This heavy recoil is then measured completely over the central pad region of the Micromegas. The energy of the heavy recoil produced in the 
$^8$B+p elastic scattering depends on the location of the interaction or c.m. energy. These scattered recoils with higher energies travel further 
through the gas and the location of the Bragg peak can be measured. Two typical events of this kind are shown in 
Figure \ref{fig:9C_Center_Energy_Deposition}. By using kinematics and energy loss, we can formulate a way to relate the maximum 
energy deposition in the Micromegas with the vertex location. This is used to determine the vertex position for the events that produced 
a proton in the central region. The vertex position determined this way is plotted against the total energy measured in the Si and CsI in 
Fig. \ref{fig:9C_VertexReconstructionRegion1}. As expected, the vertex position is further away from the Si detectors (vertex position 
going further negative) as the total energy becomes higher. This figure demonstrates that vertex location can be reliably identified even 
if an interaction occurs outside of the active region of Micromegas (negative values along the y axis in Fig. \ref{fig:9C_VertexReconstructionRegion1}).

\subsection{Vertex Reconstruction in the Side Regions}

\begin{figure}[hbt!]
	\centering
    \includegraphics[width=0.5\textwidth]{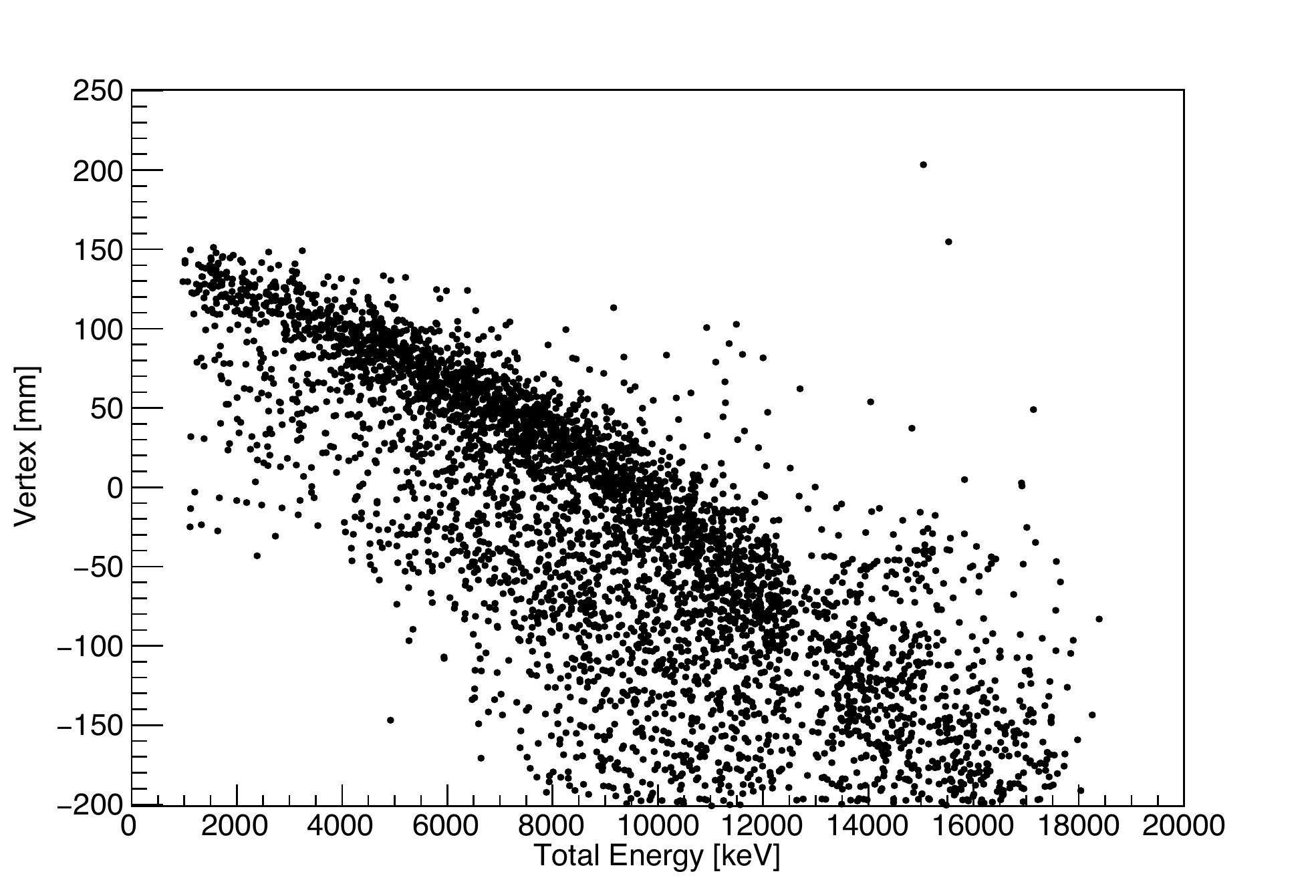}
    \caption{Vertex position vs total energy measured in the Si and CsI detectors for the outside forward detectors.}
    \label{fig:9C_VertexReconstructionRegion3}
\end{figure}

Reconstructing the event tracks in the side regions is straightforward. After matching the strips and chains as discussed above, the tracks 
can then be analyzed using the Hough transform, and traced back to the beam axis to find the vertex location. For events where the vertex 
is over the Micromegas and the incoming beam is measured, the proton track is traced to the measured incoming beam track. For higher 
energy events, the beam information is not measured in the Micromegas detector and the proton track is traced back to the ``ideal'' beam 
axis. The measured vertex position vs total energy is shown in Fig. \ref{fig:9C_VertexReconstructionRegion3}. As in the case of the central 
region, elastically scattered events occur further away for high energy events and thus higher c.m. energy events.

\section{Conclusion}

The new active target detector system - Texas Active Target - has been designed and constructed at the Cyclotron Institute at Texas A\&M 
University. The detector consists of the Time Projection Chamber readout by highly segmented Micromegas detector that provides high gas 
gains. The TPC is surrounded on five sides by a shell of Si detectors backed by the CsI(Tl) scintillators, readout out by the Si pin-diodes. 
The detector was first tested with an $\alpha$-source. The commissioning experiments with the stable $^{12}$C beam and the rare isotope 
$^8$B beam were performed. Parameters and properties of TexAT, the TexAT simulation package and simulation results and also track 
reconstruction procedures are described in this paper. Physics results of the first commissioning run, in which structure of exotic nucleus 
$^9$C was studied, are discussed in \cite{JHooker}.
 
\section*{Acknowledgements}

TexAT project was supported by the U.S. Department of Energy, Office of Science, Office of
Nuclear Science, under Award No. DE-FG02-93ER40773 and by National Nuclear Security Administration through the Center for Excellence 
in Nuclear Training and University Based Research (CENTAUR) under grant number \#DE-NA0003841.
The author G.V.R. is also acknowledge the support of the Welch Foundation (Grant. No. A-1853).

The authors are grateful to Bertrand Mehl,  Rui de Oliviera and the detector development team at CERN for their technical support. We thank 
Yuriy Penionzhkevich (JINR), M.I.Gazizov and detector group  at the Institute in Physical-Technical Problems (Dubna, Russia) for their collaboration
 to develop custom Silicon detectors for TexAT.
The authors are deeply thankful the Electronics and Detector Division  of IRFU CEA (Saclay):  Anvar Shebli, Patrick Baron, J\'er\^ome Pibernat
and  Patric Sizun for helpful discussions and their support of Micromegas and GET electronics. The authors also grateful to  Gilles Wittwer (GANIL) for useful advice.

\bibliography{NIM_TexAT}

\begin{thebibliography}{10}

\bibitem{Goldberg}
V.~Z. {Goldberg} and A.~E. {Pakhomov}.
\newblock {\em Yad.Fiz.(Rus.)}, \textbf{56}:9, 1993.

\bibitem{TACTIC}
A.M.~Laird \textit{et al.}
\newblock {\em Nuclear Instruments and Methods in Physics Research},
  \textbf{A573}:306, (2007).

\bibitem{MAYA}
C.~E.~{Demonchy} \textit{ et al.}
\newblock {\em Nuclear Instruments and Methods in Physics Research},
  \textbf{A573}:145--148, 2007.

\bibitem{AT-TPC}
D.~{Suzuki} \textit{et al.}
\newblock {\em Nuclear Instruments and Methods in Physics Research}, \textbf{A
  691}:39--54, 2012.

\bibitem{ACTAR}
C.F.~{Ginyer} \textit {et al.}
\newblock {\em SPIRAL 2 Week}, 2011.

\bibitem{ANASEN}
E.~{Koshchiy} \textit {et al.}
\newblock {\em Nuclear Instruments and Methods in Physics Research}, \textbf{A
  870}:1 -- 11, 2017.

\bibitem{MUSIC}
W.~{Christie} \textit{et al.}
\newblock {\em Nuclear Instruments and Methods in Physics Research}, \textbf{A
  255}:466--476, 1987.

\bibitem{MSTPC}
Y.~{Mizoi} \textit{et al.}
\newblock {\em Nuclear Instruments and Methods in Physics Research}, \textbf{A
  431}:112, 1999.

\bibitem{Cwiok}
\'{C}wiok \textit{et al.}
\newblock Optical time projection chamber for imaging of two-proton decay of
  $^{45}${Fe} nucleus.
\newblock {\em IEEE Trans. Nucl. Sci.}, \textbf {52}:2895, 2005.

\bibitem{Gai}
M.~{Gai} \textit{et al.}
\newblock An optical readout {TPC (O-TPC)} for studies in nuclear astrophysics
  with gamma-ray beams at {HI}${\gamma}${S}.
\newblock {\em Jour. Instr.}, \textbf{5}:12004, 2010.

\bibitem{Beceiro}
D.~{Bazin} S.~{Beceiro-Novo}, T.~{Ahn} and W.~{Mittig}.
\newblock Active targets for the study of nuclei far from stability.
\newblock {\em Progress in Particle and Nuclear Physics}, \textbf {84}:124 --
  165, 2015.

\bibitem{TPC_Review}
D.~{Gonz\'{a}lez-D\'{i}az} \textit {et al.}
\newblock {\em Nuclear Instruments and Methods in Physics Research}, \textbf{A
  878}:200--255, 2018.

\bibitem{Ayyad2018}
Y.~{Ayyad} \textit {et al.}
\newblock {\em The European Physical Journal A}, \textbf{54}(10):181, 2018.

\bibitem{MARS}
R.E. {Tribble}, R.H. {Burch}, and C.~A. {Gagliardi}.
\newblock {MARS: A Momentum Achromat Recoil Separator}.
\newblock {\em Nuclear Instruments and Methods in Physics Research}, \textbf
  {A285}:441--446, 1989.

\bibitem{SiLi}
G.A.~{Shishkina} L.A.~{Grigorieva} A.G.~{Artyukh} Yu.G.~{Teterev}
  L.A.~{Popeco}, I.M.~{Kotina} and Yu.M. {Sereda}.
\newblock Thick {S}i({L}i) coaxial detectors for registration of intermediate
  energy heavy ions.
\newblock {\em Nuclear Instruments and Methods in Physics Research}, \textbf{A
  596}:235--237, 2008.

\bibitem{Micromegas_1}
Y.~{Giomataris} \textit {et al.}
\newblock {MICROMEGAS: A High granularity position sensitive gaseous detector
  for high particle flux environments}.
\newblock {\em Nuclear Instruments and Methods in Physics Research.}, \textbf{A
  376}:29--35, 1996.

\bibitem{Micromegas_2}
S.~{Andriamonje} \textit {et al.}
\newblock {\em JINST 5}, page P02001, 2010.

\bibitem{Blum}
W.~{Blum} and L.~{Rolandi}.
\newblock {\em Particle Detection with Drift Chambers}.
\newblock Springer-Verlag, 1964.

\bibitem{GIOMATARIS2006}
I.~Giomataris at~al.
\newblock {\em Nuclear Instruments and Methods in Physics Research Section A:
  Accelerators, Spectrometers, Detectors and Associated Equipment}, 560(2):405
  -- 408, 2006.

\bibitem{Raether}
Heinz {Raether}.
\newblock {\em Electron avalanches and breakdown in gases}.
\newblock Butterworths, 1964.

\bibitem{Peskov}
V.~{Peskov} \textit{et al.}
\newblock The study and optimization of new micropattern gaseous detectors for
  high-rate applications.
\newblock {\em IEEE Trans. Nucl. Sci.}, \textbf {48}(4), 2001.

\bibitem{GMSH}
C.~{Geuzaine} and J.-F. {Remacle}.
\newblock {GMSH}: a three-dimensional finite element mesh generator with
  built-in pre- and post-processing facilities.
\newblock {\em International Journal for Numerical Methods in Engineering},
  \textbf{79}, pages =, 2009.

\bibitem{GARFIELD}
R.{Veenhof}.
\newblock {GARFIELD}: computer code.
\newblock 1998.

\bibitem{Elmer}
Elmer: computer code.
\newblock {\em http://www.csc.fi/elmer}, 2005.

\bibitem{MAGBOLTZ}
{MAGBOLTZ}: computer code, v.11.4.

\bibitem{Micron}
Micron {S}emiconductor {UK}, 2018 catalog.
\newblock {\em http://www.micronsemiconductor.co.uk}, 2018.

\bibitem{SiDetDubna}
I.M. {Gazizov}, I.V. {Nikonorov}, and A.A. {Smirnov}.
\newblock Silicon four-segment charged particle detectors.
\newblock {\em {EXON2018}, Book of {A}bstracts}, page 113, 2018.

\bibitem{Pollacco}
E.C.~Pollacco \textit {et al.}
\newblock {GET}: A generic electronics system for {TPC}s and nuclear physics
  instrumentation.
\newblock {\em Nuclear Instruments and Methods in Physics Research}, \textbf
  {A887}:81 -- 93, 2018.

\bibitem{DesignSpark}
{\em https://www.rs-online.com/designspark/pcb-software}, 2017.

\bibitem{GET}
{GET-AS-002-0009}.
\newblock {\em {CENBG CNRS} Technical Manual}.

\bibitem{MESYTEC}
{MSCF-16 F} {S}haping {A}mplifier.
\newblock {\em https://www.mesytec.com/products/nuclear-physics}.

\bibitem{OpAmp}
{LM6154} {Q}uad {O}perational {A}mplifier.
\newblock {\em Texas Instruments Incorporated
  (http://www.ti.com/lit/ds/symlink/lm6152.pdf)}.

\bibitem{Geant4}
S.~{Agostinelli} \textit {et al.}
\newblock Geant4: A simulation toolkit.
\newblock {\em Nuclear Instruments and Methods in Physics Research}, \textbf{A
  506}:250--303, 2003.

\bibitem{Duda:1972}
Richard~O. Duda and Peter~E. Hart.
\newblock Use of the hough transformation to detect lines and curves in
  pictures.
\newblock {\em Commun. ACM}, \textbf{15}(1):11--15, January 1972.

\bibitem{JHooker}
J.~{Hooker} \textit{et al.}
\newblock Structure of $^{9}\mathrm{C}$ through proton resonance scattering
  with the texas active target detector.
\newblock {\em Phys. Rev. C}, \textbf{100}:054618, 2019.

\end{thebibliography}

\end{document}